\documentclass{aa}  

\usepackage[colorlinks=true, linkcolor = blue,
            urlcolor  = blue,
            citecolor = blue,
            anchorcolor = blue, bookmarks=false]{hyperref}
\usepackage{graphicx}
\usepackage{txfonts}
\usepackage{natbib}
\usepackage{enumitem} 
\usepackage{color}
\usepackage{threeparttable}
\usepackage{tikz,amsmath}
\usetikzlibrary{shapes.geometric, arrows}
\usepackage{ulem}
\usepackage{wrapfig}
\usepackage{rotating}
\usepackage{lscape} 
\tikzstyle{process} = [rectangle, minimum width=1.5em, minimum height=3.5em, text centered, draw=blue, fill=gray!10]
\tikzstyle{process2} = [rectangle, minimum width=1.5em, minimum height=3.5em, text centered, draw=white, fill=white]
\tikzstyle{arrow} = [thick,->,>=stealth]
\usepackage{cancel}
\usepackage{ulem}
\bibpunct{(}{)}{;}{f}{}{,} 

\newcommand{\teff}{$T_{\mathrm{eff}}$}

\newcommand{\logg}{\mbox{log \textit{g}}}

\newcommand{\titan}{\textsc{Titan}}

\begin{document}

  \title{
    Observational signatures and constraints on the intermediate neutron-capture psrocess%
}
\subtitle{The case of the carbon-enhanced metal-poor star TYC~6044$-$714$-$1 (RAVE J094921.8-161722)}

   \author{R. E. Giribaldi\inst{1}
          \and
          D. Vescovi\inst{2,3}
          \and
          L. Magrini\inst{1}
          \and
          S. Cristallo\inst{2,3}
          \and
          V. D\textquotesingle Orazi\inst{4}
          \and
          L. Piersanti\inst{2,3}
          \and
          D. Cornejo Espinoza\inst{5}
          \and 
          S. Randich\inst{1}
          \and 
          M. Baratella\inst{6}
          }
   \institute{
        INAF – Osservatorio Astrofisico di Arcetri, Largo E. Fermi 5, 50125 Firenze, Italy \and 
        INAF – Osservatorio Astronomico d'Abruzzo, Via M. Maggini, 64100 Teramo, Italy \and
        INFN - Sezione di Perugia, Via A. Pascoli, 06123 Perugia, Italy \and
        Department of Physics, University of Rome Tor Vergata, via della Ricerca Scientifica 1, 00133 Roma, Italy \and
        Escuela Nacional de Estudios Superiores Unidad Morelia, Universidad Nacional Aut\'onoma de M\'exico, Morelia, 58190, M\'exico \and
        ESO - European Southern Observatory, Alonso de Cordova, 3107, Vitacura, Santiago, Chile\\
        \email{riano.escategiribaldi@inaf.it}
             }
             
    \date{Received  / Accepted }

 
  \abstract
{Observational abundances of carbon enhanced metal-Poor (CEMP) stars with patterns in between those produced by the rapid (r) and slow (s) nucleosynthesis processes ({CEMP-rs} stars) are currently invoked as evidence of synthesis via the intermediate (i) process in the early AGB evolutionary phase of metal-poor low-mass stars.
Nevertheless, discriminating between r+s- and i-process hypotheses requires high-precision abundances obtained through advanced spectral modelling techniques.
Theoretical models of the i-process have become more robust, incorporating refined stellar modelling and nuclear reaction physics, providing ranges of probable elemental abundances and isotopic ratio predictions to be confronted with observational determinations. }
{We performed a new analysis of a high-resolution and high-signal-to-noise UVES spectrum of TYC~6044-714-1, one of the best-studied {CEMP-rs} stars.}
{We derived accurate effective temperature (\teff) and highly precise atmospheric parameters, element abundances, and isotopic ratios using state-of-the-art 
one-dimensional non-local thermodynamic equilibrium (1D~non-LTE) and three-dimensional non-LTE (3D~non-LTE) spectral line modelling.
Using the latest AGB nucleosynthesis models computed with the FuNS evolutionary code, we assessed the possibility of the i-process to act aside the s-process.}
{{We find that TYC~6044-714-1 was likely born as a normal in situ halo star about 13~Gyr ago,} pre-enriched by the r process through a standard Galactic chemical-evolution pathway. Among the explored scenarios, the s$+$r model provides the best overall reproduction of the observed heavy-element abundance pattern and Ba isotopic ratios, yielding excellent agreement across all three s-process peaks. While i$+$s$+$r models with increasing overshooting efficiency improve the fit for specific elements -- particularly Nb and those between the first and second neutron-capture peaks -- they do not consistently reproduce the full abundance pattern.
}  
{Although the i+s+r models achieve statistical fits comparable to the s+r case, they require extreme and physically implausible conditions, and predict s-process Ba fractions inconsistent with those inferred from isotopic ratios of the 4934~\AA\ resonance line. We therefore conclude that the pure s+r scenario is the most plausible explanation.}



   \maketitle
%

\section{Introduction}

The synthesis of elements heavier than iron in stars is governed by neutron-capture processes, primarily classified as the slow (s-) and rapid (r-) processes, depending on the relative timescales of neutron captures and $\beta$-decays. However, observations of certain stellar populations and presolar grains reveal abundance patterns that cannot be fully explained by either of these canonical processes, pointing to the existence of an intermediate neutron-capture process, or i-process \citep{cowan1977ApJ...212..149C}. The i-process occurs under neutron densities between those that are characteristic of the s and r processes, typically $10^{13} - 10^{15}\,\mathrm{cm^{-3}}$ \citep[e.g.][]{Banerjee2018ApJ...865..120B, clarkson2018MNRAS.474L..37C}, producing a distinct nucleosynthetic signature.

Proposed sites for i-process nucleosynthesis include He-shell flashes in low-metallicity asymptotic giant branch (AGB) stars, proton ingestion events in early-generation stars, and rapidly accreting white dwarfs, among others; details of proposed scenarios are given in \cite{choplin2021A&A...648A.119C} and the literature within the paper. The study of the i-process has become increasingly important in the context of Galactic chemical evolution, as it has been invoked to explain anomalous abundance patterns. 
{Carbon-enhanced metal-poor stars \citep[CEMP\footnote{Defined as stars with [C/Fe] $> 0.7$~dex.
This notation indicates a logarithmic scale of carbon A(C) relative to that of iron A(Fe) relative to the Sun. A(X) indicates that for a given element X, the logarithmic absolute abundance is
defined as the number of atoms of element X per $10^{12}$ hydrogen
atoms, log $\epsilon$(X) = A(X) $\equiv$ log$_{10} (NX/NH) + 12.0$. 
},][]{beers2005ARA&A..43..531B, aoki2007ApJ...655..492A} are prime candidates for hosting i-process material, as they frequently exhibit simultaneous enhancements of r- and s-process elements.
In particular, based on moderately low-precision observational abundances of first peak (Sr, Y, and Zr) and heavy neutron-capture elements (from Ba to Hf), \citep{hampel2016ApJ...831..171H} reported reasonable fits with i-process nucleosynthesis models.
Further support is provided by 
the abundance pattern of  BPS~CS~31062-50 \citep{aoki2002PASJ...54..933A, johnson2004ApJ...605..462J, aoki2006ApJ...650L.127A,lai2007ApJ...667.1185L}, with moderately precise observational abundance measurements, has been announced as evidence of the i-process nucleosynthesis by \cite{wiedeking2025NatRP...7..696W}.
However, CEMP stars are not the only candidates: signatures of the i-process have also been reported in a dwarf star with solar [C/Fe], based on abundances of light trans-iron elements ($32 \leq Z \leq 47$; \citealt{roederer2016ApJ...821...37R}).
}

The i-process pattern and the combination of the s- and r-processes signatures are difficult to discern. 
The task is performed by comparing observationally-based and theoretical patterns of heavy element-to-iron ratio versus the atomic number $Z$. Among the heavy elements (i.e. those with $Z > 56$) detectable in spectra of CEMP stars, differences between theoretical models do not exceed $\sim$0.3~dex \citep[e.g.][Fig.~7]{sbordone2020A&A...641A.135S}. 
Therefore, accurate and precise abundance determinations are needed.
This goal is often compromised by three factors: (i) the difficulty of deriving accurate atmospheric parameters in CEMP stars, which mostly appear to be red giants, (ii) the lack of realistic 3D non-LTE line profile models for the majority of chemical elements, and (iii) spectra moderately crowded with carbon molecular features such as CH, CN, and C$_2$, which totally or partially blend weak atomic features.
Regarding (i), the reason is the limitation of custom methods. Namely, the excitation and ionisation balance of Fe lines from one-dimensional (1D) hydrostatic model atmospheres assuming local thermodynamic equilibrium (LTE) has been proven to strongly bias parameter determinations in very metal-poor stars ([Fe/H] $\lesssim -2$~dex). For instance, \cite{ruchti2013MNRAS.429..126R}, \cite{giribaldi2023A&A...679A.110G}, and \cite{giribaldi2023A&A...673A..18G} find typical biases of $-300$~K and $-1.0$~dex in \teff\ and \logg, respectively.  Colour calibrations based on normal FGK stars underestimate \teff\ by 100 to 500~K when applied to CEMP ones \citep[e.g.][Fig. 10]{giribaldi2023A&A...679A.110G}.
Such biases are high enough to change true abundances in a heterogeneous way, affecting the diagnose of the dominant nucleosynthesis signature the star exhibits.
Regarding (ii), only 1D non-LTE corrections are available for a few heavy elements. 
And even if available for all, there remains the fundamental issue of assessing how closely 1D~non-LTE corrections recover the true abundances, particularly in giant stars.   
As a consequence, despite recent advances in modelling and observations, the frequency, astrophysical sites, and detailed yields of the i-process remain subjects of active investigation.

TYC~6044-714-1 is an ideal target for testing the i- and r+s-process hypothesis. It is a well-characterised {CEMP-rs\footnote{Stars enhanced in neutron-capture elements that neither match the s- nor the r-processes alone. The rs notation implies $0 <$ [Ba/Eu] $0.5$~dex according to \cite{beers2005ARA&A..43..531B}, but classifications involving other elements and abundances ranges have been proposed by diverse authors.}} 
star whose kinematics and chemical composition suggest an origin within the Milky Way halo about 13~gigayear (Gyr) ago (App.~\ref{sec:dynamic} and \ref{sec:logg_sec}). It orbits the Galactic centre in the opposite direction to the Sun and its binding energy is moderate. However, its [Mg/Fe] is compatible with the [Mg/Fe]-[Fe/H] signature of the most metal-poor tail of the Milky Way halo, which likely formed in situ \citep{giribaldi2025}; further details are provided in App.~\ref{sec:dynamic}. 
TYC~6044-714-1 (also known as RAVE~J094921.8$-$161722) was identified as a CEMP-rs star by \cite{sakari2018ApJ...868..110S} and was determined to be compatible with an r+s-process signature by \cite{gull2018ApJ...862..174G}. However, this paradigm was recently challenged by \citet{choplin2022A&A...667L..13C}, who argued that a pure i-process occurring in a low-mass AGB star could explain the entire abundance pattern, including actinides.

In this work, we present new high-quality observations of TYC~6044-714-1, together with a detailed comparison to stellar evolution and nucleosynthesis models, with the aim of shedding new light on the role of the i-process in Galactic chemical evolution. The paper is organised as follows: Section~\ref{sec:data} describes our observations and data reduction, Section~\ref{sec:param} presents the determination of stellar parameters, Section~\ref{sec:cnmg} details the derivation of elemental abundances and isotopic ratios, and Section~\ref{sec:results} provides theoretical predictions for the different neutron-capture channels. Finally, Section~\ref{sec:discussion} discusses our results, and Section~\ref{sec:conclusions} summarises our conclusions.

\section{Observations and data reduction}
\label{sec:data}
We acquired spectra of TYC~6044-714-1 with the Ultraviolet and Visual Echelle Spectrograph  \citep[UVES,][]{dekker2000} mounted on the Very Large Telescope (VLT) of the European Southern Observatory (ESO).
Three spectra in the blue and the red spectral regions were acquired {using slit width $0.5^{\prime\prime}$ and binning $1 \times 1$ setup (data set ID 114.27JT.001 or 0114.B-0030(A), PI. Giribaldi).} The dates of acquisition, resolution $R = \lambda/\Delta \lambda$, and signal-to-noise (S/N) are listed in Table~\ref{tab:spectral_char}.
The data were reduced to one-dimensional (1D) spectra and wavelength-calibrated using the Phase~3 pipeline \citep{Arnaboldi2011Msngr.144...17A}.
We corrected Doppler wavelength shifts by cross-correlating the spectra with a synthetic template from the {\sc iSpec}  package \citep{blanco-cuaresma2014},
{and computed heliocentric velocities using {\it astropy} routines; see values in Table~\ref{tab:spectral_char}.
Our measurements are compatible with those compiled by \cite{gull2018ApJ...862..174G}; hence, no radial velocity variations have been detected thus far, supporting the hypothesis of a long-period binary orbit or pole-on oriented system \citep[e.g.][]{lucatello2005ApJ...625..833L,hansen2016A&A...588A...3H}.
}
We co-added the spectra. The S/N of the co-added spectrum was estimated in the blue and red regions using 14 and 12 pseudo-continuum ranges, respectively, resulting in S/N = 140 and S/N = 430.

In the analysis process, we used as a reference the  UVES archival spectra of the \titan\  star HD186478 \citep{giribaldi2023A&A...679A.110G}, whose characteristics are listed in Table~\ref{tab:parameters}. Its revised parameters \citep{giribaldi2025} are very similar to those of TYC6044-714-1, as reported in Table~\ref{tab:parameters}. The spectral processing followed exactly the same procedure as for TYC~6044-714-1.

\begin{table}
\caption{Spectral characteristics}
\label{tab:spectral_char}
\centering
\tiny 
\begin{threeparttable}
\begin{tabular}{ccccccccccc}
\hline\hline
Wavelength range [\AA] &  Obs. date & $R \equiv \lambda/\Delta\lambda$ & S/N & RV [km/s] \\ %
\hline
\multicolumn{4}{c}{ TYC 6044-714-1 (Target)}\\
\hline
3282-4563 & 03/01/2025 & 65030 & 80 & $390.62 \pm 0.13$ \\
3282-4563 & 07/01/2025 & 65030 & 70 & $389.72 \pm0.12$\\
3282-4563 & 22/01/2025 & 65030 & 40 & $390.10 \pm 0.14$ \\
4726-6834 & 03/01/2025 & 74450 & 270 & $390.76 \pm 0.20$ \\
4726-6834 & 07/01/2025 & 74450 & 248 & $390.34 \pm 0.20$\\
4726-6834 & 22/01/2025 & 74450 & 184 & $390.64 \pm 0.20$\\
\hline
\multicolumn{4}{c}{HD~186478 (Reference star)}\\
\hline
3304-4607 & 12/10/2000 & 40979 & 102 & ---\\
3304-4607 & 12/10/2000 & 40979 & 98 & ---\\
4654-6759 & 12/10/2000 & 42310 & 290 & ---\\
\hline
\end{tabular}
\begin{tablenotes}
\item{} Notes. The last column indicates heliocentric velocity.
\end{tablenotes}
\end{threeparttable}
\end{table}

\section{Stellar atmospheric parameters}
\label{sec:param}

\begin{figure*}[t]
    \centering
    \includegraphics[width=0.85\linewidth]{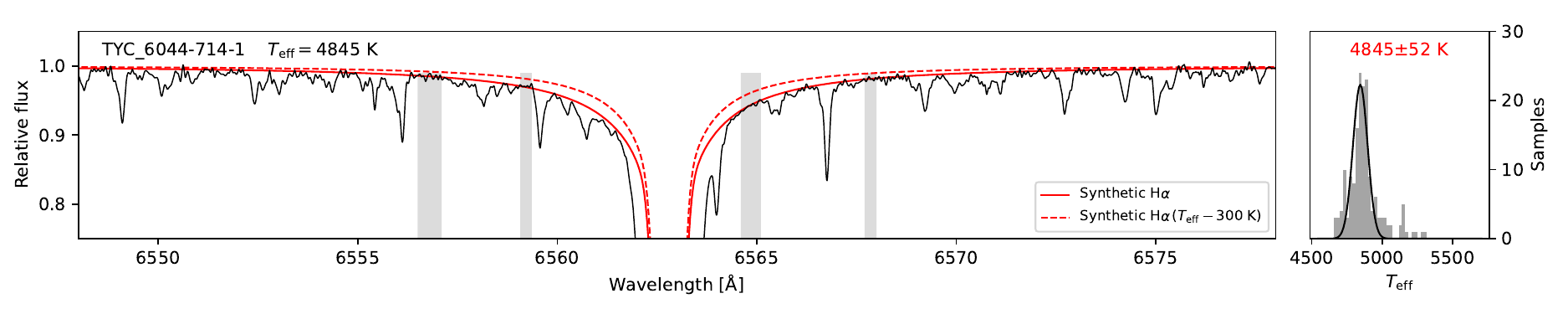}
    \includegraphics[width=0.85\linewidth]{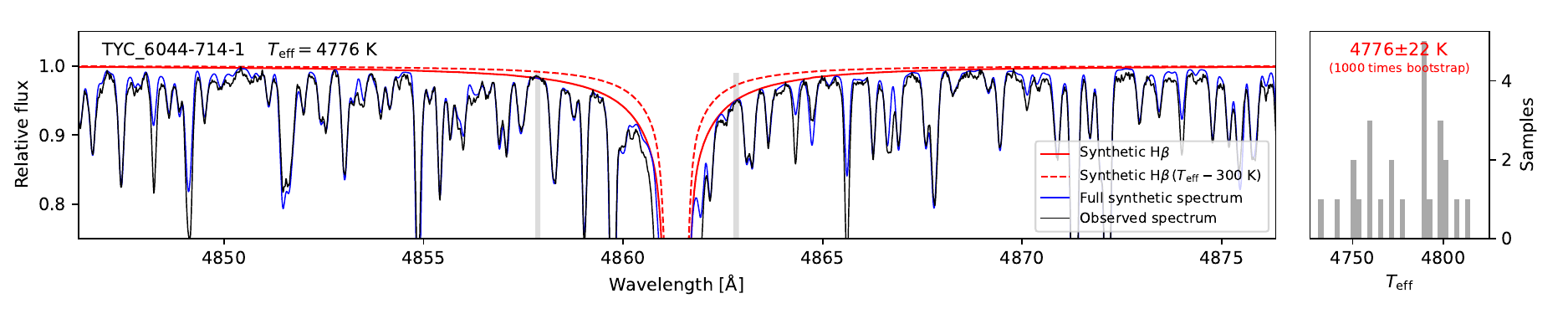}
    \caption{\tiny Fits of the H$\alpha$ and H$\beta$ Balmer lines.
    The main plots show synthetic profiles (solid red line) fitted to the observational ones (black line). 
    Synthetic lines from temperatures $300$~K cooler than the determined ones are represented by the dashed red lines.
    Shades represent the fitting regions without metal and telluric line blends.
    The panels on the right display histograms of the temperatures associated with every pixel inside of the shaded regions in the left panels.
    In the top panel, the most likely temperature and its error are given by the median of the Gaussian profile fitted to the histogram and its $\sigma$ dispersion.
    In the bottom panel, for robustness, the most likely temperature and its error are obtained by bootstrapping.
    A synthetic spectrum with molecular and atomic lines is represented in blue.
    }
    \label{fig:Halpha}
\end{figure*}

We determined the atmospheric parameters listed in Table~\ref{tab:parameters} following the methodology outlined in \cite{giribaldi2025}. Briefly, the effective temperature (\teff) was retrieved by fitting the wings of the H$\alpha$ and H$\beta$ lines
with the synthetic grids of \citet{amarsi2018}, produced with three-dimensional (3D) radiation-hydrodynamical model atmosphere simulations and considering  departures from the local thermodynamic equilibrium  (non-LTE).
The normalisation-fitting method of the observational profiles adopted in the present work  is described in \citet{giribaldi2019A&A...624A..10G}\footnote{Profile fitting code available at \url{https://github.com/RGiribaldi/Balmer-profile-fitting}}.
The accuracy of the H$\alpha$ line model has been demonstrated in \citet{giribaldi2021A&A...650A.194G} and \citet{giribaldi2023A&A...679A.110G}, showing no offset relative to the standard{\footnote{Set by the \teff\ values of the Gaia Benchmark stars \citep{heiter2015A&A...582A..49H} inferred via interferometry and the Infrared Flux Method (IRFM).}}, with a dispersion of about 50~K.
The accuracy of the H$\beta$ line has not been systematically evaluated. Therefore, we determined \teff\ from the H$\beta$ line relative to the reference star.
The profile fits are shown in Fig.~\ref{fig:Halpha}, where we also include the synthesis of molecular and atomic features for H$\beta$.
The final temperature was obtained by averaging the values derived from H$\alpha$ and H$\beta$.
Its error is given by the standard deviation of the mean ($\pm35$~K) added in quadrature  to the errors due to the uncertainties on surface gravity (\logg) and on metallicity ([Fe/H]) \citep[these are $\pm$15~K each, according to Figs. 5 and 6 in][]{giribaldi2023A&A...679A.110G}.

\begin{figure}
    \centering
    \includegraphics[width=0.8\linewidth]{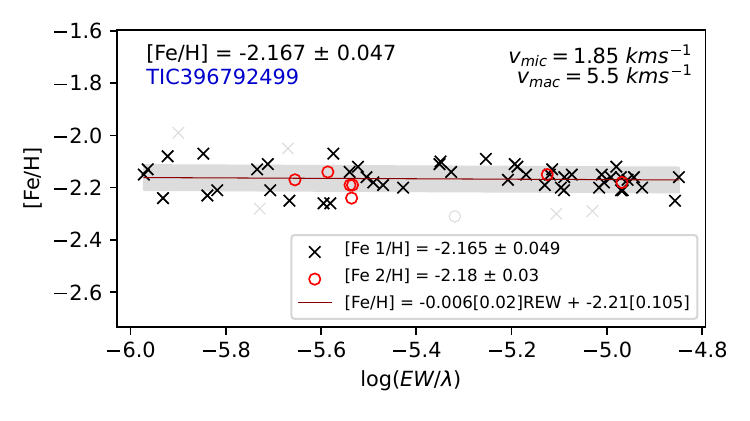}
    \caption{\tiny Determination of metallicity and $v_{mic}$.
    Crosses and  circles represent abundances from \ion{Fe}{i} and \ion{Fe}{ii} lines, respectively.
    Grey symbols are clipped outliers.
    A regression of both species is represented by the red line, the coefficients of which and corresponding errors are given in the legends.
    The shade indicates the $\sigma$ dispersion of the trend.
    Averages and standard deviations of \ion{Fe}{i} and \ion{Fe}{ii}, individually, are also given in the legends.
    }
    \label{fig:vmic}
\end{figure}

\begin{table}
\caption{Atmospheric parameters}
\label{tab:parameters}
\centering
\tiny 
\begin{threeparttable}
\begin{tabular}{lcccccccccc}
\hline\hline
Star &  \teff & \logg & [Fe/H]$_{\text{non-LTE}}$ & $v_{mic}$\\ 
& (K) & (dex) & (dex) &  (km/s)  \\
\hline
HD~186478 &  $4793 \pm 22$ & $1.72 \pm 0.15$ &  $-2.37 \pm 0.09$ & $1.94 \pm 0.20$  \\
TYC~6044-714-1  &  $4810 \pm 41$ & $1.70 \pm 0.15$ & $-2.17 \pm 0.05$ &  $1.82 \pm 0.04$ \\
\hline
\end{tabular}
\begin{tablenotes}
\item{} Notes. 
The effective temperature is  listed together with its total uncertainty. 
{The uncertainty in 
\logg\ corresponds to the range of values for which the slope of the trend in Fig.~\ref{fig:vmic} remains within 1$\sigma$.}
The metallicity and $v_{mic}$ are given with their internal uncertainties. We adopted the reference solar iron abundance 
A(Fe) = 7.45~dex from \cite{lodders2021SSRv..217...44L}.
The \teff\ value for HD~186478 was taken from \cite{giribaldi2025}, whereas all other stellar parameters were re-derived in this work.
\end{tablenotes}
\end{threeparttable}
\end{table}

The surface gravity and metallicity  were derived by assuming the excitation equilibrium of Fe lines under 1D~non-LTE assumption. 
An assessment of \logg\ is presented in Sect.~\ref{sec:logg_sec}.
We applied line-by-line synthesis running the radiative transfer code Turbospectrum \footnote{\url{https://github.com/bertrandplez/Turbospectrum_NLTE}}
\citep{gerber2023} with MARCS model atmospheres \citep{gustafson2008} considering the line list with atomic parameters in \cite{heiter2021A&A...645A.106H}. We extracted iron departure coefficients based on the model atom developed in \cite{bergemann2012} and \cite{semenova2020}. 
We applied the differential approach with respect to the reference star in order to optimise the [Fe/H] precision; this method minimises the error related to the atomic parameters and normalisation in twin stars \citep[e.g.][]{melendez2014ApJ...791...14M}. 
To ensure consistency, the spectra of the target star (TYC~6044-714-1) and the reference star (HD~186478) were analysed following the same procedure, adopting an identical line list and anchoring the normalisation to common pseudo-continuum points.
For this latter step, we performed simultaneous local normalisations relative to a synthetic spectrum generated using the atmospheric parameters of the reference star and the abundances reported in \cite{gull2018ApJ...862..174G}.
We also avoided using Fe lines blended with molecular features by compiling an adapted line list based on the Fe line compilations from \cite{giribaldi2025A&A...702A..65G}, created through simultaneous visual inspection of the reference star and target spectra.
Our final Fe line list is provided on Zenodo\footnote{\url{https://doi.org/10.5281/zenodo.20159108}.}.

For both the reference  and  target stars, we applied spectral synthesis fixing \teff\ and \logg\ to the 
values in Table~\ref{tab:parameters}. 
We fixed the projected rotational velocity to 0 km~s$^{-1}$, and allowed the microturbulence ($v_{mic}$) and the macroturbulence ($v_{mac}$) to vary freely. For the reference star, we set an [$\alpha$/Fe] enhancement of +0.35 \citep{giribaldi2025}; whereas for the latter we set [$\alpha$/Fe] = +0.25 according to a preliminary Mg abundance determination, obtained considering the 1D~non-LTE corrections in Fig.~6 of \cite{giribaldi2025}. Our 1D~non-LTE determination presented in Sec.~\ref{sec:cnmg} further  validated this value. 
We constrained the most likely [Fe/H] by requiring
the absence of correlation between [Fe/H] and the reduced equivalent width REW = log$(EW/\lambda)$\footnote{$EW$ is the equivalent width in \AA\ and $\lambda$ is the wavelength in \AA.}. We set the REW limit to values lower than $-4.8$ to exclude excessively saturated lines.
Figure~\ref{fig:vmic} shows the outcome of our differential analysis.

Following the results of \cite{giribaldi2025A&A...702A..65G}, we derived the $v_{mic}$ value in Table~\ref{tab:parameters} by averaging the values obtained from 1D non-LTE analysis and from the 
\teff–\logg-dependent relation of \cite{dutra-ferreira2016A&A...585A..75D}, which is based on 3D model atmospheres. The individual estimates are $1.85 \pm 0.07$~km~s$^{-1}$ for the former and $1.78 \pm 0.05$~km~s$^{-1}$ for the latter.

\section{Elemental abundances and isotopic ratios from spectral analysis}
\label{sec:cnmg}

We derived elemental abundances   
by spectral synthesis using Turbospectrum with MARCS model atmospheres.
We used the atomic parameters from \cite{heiter2021A&A...645A.106H} for lines between 4200~\AA\ and redder wavelength limit, while for lines at bluer wavelengths we relied on the VALD~3 database \citep{vald32015PhyS...90e4005R}.  For all elements except C, Ba, and Eu, the solar isotopic abundance fractions were kept fixed at their default values\footnote{Corresponding to the solar-system isotopic abundances compiled by \cite{isotopic2005JPCRD..34...57B}, as specified in the Turbospectrum file atomicweights.dat.}.
We modelled the molecular bands using the following line lists: 
$^{12}$CH and $^{13}$CH \citep{masseron2014A&A...571A..47M}, 
$^{12}$C$^{14}$N \citep{brooke2014ApJS..210...23B},
$^{12}$C$_{2}$ \citep{brooke2013JQSRT.124...11B},
$^{12}$C$^{13}$C \citep{Ram2014ApJS..211....5R},
$^{13}$C$^{14}$N \citep{sneden2014ApJS..214...26S}, and
$^{24}$MgH \citep{skory2003ApJS..148..599S}.

To derive elemental abundances we use used the lists of \cite{gull2018ApJ...862..174G} and \cite{da_silva2025A&A...696A.122D} as initial references. We performed a visual inspection comparing our observational spectrum with synthetic spectra searching for all possible detectable lines.
Considering all elements, the median of the abundance error associated with the microturbulence uncertainty ($\sigma_{vmic}$) is 0.01~dex.
Ba and Yb are exceptions because their lines are saturated, their $\sigma_{vmic}$ are 0.10 and 0.12~dex, respectively. 
We computed the total error budget by adding in quadrature the following sources: $\sigma_{vmic}$, the dispersion of the line-by-line measurement ($\sigma_{stat}$) or the error related to the noise in case  only one line is available, the error related to \teff\ ($\sigma_{T}$), the error related to \logg\ ($\sigma_{g}$), and the error related to [Fe/H] ($\sigma_{[Fe/H]}$).
The errors $\sigma_{vmic}$, $\sigma_{T}$, $\sigma_{g}$, and  $\sigma_{[Fe/H]}$ were computed by changing the corresponding parameters by their errors in Table~\ref{tab:parameters}.
Figure~\ref{fig:errors} shows the impact of typical parameter errors in the element abundances.
Table~\ref{tab:abundances} lists our abundances and the number of lines used. We provide the entire line list in Zenodo\footnote{\url{https://doi.org/10.5281/zenodo.20159609}}. Errors related to line-by-line abundance dispersions and total ones are listed separately.
{Nb and Th abundance determinations are presented in Sect.~\ref{sec:Nb}. The former is relevant for the discussion Sect.~\ref{sec:discussion}.}

1D non-LTE corrections were applied when available. 
The adopted corrections for each element are listed in Table~\ref{tab:abundances} and were obtained from the following sources: 
Si \citep{bergemann2013ApJ...764..115B, amarsi2017MNRAS.464..264A}, Ca \citep{mashonkina2007}, Ti \citep{bergemann2011}, Cr \citep{bergemann2010A&A...522A...9B}, Mn \citep{bergemann2008A&A...492..823B}, Co \citep{bergemann2010MNRAS.401.1334B}, 
Ni \citep{Eitner2023A&A...677A.151E}, Zn \citep{Sitnova2022}, Sr \citep{Mashonkina2022}
 and Pb and Th \citep{mashonkina2012A&A...540A..98M}.
For Al, we employed 3D non-LTE corrections in \cite{nordlander2017A&A...607A..75N}.
The online tools  MPIA\footnote{\url{https://nlte.mpia.de/gui-siuAC_secE.php}} \citep{NLTE_MPIA} and NL{\sc i}TE\footnote{\url{https://nlite.pythonanywhere.com/}} \citep{Koutsouridou2025A&A...699A..32K} were used when interpolations are implemented. Otherwise, corrections are extracted from tables in the papers themselves.
In the following sections, we provide details about the analysis of some specific elements that are more challenging or critical. 

\begin{table}
\caption{Elemental abundances.}
\label{tab:abundances}
\centering
\tiny
\begin{tabular}{|l|l|c|c|c|c|c|}
\hline\hline
Z & Element &  A(X) & [X/Fe]+cor & $\sigma_{stat}$ & $\sigma_{Tot}$ & N.\\
\hline
6 & C~{\sc i} & 7.30 & 1.00 & 0.03 & 0.06 & 2 \\
6 & $^{12}$C$_{2}$ & 7.25 & 0.98 & 0.02 & 0.07 & 2 b \\
6 & $^{12}$C/$^{13}$C & 19$^\clubsuit$ & ---  & 5.4$^\clubsuit$ & --- & 1 b \\
7 & N~{\sc i} & 6.48 & 0.81 & 0.03 & 0.13 & 2 b \\
8 & O~{\sc i} & 7.15 & 0.62 & 0.04 & 0.06 & 2 \\
11 & Na~{\sc i} & 4.02 & $-0.06$ & 0.06 & 0.07 & 4 \\
12 & Mg~{\sc i} & [$5.66$] & $0.42 - 0.14$ & 0.03 & 0.04 & 2 \\
13 & Al~{\sc i} & [4.20]$\dagger$ & $-0.42 + 0.33\dagger$ & 0.04 & 0.13 & 1 \\
14 & Si~{\sc i} & [5.69] & $0.24 + 0.10$ & 0.08 & 0.08 & 3 \\
20 & Ca~{\sc i} & [4.42] & $0.16 + 0.08$ & 0.05 & 0.06 & 12 \\
21 & Sc~{\sc i} \& {\sc ii} & 1.13 & $0.14$ & 0.04 & 0.09 & 6 \\
22 & Ti~{\sc i} & [2.99] & $0.16 + 0.04$ & 0.06 & 0.08 & 5 \\
22 & Ti~{\sc ii} & [3.01] & $0.22 + 0$ & 0.01 & 0.05 & 2\\
23 & V~{\sc ii} & 1.91 & 0.14 & 0.06 & 0.10 & 6\\
24 & Cr~{\sc i} & [3.44] & $-0.38 + 0.34$ & 0.03 & 0.06 & 6 \\
25 & Mn~{\sc i} & [3.24] & $-0.68 + 0.65$ & 0.03 & 0.06 & 4 \\
26 & Fe~{\sc i} & [5.28] & --- & 0.05 & 0.07 & 69 \\
26 & Fe~{\sc ii} & [5.28] & --- & 0.03 & 0.05 & 8 \\
27 & Co~{\sc i} & [3.05] & $-0.18 + 0.41$ & 0.12 & 0.13 & 8 \\
28 & Ni~{\sc i} & [4.08] & $-0.13 + 0.15$ & 0.04 & 0.06 & 4 \\
30 & Zn~{\sc i} & [2.53] & $-0.01 + 0.14$ & 0.04 & 0.06 & 1\\
38 & Sr~{\sc ii} & [1.00] & $0.10 +0.19$ & 0.07 & 0.10 & 3 \\
39 & Y~{\sc ii} & 0.12 & $0.07$ & 0.08 & 0.11 & 5 \\
40 & Zr~{\sc ii} & 0.72 & 0.30 & 0.06 & 0.10 & 10 \\
41 & Nb~{\sc ii} & $0.02$ & 0.73 & 0.18 & 0.20 & 2 \\
44 & Ru~{\sc i} & 0.13 & 0.54 & 0.04 & 0.10 & 4 \\
46 & Pd~{\sc i} & $-0.16$ & $0.44$ & 0.20 & 0.22 & 1 \\
56 & Ba~{\sc ii} & 0.79 & $0.77$ & 0.03 & 0.10 & 2\\
57 & La~{\sc ii} & $-0.20$ & 0.86 & 0.03 & 0.08 & 5 \\
58 & Ce~{\sc ii} & $0.20$ & 0.78 & 0.02 & 0.08 & 4\\
59 & Pr~{\sc ii} & $-0.61$ & 0.83 & 0.02 & 0.09 & 2 \\
60 & Nd~{\sc ii} & $0.03$ & 0.77 & 0.03 & 0.08 & 14 \\
62 & Sm~{\sc ii} & $-0.53$ & 0.67 & 0.05 & 0.09 & 2\\
63 & Eu~{\sc ii} & [$-1.19$] & $0.40 + 0.05$ & 0.02 & 0.08 & 4 \\
64 & Gd~{\sc ii} & $-0.47$ & 0.62 & 0.05 & 0.09 & 4 \\
65 & Tb~{\sc ii}& $-1.52$ & 0.34 & 0.06 & 0.13 & 3 \\
66 & Dy~{\sc ii}& $-0.52$ & 0.54 & 0.08 & 0.11  & 8 \\
68 & Er~{\sc ii} & $-0.60$ & 0.64 & 0.10 & 0.14 & 7 \\
70 & Yb~{\sc ii} & $-0.02$$\dagger$ & 1.30 & 0.02 & 0.14 & 2 \\
72 & Hf~{\sc ii} & $-0.33$ & 0.98 & 0.04 & 0.09 & 2 \\
76 & Os~{\sc i} & $-0.42$ & 0.34 & 0.20 & 0.33 & 1\\
77 & Ir~{\sc i} & $-0.46$ & 0.32 & 0.06 & 0.14 & 1 \\
82 & Pb~{\sc i} & [1.35] & 1.39 + 0.37 & 0.04 & 0.09 & 1 \\
90 & Th~{\sc ii} & $[-1.77]$ & 0.28 + 0.09 & $\pm^{0.18}_{0.29}$ & $\pm^{0.23}_{0.32}$ & 1\\
\hline
\end{tabular}
\begin{tablenotes}
\item{} Notes. The symbol $\dagger$ indicates an unreliable abundance derived from an intense line. The symbol $^\clubsuit$ indicates isotopic fractions instead of element abundance. Fractions [X/Fe] were computed considering the solar photospheric abundances in \cite{asplund2009ARA&A..47..481A}; except for Mg, which adopts the solar value A(Mg)$_{\odot}$ = 7.55~dex of \cite{asplund2021A&A...653A.141A}.
Abundances A(X) within brackets indicate determinations that consider 1D~non-LTE or 3D~non-LTE corrections according to Sect.~\ref{sec:cnmg}. Element-to-iron ratios [X/Fe] of the same elements are listed in 1D~LTE along with their corresponding 1D~non-LTE corrections. Although Mg and Eu were determined by line profile fitting under 1D~non-LTE on the fly, their ratios with respect to iron under 1D~LTE are listed for comparison. The last column indicates the number of features used to obtain the abundances. Numbers accompanied by ‘b’ indicate wide wavelength bands. Errors
related to line-by-line abundance dispersions ($\sigma_{stat}$) and total ones are listed in columns five and six. 
For Nb, the error related to noise is listed instead of $\sigma_{stat}$ to account for likely noise-induced biases. 
\end{tablenotes}
\end{table}

\subsection{Abundances and isotopic ratios of carbon, nitrogen, and magnesium}
\label{sec_mg}
We derived the carbon abundance using the atomic C~{\sc i} lines at 5052.145 and 5380.320~\AA, as well as the Swan $^{12}$C$_{2}$ band in the 5140–5166~\AA\ region, which exhibits moderately weak molecular features. These diagnostics yielded A(C) = 7.33, 7.27, and 7.25~dex, respectively.
The CH~G band near 4300~\AA\ was not used in the final determination because its strong molecular features may introduce systematic biases. For reference, we obtain A(C) = 7.12~dex with that band. 
We determined the $^{12}$C/$^{13}$C  isotopic ratio using the molecular features in the 4155-4240~\AA\ region, a section of which is shown in Fig.~\ref{fig:c13}.
Nitrogen abundances were derived by fitting the CN bands at 3556–3590 and 3787–3812~\AA, which yielded A(N) = 6.53 and 6.46~dex, respectively.

The internal errors of the C and N abundances correspond to flux variations caused by noise. The internal errors of the $^{12}$C/$^{13}$C ratio were estimated from the dispersion of the results of the most prominent individual features (31 features were selected for this error estimate) within the specified wavelength range.
Namely, we computed the average of the $25\% - 50\%$ and $75\% - 50\%$ quantiles. 
We derived magnesium as described in Sec.~4.1 of \cite{giribaldi2025}, by averaging the outcomes of the fits of the the lines at 5528 and 5711~\AA, assuming 1D~non-LTE departures in \cite{bergemann2017_}; each line yields A(Mg) = 5.63 and 5.69~dex, respectively.
We iterated the determination of the elemental abundances (including Fe) until we obtained consistent results.

\subsection{Abundances and isotopic ratios of barium and europium}
\label{sec:Ba_Eu_iso}

The barium abundance and its isotopic fractions were determined following \cite{giribaldi2026_barium}, where the method was calibrated with 72 stars and validated with Galaxy evolution models.
It consists of deriving A(Ba) via 1D~LTE spectral synthesis of the subordinate lines at 6141.715 and 6496.900~\AA, while  deriving isotopic fractions from 1D non-LTE synthesis of the resonance line \ion{Ba}{ii} 4934~\AA\ fixing the  A(Ba) from 1D~LTE. 
The accurate log~$gf$ values from \cite{de_munshi2015PhRvA..91d0501D, dutta2016NatSR...629772D}, as in \cite{gallagher2020A&A...634A..55G} are required.

\begin{figure}
    \centering
    \includegraphics[width=0.49\linewidth]{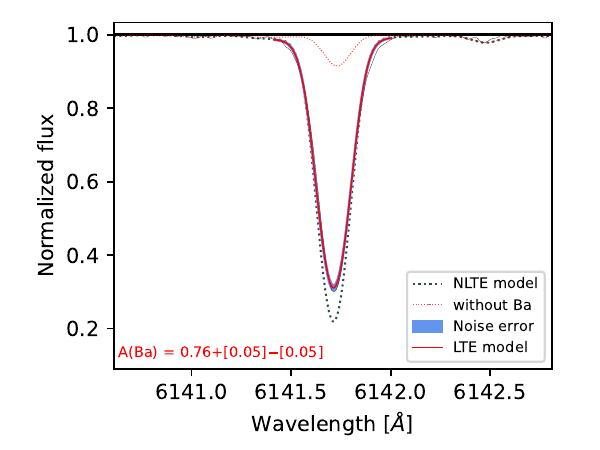}
    \includegraphics[width=0.49\linewidth]{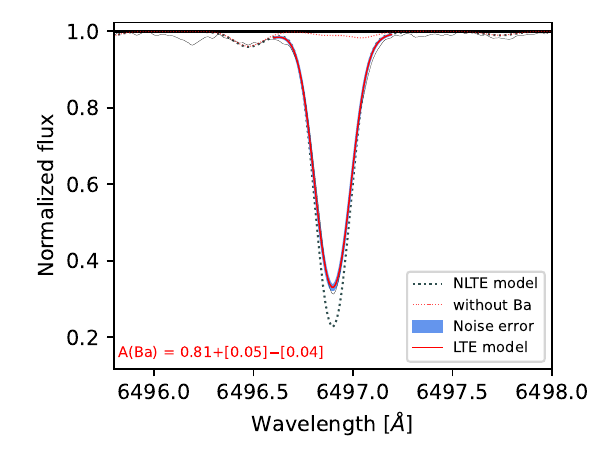}
    \caption{\tiny Profile fits of Ba lines. The observed spectrum is shown in black, while the modelled line profiles in 1D~LTE are colour-coded according to the legend. The associated abundances and noise-related uncertainties are indicated in each panel. 1D~non-LTE line profiles synthesised with the same abundances obtained from 1D~LTE fits are shown for comparison.}
    \label{fig:ba}
\end{figure}

\cite{giribaldi2026_barium} shows that 1D~LTE yields equivalent A(Ba) to prototype 3D~non-LTE\footnote{Computed by the code Linfor3D \citep{Linfor3D} \url{https://www.chetec-infra.eu/3dnlte/abundance-corrections/barium/}.} calculations within 0.1~dex for every star  \cite[see also a test case in][]{giribaldi2025A&A...702A..65G}.
On the other hand, 1D non-LTE calculations applied to subordinate lines underestimate A(Ba) by $-0.2$~dex for the Sun \citep{gallagher2020A&A...634A..55G}, and by $-0.3$ to $-0.4$~dex for stars with [Ba/Fe]~$> 0.3$~dex \citep{giribaldi2026_barium}.
Figure~\ref{fig:ba} shows the 1D~LTE fits and the corresponding 1D~non-LTE profiles; the latter are systematically deeper and would yield lower abundances.

\begin{figure}
    \centering
    \includegraphics[width=0.9\linewidth]{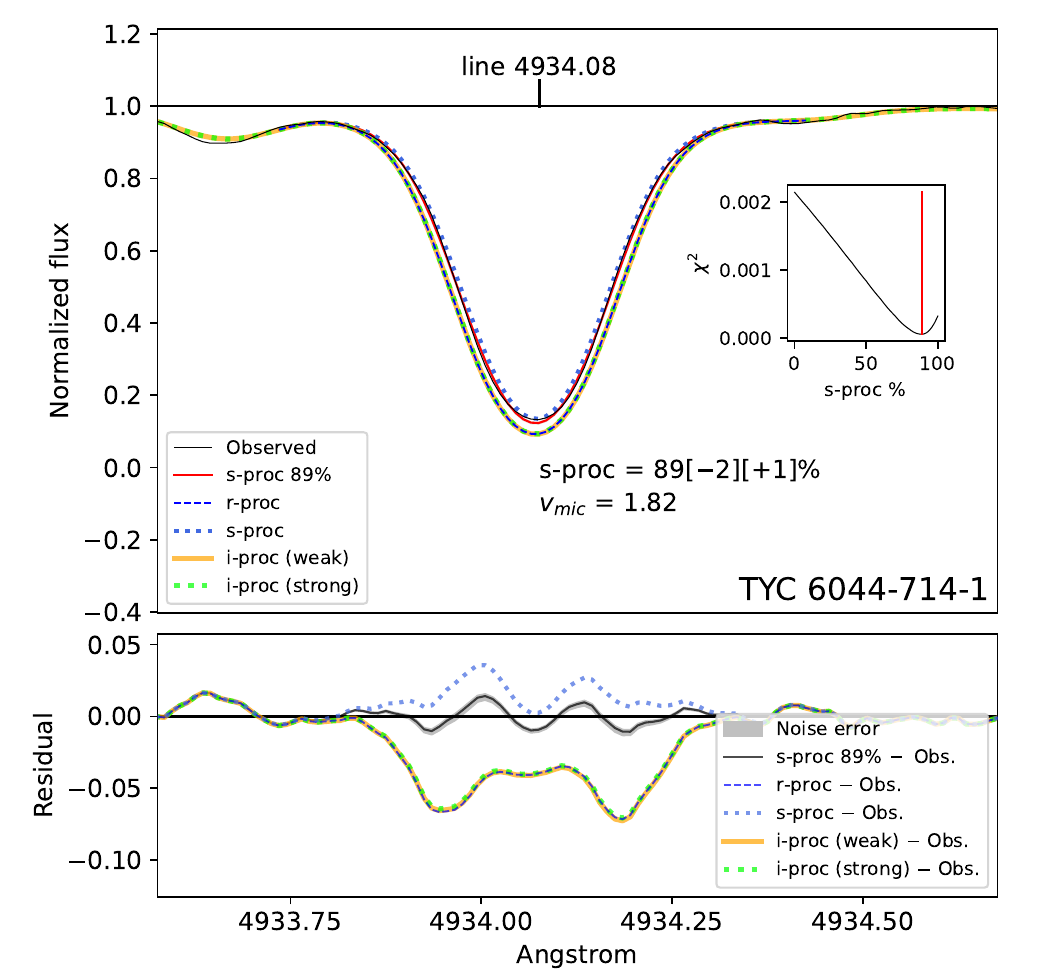}
    
    \caption{\tiny Fit of the Ba resonance line at $\lambda$4934~\AA. {\it Top panel:} Observational profile is represented by the black line. The synthetic profile that best fit the observational profile is represented in red. Its isotopic ratio in terms of an s-process percentage is noted in the plot along with its error due to the spectral noise.
    Synthetic profiles with isotopic ratio combinations in Table~\ref{tab:ratios} are included as indicated in the legends. 
    The inner plot shows the $\chi^2$ related to the fit, where the determined percentage corresponding to the minimum $\chi^2$  is represented by the bar.
    {\it Bottom panel:} Black line represents the difference between the best fit and the observational profile. The grey shade represents the noise error in terms of flux.
    Residuals of the r-, s-, and i-process profiles minus the observational one are represented by the same line style as in the top panel.
    }
    \label{fig:Ba_iso}
\end{figure}

Isotopic fractions were obtained using 1D non-LTE departure coefficients \citep{gallagher2020A&A...634A..55G} and hyperfine-structure data in \cite{giribaldi2025A&A...702A..65G}. 
The best fit corresponds to an 89\% s-process contribution (Figure~\ref{fig:Ba_iso}). The main uncertainty comes from \teff\ (12\% per 50~K), while flux noise and microturbulence have negligible impact. 
The remaining 11\% of the abundance is attributed to the r-process, to the i-process, or a combined r+i contribution. 
From Table~\ref{tab:abundances}, the abundance fraction corresponding to the s-process is of A(Ba)$_{s-proc} = 0.74$~dex, and the complement corresponding to r-, i-, or r+i- is of A(Ba)$_{r/i-proc} = -0.17 \pm ^{0.15}_{\infty}$~dex.

The europium abundance was determined under 1D~non-LTE \citep[][]{storm2024} from the lines at 4129.72 and 4205.03~\AA, and under 1D~LTE from the lines at 6437.64 and 6645~\AA. 
For the lines at 4129.72 and 4205.03~\AA, the hyperfine structure (HFS) and its corresponding atomic parameters are taken from \cite{da_silva2025A&A...696A.122D}, while isotopic ratios related to the r- and i-processes are listed in Table~\ref{tab:ratios}. Both isotopic combinations produce nearly identical line profiles and equivalent abundances; see Fig.~\ref{fig:Eu}.

\begin{table}
\caption{Isotopic ratios}
\label{tab:ratios}
\centering
\tiny 
\begin{threeparttable}
\begin{tabular}{l|ccccccccc}
\hline\hline
Process & $^{134}$Ba & $^{135}$Ba & $^{136}$Ba & $^{137}$Ba & $^{138}$Ba \\
\hline
Slow  & 0.0081 & 0.0175 & 0.0558  & 0.0611  & 0.8586 \\
Rapid  & 0.0000 & 0.4177 & 0.0000 & 0.3341 & 0.2482 \\
Intermediate (weak) & 0.0003 & 0.9296 & 0.0028 & 0.0374 & 0.0299 \\
Intermediate (strong) & 0.0001 & 0.7246 & 0.0077 & 0.1342 & 0.1334 \\

\hline
 & $^{151}$Eu & $^{153}$Eu \\
 \hline
Slow  & 0.4673 & 0.5327 \\
Rapid  & 0.5008 & 0.4992 \\
Intermediate (weak) & 0.5291 & 0.4709 \\
Intermediate (strong) & 0.5151 & 0.4849 \\

\hline
\end{tabular}
\begin{tablenotes}
\item{} Notes. Quantities related the r-process are inferred from \cite{Prantzos20}. Quantities related to the s-process and i-process were inferred from a model of 1.5$M_{\odot}$ with [Fe/H] $=-2.27$~dex (see Section~\ref{sec:results}); for the i-process, both a weak case ($f_{\mathrm{top}}=0.05$) and a strong case ($f_{\mathrm{top}}=0.15$, after the second PIE) are shown. 
For Ba, the contributions of $^{135}$Ba and $^{137}$Ba include the decay of $^{135}$Cs and $^{137}$Cs.
\end{tablenotes}
\end{threeparttable}
\end{table}

\subsection{Prior Eu and Ba abundance enrichment of the binary system}
\label{sec:prior}
Assuming that TYC~6044-714-1 became a CEMP star through binary interaction, it is reasonable to consider that, prior to its current evolutionary stage, it was a normal field star whose heavy-element composition reflected the Galactic r-process background. This assumption can be assessed observationally, independent of nucleosynthesis models, by comparing the Eu and Ba abundances of TYC~6044-714-1 with those of typical field stars at similar metallicity.

Figure~\ref{fig:AEu} shows the distribution of A(Eu) as a function of [Fe/H] for the \titan\ stars \citep{giribaldi2021A&A...650A.194G,giribaldi2023A&A...679A.110G}, including isotopic fractions from \cite{giribaldi2026_barium}. TYC~6044-714-1 and the CEMP HD~196944 closely follow the trend defined by dwarf field stars \citep[a similar trend was reported by][]{simmerer2004ApJ...617.1091S}, indicating an r-process enrichment consistent with the general field-star population and showing no evidence of anomalous enhancement.
{The CEMP BPS~CS~22892-052, and BPS~CS~31082-001 (enhanced in barium) are clearly enriched in Eu.}
To further verify if the actinide content of TYC~6044-714-1 is consistent with this r-process background, we compared the Th abundance with the Eu estimate assuming a scaled solar r-process pattern. Adopting the r-process fractions from \citet{Prantzos20}, the expected solar r-process ratio is $A \text(\text{Th/Eu})_{r} \approx -0.35$. At the metallicity of TYC~6044-714-1, the field-star trend suggests $A(\text{Eu}) \approx -1.4$, which would correspond to an expected r-process baseline of $A(\text{Th}) \approx -1.75$. 
Our spectroscopic measurement for the star is $A(\text{Th}) = -1.77 \pm^{0.23}_{0.32}$. By accounting for radioactive decay over the system's age ($\approx +0.2$~dex), we derived an initial abundance of $A(\text{Th})_{\text{ini}} \approx -1.57\pm^{0.23}_{0.32}$. This value is in good agreement with the predicted r-process models, considering the observational errors and the intrinsic scatter of Galactic r-process enrichment. The slight observed enhancement of Eu in TYC~6044-714-1 relative to the mean trend {($\approx +0.2$~dex) is separately explained by a combination of observational uncertainties ($0.1$~dex) and a minor $s$-process contribution from the AGB companion (see Sect.~\ref{sec:results}). Indeed, the accretion of $\approx 3\%$ of AGB material, which is instrinsecally enriched in s-process europium (${\rm [Eu/Fe]_{agb}} \simeq 1.3$), would shift the initial r-process Eu abundance by $\approx +0.1$~dex, without requiring an anomalous actinide-to-lanthanide ratio. These results confirm that the Th enrichment in this star follows a scaled-solar r-process distribution.

In the A(Ba)–[Fe/H] plane (Fig.~\ref{fig:ABa}), TYC~6044-714-1 lies above the lower-envelope trend traced by dwarf stars, as seen for other CEMP stars and the Ba enhanced star BPS~CS~31082-001 \citep{hill2002A&A...387..560H}. At its metallicity, this trend corresponds to A(Ba)$ = -0.25 \pm 0.15$~dex, which we adopt as a plausible estimate of the star’s initial barium abundance. This baseline is consistent with the r-process contribution inferred from isotopic ratios (Sect.~\ref{sec:Ba_Eu_iso}), A(Ba)$_{r/i-proc} = -0.17^{+0.15}_{-\infty}$~dex (blue point and error bar in the figure). Together, the Eu and Ba abundances, along with the Ba isotopic ratios, indicate that the pre-enrichment of TYC~6044-714-1 shows no evidence of abnormal nucleosynthetic enrichment prior to the binary interaction.

\begin{figure}
    \centering
    \includegraphics[width=0.9\linewidth]{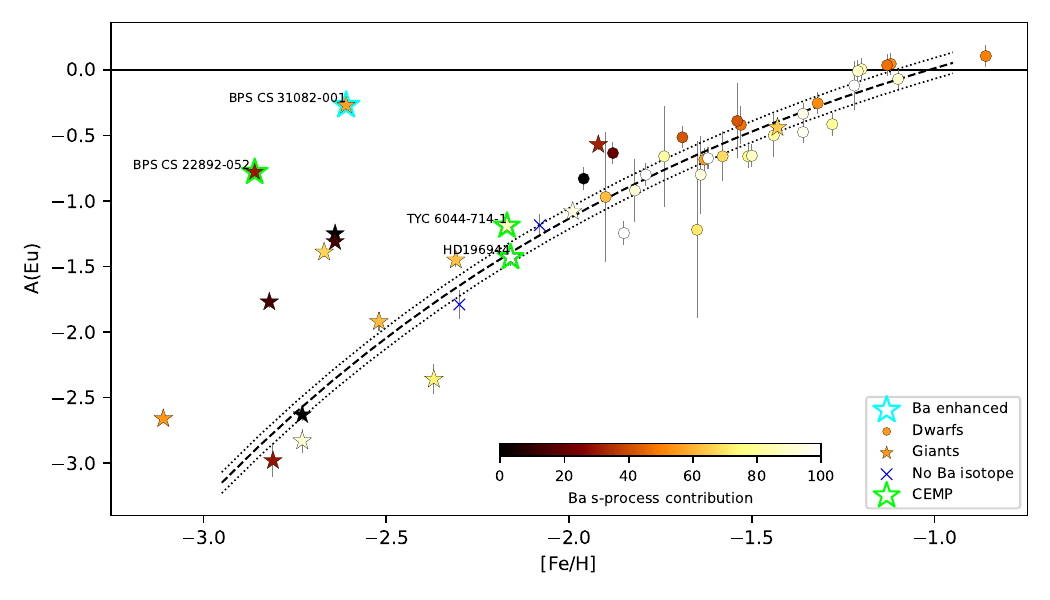}
    \caption{\tiny A(Eu) versus [Fe/H] of the \titan\ stars.
    Dwarfs and giants are shown as circles and stars, respectively. The s-process contribution to the barium abundance, derived from isotopic ratios, is colour-coded as indicated by the colour bar.
    TYC~6044-714-1, other CEMP stars, and a {star enhanced in barium} are highlighted as indicated in the legend.
    Stars without Ba isotopic ratio measurement are represented by blue crosses.
    An exponential trend fitted to the dwarfs is shown by the dashed line, with its dispersion of A(Eu)~$\pm0.08$~dex indicated by the dotted lines. 
}
    \label{fig:AEu}
\end{figure}

\begin{figure}
    \centering
    \includegraphics[width=0.9\linewidth]{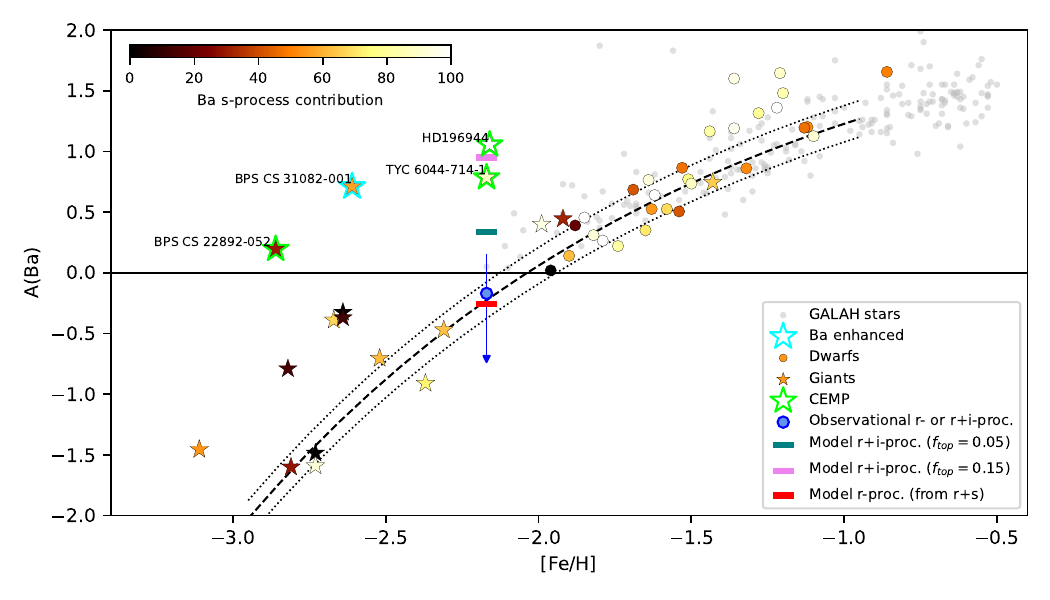}
    \caption{\tiny  
    A(Ba) versus [Fe/H] of the \titan\ stars.
    The elements of the plots are the same as in Fig.~\ref{fig:AEu}.
    Grey dots represent dwarf stars from the GALAH survey \citep{buder2021galah} and are shown for reference. 
    The blue circle and its error bar indicate A(Ba)$_{r/i-proc}$, according to the isotopic ratio analysis presented in Sect.~\ref{sec:Ba_Eu_iso}.
    Ba abundances predicted by the nucleosynthesis models shown in Fig.~\ref{fig:ele_fit_r}, corresponding to the total barium abundance minus the s-process contribution (i.e., the r- or r$+$i-process components), 
    are indicated by the dashes according to the legends.
    An exponential trend fitted to the dwarf \titan\ and GALAH stars is shown by the dashed line, with its dispersion of A(Ba)~$\pm0.15$~dex indicated by the dotted lines.
}
    \label{fig:ABa}
\end{figure}

\section{Nucleosynthesis modelling}
\label{sec:results}

In this section, we present our theoretical predictions for the s- and  r-process contributions to the observed abundances and assess the possible presence of an i-process component. We compare our abundance results with a set of nucleosynthesis models  varying contributions from the s-, i-, and r-processes.

\subsection{AGB models}
Most CEMP-s and CEMP-rs stars are found in binary systems \citep[e.g.][]{lucatello2005ApJ...625..833L,hansen2016A&A...588A...3H}; thus, their chemical configurations are commonly interpreted as the outcome of mass-transfer events, in which material enriched in neutron-capture elements is accreted via stellar winds from a former AGB companion \citep[e.g.][]{masseron10,bisterzo2010,bisterzo2011,bisterzo2012}. Therefore, the common practice is to compare their observed abundance patterns with those predicted for the nucleosynthetic yields of AGB stars.

Here, the observed abundance pattern of TYC~6044-714-1 is compared with theoretical AGB nucleosynthesis predictions computed using the FuNS evolutionary code \citep{straniero06,cristallo09a,vescovi21a}. To model s-process nucleosynthesis, we adopted an extended nuclear network including nearly 500 isotopes connected by more than 800 nuclear reactions \citep{vescovi21b}. For the purposes of the present study, the network has been further expanded in order to properly follow the i-process nucleosynthesis path, resulting in a network comprising more than 1300 isotopes and 2500 reactions.
The baseline nuclear network is extended by adopting neutron-capture reaction rates from the TENDL-astro 2023 database~\citep{rochman25} and from~\citet{spyrou24} for the $\mathrm{^{139}Ba(n,\gamma)^{140}Ba}$ reaction rate, while $\beta$-decay rates are taken from the NUBASE2020 evaluation~\citep{kondev21}, complemented with theoretical rates from~\citet{marketin16}.
Neutron-induced fission rates for trans-lead nuclei are from~\citet{panov10}.

During proton ingestion episodes (PIEs), the characteristic timescales of nuclear burning and convective transport become comparable \citep[e.g.,][]{iwamoto04}. As a consequence, nucleosynthesis and mixing equations are solved simultaneously in our calculations. A diffusive mixing scheme was adopted.
We computed AGB models with an initial mass of $1.5M_\odot$ and a metallicity of $\mathrm{[Fe/H]} = -2.27$~dex, scaling A(Fe) to the solar iron abundance \citep{lodders2021SSRv..217...44L}. The initial chemical composition assumes an $\alpha$ enhancement of $\mathrm{[O/Fe]} = 0.7$~dex and $\mathrm{[\alpha/Fe]} = 0.4$~dex for the other $\alpha$ elements, consistent with the abundance patterns observed in low-metallicity Galactic halo stars \citep[e.g.,][]{Lind2024ARA&A..62..475L}. According to current theoretical models, AGB stars of this mass and metallicity are not expected to experience PIEs, as the entropy barrier associated with the H-burning shell prevents efficient mixing between the He- and H-rich layers \citep{fujimoto2000,iwamoto04,cristallo09b,cristallo16,choplin2022A&A...667A.155C}. Consistently, our standard models do not undergo PIEs and only experience s-process nucleosynthesis.
In these models, the formation of the $^{13}$C pocket is induced by magnetic buoyancy-driven mixing, following the prescription described in \cite{vescovi20,vescovi21a}. As expected for s-process nucleosynthesis at low metallicity, this results in strong lead production and significant enhancements of both the first ([ls/Fe]\footnote{[ls/Fe]=([Sr/Fe]+[Y/Fe]+[Zr/Fe])/3}) and second s-process peaks ([hs/Fe]\footnote{[hs/Fe]=([Ba/Fe]+[La/Fe]+[Ce/Fe]+[Pr/Fe]+[Nd/Fe])/5}), with representative values of $[\mathrm{ls/Fe}] = 1.49$~dex, $[\mathrm{hs/Fe}] = 2.24$~dex, and $[\mathrm{Pb/Fe}] = 3.37$~dex (see Fig.~\ref{fig:ele_fe}) 

However, the inclusion of additional extra mixing at the top of the convective thermal pulse can facilitate the occurrence of PIEs \citep{choplin24}. 
In stellar evolution calculations, such mixing is commonly In stellar evolution calculations, this extra mixing is commonly described by assuming an exponential decay of the diffusion coefficient beyond the formal convective boundary, following
\begin{equation}
D_{\rm over}(z) = D_{\rm cb} \exp\left(-\frac{2z}{f_{\mathrm{top}} H_{P,{\rm cb}} }\right),
\end{equation}
where $z$ is the distance from the convective boundary defined by the Schwarzschild criterion, $D_{\rm cb}$ and $H_{P,{\rm cb}}$ are the diffusion coefficient and the local pressure scale height at the boundary, respectively, and $f_{\mathrm{top}}$ is a free parameter that controls the efficiency of the overshoot at the top of the convective thermal pulse.
We find that, for a $1.5\,M_\odot$ model, adopting an overshooting parameter of $f_{\mathrm{top}} = 0.05$ at the upper boundary of the convective thermal pulse is sufficient to trigger a PIE during the third thermal pulse, whereas lower values of $f_{\mathrm{top}}$ do not lead to proton ingestion. 
We also computed additional models with initial masses of 1 and $1.2\,M_\odot$ adopting the same overshooting efficiency ($f_{\mathrm{top}} = 0.05$).
These models do not show significant differences in either the occurrence of PIEs or the resulting nucleosynthetic signatures with respect to the $1.5\,M_\odot$ case.
For this reason, in the following we focus on the $1.5\,M_\odot$ models, exploring the effects of varying the overshooting parameter at the top of the convective thermal pulse, considering values of $f_{\mathrm{top}} = 0.05$, $0.10$, and $0.15$.

In all cases in which a PIE occurs, the resulting nucleosynthesis is characterised by i-process conditions, with neutron densities exceeding $10^{15}\,\mathrm{cm^{-3}}$, intermediate between those typical of the s- and r-processes.
After the PIE, the convective shell merges with the envelope, leading to a strong surface enrichment in heavy elements.
For elements up to Ba, the resulting abundance patterns are broadly similar among the different $f_{\mathrm{top}}$ cases (see upper panel of Figure~\ref{fig:ele_fe}).
However, increasing the overshooting efficiency leads to a higher neutron exposure, with elements beyond Ba reaching very high surface overabundances.
In particular, models with $f_{\mathrm{top}} = 0.10$ and $0.15$ are able to efficiently populate the Pb--Bi region.
In these models, the nucleosynthetic flow also proceeds towards heavier nuclei to produce non-negligible amounts of actinides~\citep{choplin2022A&A...667L..13C}.
For $f_{\mathrm{top}} = 0.15$, two PIEs occur.
While the first ingestion episode produces abundance patterns similar to those obtained for $f_{\mathrm{top}} = 0.10$, the second PIE substantially enhances the neutron exposure, resulting in an even more efficient production of Pb and actinide nuclei.

In addition to heavy elements, large amounts of $^{12}$C, $^{13}$C, and $^{14}$N are mixed into the envelope following a PIE.
The enrichment in $^{12}$C is mainly associated with the occurrence of a deep third dredge-up episode, which transports the products of $3\alpha$ burning to the stellar surface.
Conversely, $^{13}$C and $^{14}$N are produced by incomplete H burning within the H-rich convective shell formed during the ingestion event.
The resulting increase in the CN abundance generally inhibits further proton ingestion events. 
The enhanced metallicity boosts the efficiency of H-shell burning in subsequent interpulse phases, strengthening the entropy barrier at the H/He interface and preventing the convective He-shell from penetrating into proton-rich regions during later thermal pulses.
The star therefore resumes a standard AGB evolution, experiencing regular third dredge-up episodes and s-process nucleosynthesis during the following thermal pulses \citep{cristallo09b}.
This behaviour is observed in the model with $f_{\mathrm{top}} = 0.05$, which develops a single PIE.
In this case, the final surface composition displays a genuinely mixed i+s chemical signature that can be approximately decomposed into two distinct contributions: elements lighter than Ba ($Z < 56$) are predominantly shaped by the i-process nucleosynthesis associated with the PIE, while heavier nuclei mainly reflect the subsequent s-process production (see lower panel of Figure~\ref{fig:ele_fe}).
The model with $f_{\mathrm{top}} = 0.10$ also experiences a single PIE, but the substantially higher neutron exposure significantly alters the nucleosynthetic outcome, efficiently synthesising heavy elements from Xe all the way through to the Pb peak. Consequently, the s-process contribution becomes secondary, and the final abundance pattern is predominantly determined by i-process nucleosynthesis across the entire heavy-element range.
For larger overshooting efficiencies, however, multiple PIE episodes may occur~\citep{choplin24}.
In particular, the model with $f_{\mathrm{top}} = 0.15$ experiences a second PIE despite the prior CN enrichment of the envelope.
As a consequence, the heavy-element nucleosynthesis is further enhanced by an additional i-process episode, leading to a final abundance pattern that is largely dominated by the i-process contribution over the whole mass range, including Pb and trans-lead elements.
In both $f_{\mathrm{top}} = 0.10$ and $f_{\mathrm{top}} = 0.15$ models, the subsequent s-process nucleosynthesis during the late AGB phase has only a minor impact on the final surface composition.

\subsection{Nucleosynthesis models versus measured abundances}
\label{sec:patterns}
\begin{figure*}[!htbp]
\centering
\includegraphics[width=0.85\linewidth]{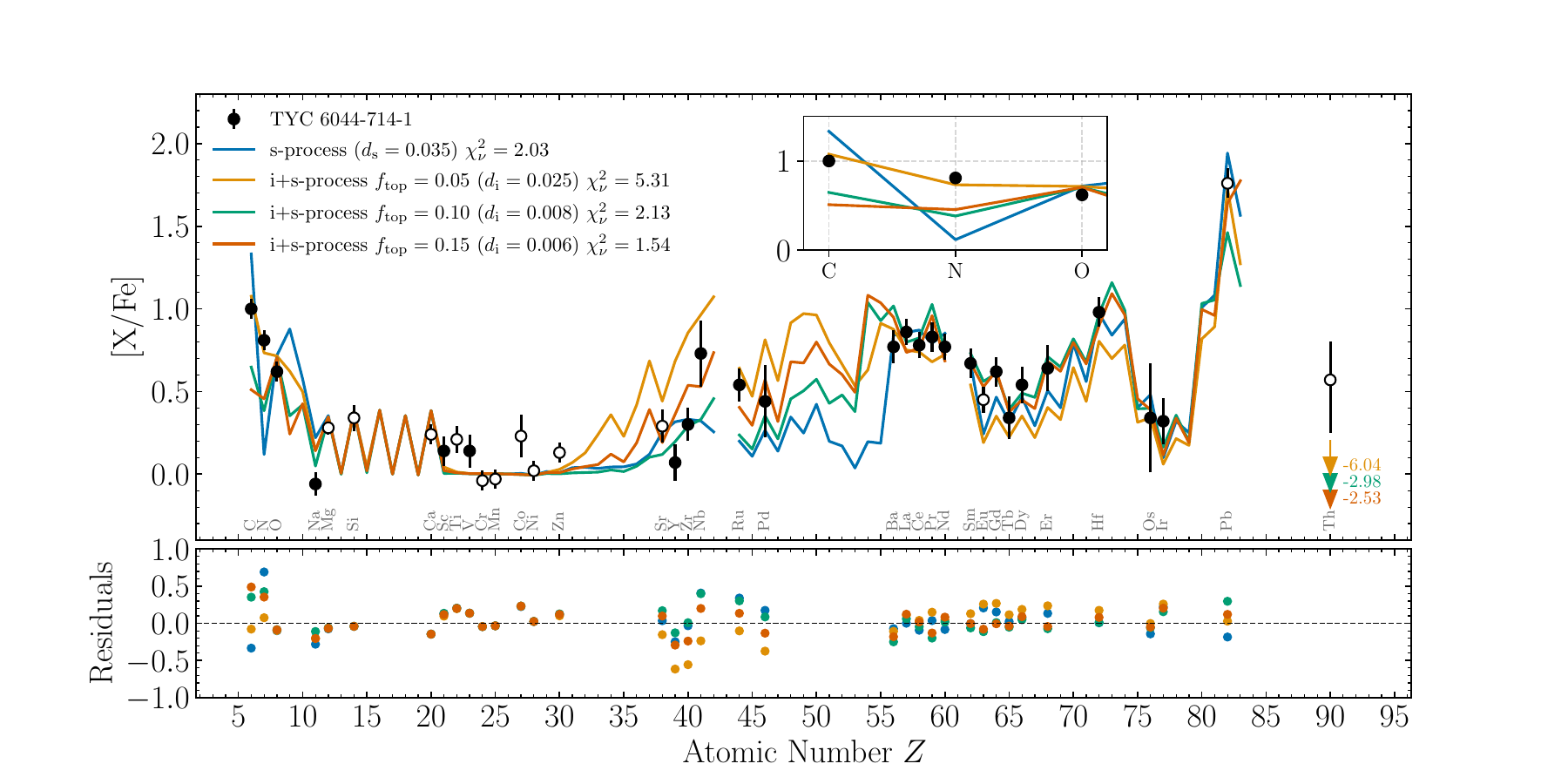}
\caption{{\it Upper panel}: Comparison between the observed abundance pattern of TYC~6044-714-1 and AGB nucleosynthesis predictions without an initial r-process enrichment. 
Observed abundances are shown as open white circles for elements corrected for non-LTE effects, and as filled black circles when corrections are not applied.
Results for a s-process model (blue line) and for three mixed i+s models with $f_{\mathrm{top}} = 0.05$ (yellow line), $f_{\mathrm{top}} = 0.10$ (green line), and $f_{\mathrm{top}} = 0.15$ (orange line), after dilution with the envelope of the companion star, are shown. 
Th abundance has been increased by 0.2~dex to account for radioactive decay over $\sim$10 Gyr (see text for details).
The inset provides a zoom on the CNO region.
The $\chi^2_\nu$ values and the corresponding dilution factors are computed excluding Th, whose predicted abundance in the AGB models is virtually zero.
{\it Lower panel}: Residuals between the observed and theoretically predicted abundance patterns, with symbol colours corresponding to the models shown in the upper panel. Residuals associated with the extra-mixing model are indicated by triangular symbols.}
\label{fig:ele_fit_is}
\end{figure*}
To compare the chemical abundances of TYC~6044-714-1 with predictions from AGB models, we adopted a $\chi^2$ fitting procedure. In particular, the model that best reproduces the observed abundances was identified by minimising the reduced $\chi^2$ value ($\chi^{2}_{\nu}$), which quantifies the agreement between observations and theoretical predictions. The minimisation was achieved by combining the material ejected by the AGB star with the envelope of the companion star. 
If $m_{\rm{agb}}$ denotes the mass accreted by the companion from the AGB ejecta and $m_{\mathrm{com}}$ the mass of the companion's envelope involved in the mixing, the resulting mass fraction of isotope $i$ of element $j$, $X_{i,j}$, can be expressed as
\begin{equation}\label{eq:mixing}
X_{i,j}
= X_{i,j}^{\rm{agb}}\, d_{\rm{agb}}
+ X_{i,j}^{\mathrm{com}} \left(1 - d_{\rm{agb}}\right),
\end{equation}
where $X_{i,j}^{\rm{agb}}$ and $X_{i,j}^{\mathrm{com}}$ denote the isotopic mass fractions in the AGB ejecta and in the envelope of the companion star, respectively. The dilution factor is defined as $d_{\rm{agb}} = m_{\rm{agb}} / m_{\rm{tot}}$, where the total mixed mass is
$m_{\rm{tot}} = m_{\rm{agb}} + m_{\mathrm{com}}$.

In principle, the initial composition of the companion star could be assumed to consist of pristine, $\alpha$-enhanced material only. On the other hand, the presence of an additional pre-enrichment by a pure r-process contribution in the companion star can also be considered. In this more general case, Eq.~\ref{eq:mixing} becomes
\begin{equation}\label{eq:mixing_all}
X_{i,j}
= X_{i,j}^{\rm{agb}}\, d_{\rm{agb}}
+ X_{i,j}^{\rm r}\, d_{\rm r}
+ X_{i,j}^{\alpha}\, d_{\alpha},
\end{equation}
where $X_{i,j}^{\alpha}$ and $X_{i,j}^{\rm r}$ denote the isotopic mass fractions associated with the $\alpha$-enhanced pristine material and with the r-process component, respectively. The corresponding dilution factors are defined as $d_{\alpha} = m_{\alpha}/m_{\rm{tot}}$ and $d_{\rm r} = m_{\rm r}/m_{\rm{tot}}$, with the total mass now given by
$m_{\rm{tot}} = m_{\rm{agb}} + m_{\alpha} + m_{r}$.
By construction, the dilution factors satisfy
$d_{\rm{agb}} + d_{\alpha} + d_{\rm r} = 1$.

The resulting overabundance of element $j$ relative to iron is then expressed as
\begin{equation}\label{eq:dilution_ele}
\left[ \frac{{\rm X}}{\mathrm{Fe}} \right]
= \log \left( \frac{X_j}{\mathrm{Fe}} \right)
- \log \left( \frac{X_j}{\mathrm{Fe}} \right)_\odot ,
\end{equation}

where the elemental abundance $X_j$ was obtained by summing over the individual isotopic mass fractions.
Only elements heavier than Zn ($Z = 30$) were considered in the $\chi^{2}_{\nu}$ calculation.

\paragraph{The s-process and mixed i+s models.}
We first compare pure s-process models with mixed i+s models in Figure~\ref{fig:ele_fit_is}. Among them, the best agreement with the observed abundance pattern is obtained for the i+s model with $f_{\mathrm{top}} = 0.15$, which yields the lowest reduced $\chi^2$ value. 
In particular, this model successfully reproduces the abundances of elements such as Eu and Gd, which are severely underproduced in the pure s-process scenario and are known to receive a dominant contribution from the r-process \citep{Prantzos20}.
Despite this overall improvement, the i+s model still fails to reproduce the observed Th abundance, which remains underpredicted by about $\sim 3$~dex with respect to the measured value.
It is important to note that the abundances of trans-lead elements are strongly affected by nuclear-physics uncertainties. In particular, \citet{Martinet24} showed that correlated model uncertainties can induce surface abundance variations of up to $\sim 3$~dex for Th and U. 
However, even when considering these uncertainties in the most favourable case, the predicted actinide abundances remain only marginally consistent with the observations, which show $[\mathrm{Th/Fe}] > 0$.
For this comparison, the observed Th abundance has been increased by 0.2~dex to account for radioactive decay over a timescale of at least $\sim$10~Gyr.
Overall, this indicates that while a strong i-process contribution significantly improves the agreement for most heavy elements, an initial r-process enrichment is still required to fully account for the actinide abundances observed in TYC~6044-714-1.

\begin{figure*}[!htbp]
\centering
\includegraphics[width=0.85\linewidth]{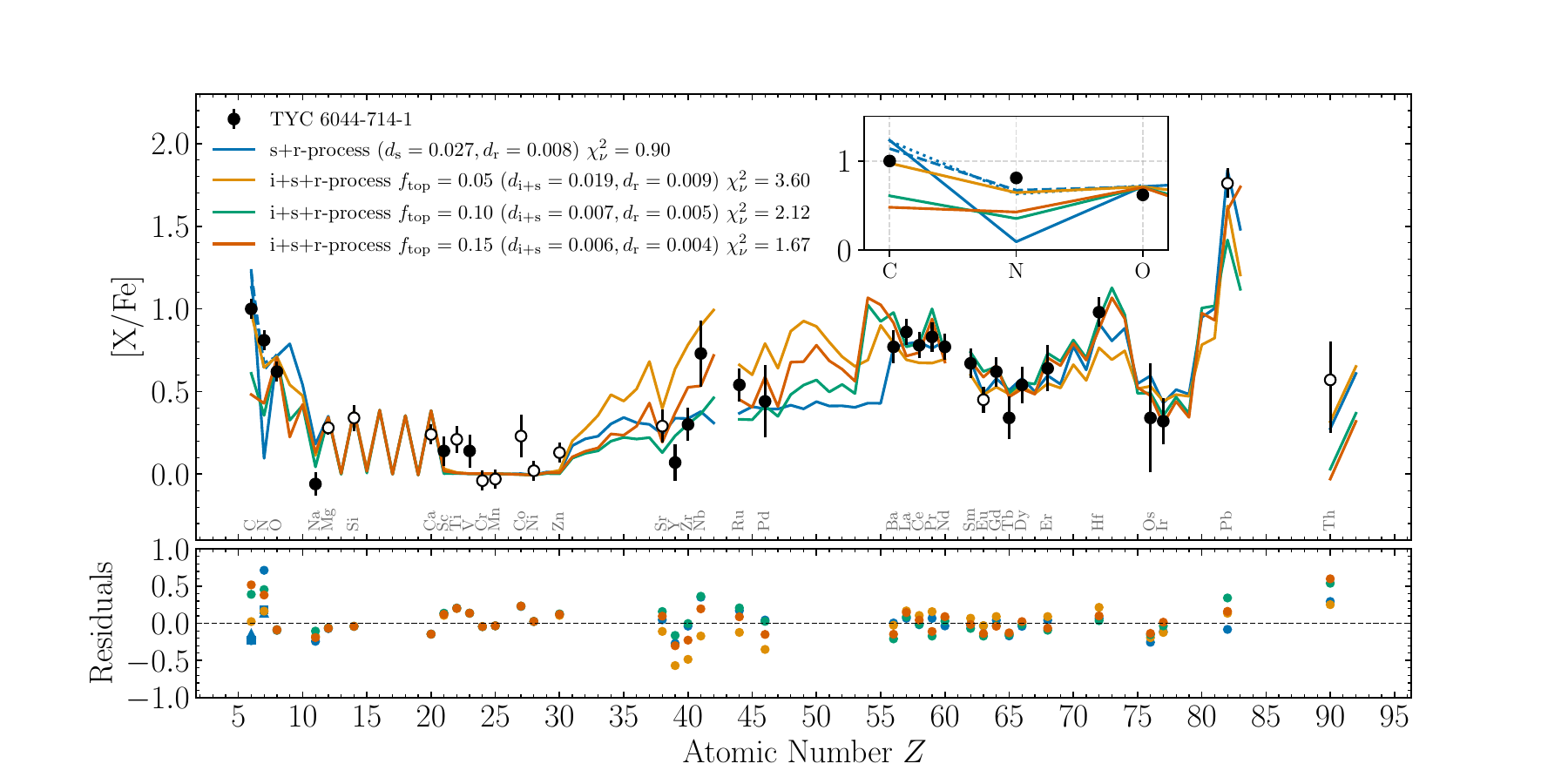}
\caption{{\it Upper panel}: Comparison between the observed abundance pattern of TYC~6044-714-1 and the best-fitting AGB nucleosynthesis models including an initial r-process enrichment. Observed abundances are shown as open white circles for elements corrected for non-LTE effects, and as filled black circles when corrections are not applied.
Results for an s+r model (solid brown line)(blue line) and for three mixed i+s+r models with $f_{\mathrm{top}} = 0.05$ (yellow line), $f_{\mathrm{top}} = 0.10$ (green line), and $f_{\mathrm{top}} = 0.15$ (orange line), after dilution with the envelope of the companion star, are shown. 
Additional models including extra mixing during the AGB (primary star) or tht RGB (secondary star) phase are shown as a dashed and dotted blue curve (see the CN region in the [X/Fe] distributions and the corresponding residuals), respectively.
The r-process contribution is required to reproduce the Th abundance.
{\it Lower panel}: Residuals between the observed and theoretically predicted abundance patterns, with symbol colours corresponding to the models shown in the upper panel. Residuals associated with the extra-mixing model are indicated by triangular (AGB) and square (RGB) symbols.}
\label{fig:ele_fit_r}
\end{figure*}
\paragraph{The s-process and mixed i+s models with the r pre-enrichment.}
Figure~\ref{fig:ele_fit_r} shows the comparison between the observed abundance pattern and the predictions obtained by combining AGB yields with an additional r-process contribution. 
In these models, the r-process input follows a scaled-solar distribution \citep{Prantzos20}. This choice is justified by the observed universality of the main r-process for elements heavier than Ba in metal-poor stars \citep[e.g.,][]{Sneden2008ARA&A..46..241S}. Lighter r-elements can show variations \citep[e.g.,][]{hansen2012A&A...545A..31H}, but their impact on our fit is minimal as their abundances are primarily determined by the yields from the AGB companion. The absolute scale of the r-process component, $d_{\rm r}$ (Eq.~\ref{eq:mixing_all}), is treated as a free parameter optimised through a global $\chi^2$ minimisation including all measured elements with $Z > 30$. Although the fitting is performed over the entire heavy-element range, the value of $d_{\rm r}$ is principally constrained by those species with a significant r-process contribution, such as Eu, Gd, and the actinides (Th).

We consider the same set of AGB models discussed above, namely a pure s-process model (s+r) and three mixed i+s models with an increasing overshooting efficiency ($f_{\mathrm{top}} = 0.05$, $0.10$, and $0.15$), all combined with an initial r-process enrichment (i+s+r). In these cases, the dilution factor associated with the AGB ejecta is denoted as $d_{\rm s}$ and $d_{\rm i+s}$, respectively, while the r-process dilution factor $d_{\rm r}$ is treated as an independent free parameter.

For the mixed i+s models with $f_{\mathrm{top}} = 0.10$ and $0.15$, the inclusion of an r-process component mainly affects the predicted actinide abundances, bringing Th into agreement with the observations, while the abundances of lighter neutron-capture elements remain almost unchanged. 
In contrast, for the model with $f_{\mathrm{top}} = 0.05$, the addition of the r-process contribution leads to a substantial improvement not only for Th, but also for elements in the Eu--Er region, although Hf remains significantly underproduced.
The pure s+r model also shows a marked improvement with respect to the s-only case, successfully reproducing the observed abundances of typical r-process tracers such as Eu and Gd, as well as the Th abundance. 
From a global perspective, the s+r model and the i+s+r model with $f_{\mathrm{top}} = 0.15$ provide the best overall agreement with the observations, yielding comparable and significantly lower $\chi^2_\nu$ values than the other cases.

\paragraph{Abundance percentages from different processes.}
Figure~\ref{fig:percentage} shows the elemental abundance fractions derived from the nucleosynthesis models presented in Fig.~\ref{fig:ele_fit_r}. For barium, the s$+$r model predicts contributions of $96\%$ and $4$\% from the s- and r-processes, respectively. For the mixed i$+$s$+$r models, the corresponding fractions are $31+65+4$\% for $f_{\mathrm{top}} = 0.05$ and $90+9+1$\% for $f_{\mathrm{top}} = 0.15$, reflecting the increasing dominance of the i-process contribution with stronger PIEs.

The observed A(Ba)$_{r/i-proc}$ value (blue dot in Fig.~\ref{fig:ABa}) agrees with the s$+$r theoretical prediction, A(Ba)$ = -0.26$~dex, obtained by subtracting the s-process contribution from the total barium abundance (red dash in Fig.~\ref{fig:ABa}).
When the i-process is included, the combined r$+$i contribution yields A(Ba)~$= 0.33$~dex for $f{_{\mathrm{top}}} = 0.05$ (green dash in Fig.~\ref{fig:ABa}) and A(Ba)~$= 0.95$~dex for $f_{\mathrm{top}} = 0.15$ (pink dash in Fig.~\ref{fig:ABa}).
Overall, the r$+$s-process model reproduces A(Ba)$_{r/i\text{-}proc}$ accurately, whereas the r$+$i-process predictions fall outside the $1\sigma$ uncertainty.

\section{Discussion}
\label{sec:discussion}

Our analysis yields stellar parameters and abundance patterns in excellent agreement with those reported by \citet{gull2018ApJ...862..174G}. This similarity provides a consistent baseline to address the ongoing debate regarding the nucleosynthetic origin of the heavy elements in TYC~6044-714-1.
While \citet{choplin2022A&A...667L..13C} recently suggested a pure $i$-process origin for this star, our results support the original r+s interpretation by \citet{gull2018ApJ...862..174G}. We encounter significant difficulties in reconciling a pure i-process or a mixed i+s-process scenario with the observed pattern and, crucially, with our new isotopic constraints.
Among the explored scenarios, the s+r model yields the lowest reduced $\chi^2_\nu$, providing an excellent fit to almost all elements across the three s-process peaks. The i+s+r models show a progressive improvement in the statistical quality of the fit as the overshooting efficiency increases, but they still present critical physical challenges.

In our calculations, which account for the $\alpha$ enhancement typical of the Galactic halo at this metallicity and is consistent with the specific abundances observed in TYC~6044-714-1, we applied an extra-mixing (overshooting) at the upper border of the convective thermal pulse to trigger the proton ingestion and the i-process nucleosynthesis. Despite a systematic exploration of the overshooting efficiency ($f_{\mathrm{top}}$), the resulting i+s+r models do not consistently improve the fit compared to the standard s+r scenario. Specifically, the $f_{\mathrm{top}} = 0.05$ case performs significantly worse than the s+r model. Although it reproduces some elements between the first and second neutron-capture peaks, this comes at the cost of overproducing Y and Zr, and underproducing elements between La and Nd as well as Hf. The $f_{\mathrm{top}} = 0.10$ model shows an improved $\chi^2_\nu$ relative to the $f_{\mathrm{top}} = 0.05$ case, but it overestimates Ba and Eu while underestimating Pb and Th. The $f_{\mathrm{top}} = 0.15$ model achieves a reduced $\chi^2_\nu$ comparable to the s+r case. This model solves several weaknesses of the lower overshooting scenarios: it reproduces Nb well and fits the Sr-Y-Zr region and the lanthanides satisfactorily, but it predicts a too low Th abundance. In addition, it predicts an i-process Ba fraction (90\%, see Fig.~\ref{fig:percentage}) that largely exceeds the observational one (11\% at most) derived from isotopic ratios (Sect.~\ref{sec:Ba_Eu_iso}). 

Despite the statistical equivalence between the s+r and i+s+r $f_{\mathrm{top}} = 0.15$  models, we consider the s+r interpretation more physically plausible. The $f_{\mathrm{top}} = 0.15$ model requires extreme conditions, specifically a violent extra-mixing process at the thermal pulse border, making it a rather exotic scenario. The improved performance of the i+s+r models for Nb deserves particular attention, as this element benefits from the additional i-process contribution. However, no 1D non-LTE or 3D non-LTE corrections are available for Nb in the literature, so constraints from this element should be interpreted with caution. Overall, while including an i-process contribution improves the reproduction of some elements, it does not yield a better global fit than the s+r model. 
We acknowledge the theoretical capability of the i-process to synthesize actinides under certain conditions \citep{choplin2022A&A...667L..13C}. However, the complete chemical signature of TYC~6044-714-1, including its isotopic composition, is much more consistently explained by mass transfer from an AGB companion dominated by standard s-process nucleosynthesis, superimposed on a pre-existing r-process enrichment.

A potential concern is whether the initial r-process enrichment of the birth cloud ([r/Fe]~$\approx 0.4$~dex) could have impacted the nucleosynthesis in the primary AGB or the final pattern observed in the secondary. We find this to be effect negligible. In the primary AGB star, freshly synthesised material reaches intrinsic levels of $\rm [El/Fe] \gtrsim 1.5$~dex, dominating over the initial r-seeds by more than an order of magnitude. 
Therefore, after strong dilution in the secondary ($d_\text{agb} \approx 0.03$), the s-process signature remains the only detectable contribution.
The final chemical signature of TYC 6044-714-1 is thus a simple combination of the secondary's own birth-cloud r-process enrichment and the accreted s-process material.

Having discussed the heavy-element abundances, we now turn to the constraints provided by carbon and nitrogen. Under the simplified assumption that no extra-mixing processes operate, i+s+r models experiencing PIEs exhibit high C and N envelope abundances. In these models, the proton ingestion leads to local production of large amounts of $^{13}$C and $^{14}$N \citep{cristallo09b,choplin2021A&A...648A.119C}. Once this enriched material is mixed into the envelope, all models experiencing PIEs predict similarly low carbon isotopic ratios, $^{12}$C/$^{13}$C $\simeq 4$, together with strong carbon and nitrogen enhancements ($\gtrsim 2.2$, see Figure~\ref{fig:ele_fit_is}). Subsequent standard TDU episodes progressively increase the surface $^{12}$C abundance, raising the $^{12}$C/$^{13}$C ratio while maintaining high C and N enhancements. This behaviour is common to the models with $f_{\mathrm{top}} = 0.05$, $0.10$, and $0.15$.

Models with larger overshooting parameters experience progressively stronger heavy-element enrichment during the PIEs. Although these models produce large absolute C and N enhancements, their C and N abundances are low relative to the neutron-capture elements, which dominate the $\chi^2$ minimisation. Consequently, when the dilution factor is set to reproduce the observed heavy-element abundances, carbon and nitrogen are over-diluted and underpredicted relative to observations. This effect becomes more pronounced for the $f_{\mathrm{top}} = 0.10$ and $0.15$ models. In this respect, the $f_{\mathrm{top}} = 0.05$ model provides the best agreement, yielding C and N abundances closest to the observed values after dilution.

The pure s+r model, which does not experience PIEs, predicts much lower nitrogen enrichment and extremely high carbon isotopic ratios, with $^{12}$C/$^{13}$C exceeding $8 \times 10^{3}$. Nevertheless, low-mass stars ($\lesssim 2 M_\odot$) are known to undergo extra mixing processes during both the RGB and AGB phases \citep[e.g.,][]{Gratton2000A&A...354..169G,Denissenkov2003ApJ...593..509D,Nollett2003ApJ...582.1036N,Busso2010ApJ...717L..47B}. Such processes can significantly lower $^{12}$C/$^{13}$C and $[\mathrm{C/Fe}]$, while increasing $[\mathrm{N/Fe}]$, thereby improving agreement with the observed CN abundances.
In the context of a binary system, such processes could have operated in the primary AGB donor, in the secondary star during its current RGB ascent \citep[e.g.,][]{Placco2014ApJ...797...21P}, or as a combination of both.

To illustrate this, we computed two independent exploratory models. In the first case, following previous works \citep[e.g.,][]{Abia2017A&A...599A..39A}, extra mixing was included only during the AGB phase of the primary star assuming a constant mixing velocity $v_{\rm mix} = 100~\mathrm{cm~s^{-1}}$ below the convective envelope, extending down to layers where the temperature reaches $T_{\rm mix} = 36$~MK. As a result, nitrogen abundance is enhanced, while the carbon isotopic ratio is lowered before the mass transfer occurs. Carbon elemental abundance in only slightly depleted.
After dilution with the companion’s envelope, the resulting surface abundances are in good agreement with the observed values of TYC~6044-714-1 (see Fig.~\ref{fig:ele_fit_r}).
In the second case, we modelled the internal evolution of the secondary star only ($0.8\,M_\odot$), assuming extra mixing ($v_{\rm mix} = 1~\mathrm{cm~s^{-1}}$, $T_{\rm mix} = 23$~MK) during the RGB phase, specifically after the luminosity bump.
As shown in Fig.~\ref{fig:cn_logg}, this second scenario effectively reprocesses the accreted carbon into nitrogen, matching the observed [N/Fe] and $^{12}$C/$^{13}$C values at the current evolutionary stage of TYC~6044-714-1 ($\log g \approx 1.7$). In this case, carbon remains slightly overproduced compared to observations due to the high carbon content originally predicted for the primary AGB.
While we cannot uniquely determine whether the mixing occurred in the primary, the secondary, or both, these tests demonstrate that a modest amount of extra mixing -- a standard physical process in metal-poor giants -- suffices to reproduce the observed CN abundances. Consequently, light elements do not provide a decisive diagnostic against the s+r scenario; on the contrary, i-process models with $f_{\mathrm{top}} = 0.10$ and $0.15$ are further disfavoured because they underpredict C and N after dilution. Since any subsequent extra mixing would further reduce [C/Fe] while increasing [N/Fe], it would accentuate the discrepancy in these models, whereas it provides a consistent physical solution for the s+r case.

\section{Summary and conclusion}
\label{sec:conclusions}

We have analysed the chemical abundances of the {CEMP-rs} star TYC~6044-714-1 using high-quality observations combined with detailed nucleosynthesis modelling. Among the explored scenarios, the s+r model provides the best overall fit, achieving the lowest reduced $\chi^2_\nu$ and reproducing the heavy-element pattern across all three s-process peaks. Although i+s+r models with increasing overshooting efficiency ($f_{\mathrm{top}} = 0.05$–0.15) improve the fit for certain elements, particularly Nb and those between the first and second neutron-capture peaks, they generally fail to reproduce the full abundance pattern consistently. The $f_{\mathrm{top}} = 0.15$ model achieves a statistical fit comparable to s+r, but requires extreme and physically unlikely conditions, making the pure s+r scenario more plausible.
In addition, it predicts an s-process Ba fraction of 9\%, which is inconsistent with the value of 
$89\pm12$\% inferred from isotopic ratios.

Carbon and nitrogen abundances provide additional constraints. i+s+r models experiencing PIEs predict large absolute enhancements of C and N, but when diluted to match heavy-element abundances, they underpredict the observed C and N. Among these models, the $f_{\mathrm{top}} = 0.05$ case yields the closest agreement with the observed CN abundances. In contrast, the s+r model, which does not include PIEs, predicts very high $^{12}$C/$^{13}$C ratios and low N enrichment; however, the inclusion of modest extra-mixing processes during the AGB phase reproduces the observed C and N abundances, consistent with expectations for low-mass stars.

Overall, our results indicate that the chemical pattern of TYC~6044-714-1 is most consistently explained by mass transfer from an AGB companion dominated by s-process nucleosynthesis, superimposed on a pre-existing r-process enrichment from standard Galactic chemical-evolution. While the i-process may contribute to certain elements, its inclusion does not improve the global fit, and extreme i-process scenarios are disfavoured. Light elements alone cannot provide a decisive diagnostic, but the combination of heavy-element abundances and isotopic ratios supports a dominant s+r origin for this star’s enrichment.

In stars with noisier blue spectra ($\lambda \lesssim 4800$~\AA) or stronger blending from carbon features, the uncertainties in elemental abundances may easily increase by a factor of two to three relative to those reported here. Under such conditions, discriminating between the s$+$r and i$+$s$+$r nucleosynthetic scenarios solely through comparisons between observed and model-predicted abundances becomes nearly impossible. We demonstrate that a careful analysis of the \ion{Ba}{ii} resonance line at 4934~\AA, aimed at determining the s-process contribution to the total barium abundance, can substantially improve the ability to discriminate between the nucleosynthetic processes responsible for the observed chemical pattern.

The UVES spectrum employed in this study closely reproduces the wavelength coverage and resolving power foreseen for HRMOS, a very high-resolution spectrograph that will be proposed as a future instrument for the ESO VLT. It thus provides an empirical benchmark for assessing the expected performance of one of its proposed science cases \citep{magrini2023arXiv231208270M}. Given that the current requirement for HRMOS is a multiplex capability of 50-80, this same analysis could in principle be extended to comparably large stellar samples in a single observational set-up.

\begin{acknowledgements}
The authors thank A. Amarsi for discussions and support to the manuscript.
The authors thank the anonymous referee for constructive criticism that significantly improved the quality of the paper.
R.E.G. and L.M. acknowledge support from INAF through the Large Grants EPOCH and WST, funding for the WEAVE project, the Mini-Grants Checs (1.05.23.04.02), and financial support under the National Recovery and Resilience Plan (PNRR), Mission 4, Component 2, Investment 1.1, Call for tender No. 104 published on 2 February 2022 by the Italian Ministry of University and Research (MUR), funded by the European Union – NextGenerationEU, through the Project ‘Cosmic POT’ (Grant Assignment Decree No. 2022X4TM3H, MUR).
D.V. and S.C. acknowledge funding by the European Union – NextGenerationEU RFF M4C2 1.1 PRIN 2022 project "2022RJLWHN URKA" and by INAF 2023 Theory Grant ObFu 1.05.23.06.06 “Understanding R-process \& Kilonovae Aspects".
L.P. acknowledges partial financial support from the INAF Minigrant 2023 Self-consistent Modeling of Interacting Binary Systems (1.05.23.04.02 )
Use was made of the Simbad database, operated at the CDS, Strasbourg, France, and of NASA’s Astrophysics Data System Bibliographic Services. This publication makes use of data products from the Two Micron All Sky Survey, which is a joint project of the University of Massachusetts and the Infrared Processing and Analysis Center/California Institute of Technology, funded by the National Aeronautics and Space Administration and the National Science Foundation. This research used Astropy \url{http://www.astropy.org} a community-developed core Python package for Astronomy \citep{astropy:2018}. This work presents results from the European Space Agency (ESA) space mission Gaia. Gaia data are processed by the Gaia Data Processing and Analysis Consortium (DPAC). Funding for the DPAC is provided by national institutions, in particular the institutions participating in the Gaia MultiLateral Agreement (MLA). The Gaia mission website is \url{https://www.cosmos.esa.int/gaia}. The Gaia archive website is \url{https://archives.esac.esa.int/gaia}. Based on observations made with ESO Telescopes at the La Silla Paranal Observatory under programme IDs 114.27JT.001 or 0114.B-0030(A) and 165.N-0276(A).
\end{acknowledgements}

\bibliographystyle{aa.bst}

\bibliography{Faint2}

@ARTICLE{Placco2014ApJ...797...21P,
       author = {{Placco}, Vinicius M. and {Frebel}, Anna and {Beers}, Timothy C. and {Stancliffe}, Richard J.},
        title = "{Carbon-enhanced Metal-poor Star Frequencies in the Galaxy: Corrections for the Effect of Evolutionary Status on Carbon Abundances}",
      journal = {\apj},
         year = 2014,
        month = dec,
       volume = {797},
       number = {1},
          eid = {21},
        pages = {21},
          doi = {10.1088/0004-637X/797/1/21},
archivePrefix = {arXiv},
       eprint = {1410.2223},
 primaryClass = {astro-ph.SR},
       adsurl = {https://ui.adsabs.harvard.edu/abs/2014ApJ...797...21P}
}

@ARTICLE{giribaldi2026_barium,
       author = {{Giribaldi}, Riano E. and {Palla}, Marco and {Magrini}, Laura and {Rizzuti}, Federico and {Cescutti}, gabriele and {Vescovi}, Deigo and {Cristallo}, Sergio and {Belmonte}, Maria Teresa and {Randich}, Sofia},
        title = "{The nucleosynthesis of Ba in the Early Universe. Constraints from elemental abundances and isotopic ratios}",
      journal = {arXiv e-prints},
         year = 2026,
        month = may,
          eid = {arXiv:2605.25201},
        pages = {arXiv:2605.25201},
archivePrefix = {arXiv},
       eprint = {2605.25201},
 primaryClass = {astro-ph.SR},
       adsurl = {https://ui.adsabs.harvard.edu/abs/2026arXiv260525201G}
}

@ARTICLE{hansen2012A&A...545A..31H,
       author = {{Hansen}, C.~J. and {Primas}, F. and {Hartman}, H. and {Kratz}, K.-L. and {Wanajo}, S. and {Leibundgut}, B. and {Farouqi}, K. and {Hallmann}, O. and {Christlieb}, N. and {Nilsson}, H.},
        title = "{Silver and palladium help unveil the nature of a second r-process}",
      journal = {\aap},
         year = 2012,
        month = sep,
       volume = {545},
          eid = {A31},
        pages = {A31},
          doi = {10.1051/0004-6361/201118643},
archivePrefix = {arXiv},
       eprint = {1205.4744},
 primaryClass = {astro-ph.SR},
       adsurl = {https://ui.adsabs.harvard.edu/abs/2012A&A...545A..31H}
}

@ARTICLE{Sneden2008ARA&A..46..241S,
       author = {{Sneden}, C. and {Cowan}, J.~J. and {Gallino}, R.},
        title = "{Neutron-capture elements in the early galaxy.}",
      journal = {\araa},
         year = 2008,
        month = sep,
       volume = {46},
        pages = {241-288},
          doi = {10.1146/annurev.astro.46.060407.145207},
       adsurl = {https://ui.adsabs.harvard.edu/abs/2008ARA&A..46..241S}
}

@ARTICLE{Gratton2000A&A...354..169G,
       author = {{Gratton}, R.~G. and {Sneden}, C. and {Carretta}, E. and {Bragaglia}, A.},
        title = "{Mixing along the red giant branch in metal-poor field stars}",
      journal = {\aap},
         year = 2000,
        month = feb,
       volume = {354},
        pages = {169-187},
       adsurl = {https://ui.adsabs.harvard.edu/abs/2000A&A...354..169G}
}

@ARTICLE{Abia2017A&A...599A..39A,
       author = {{Abia}, C. and {Hedrosa}, R.~P. and {Dom{\'\i}nguez}, I. and {Straniero}, O.},
        title = "{The puzzle of the CNO isotope ratios in asymptotic giant branch carbon stars}",
      journal = {\aap},
         year = 2017,
        month = mar,
       volume = {599},
          eid = {A39},
        pages = {A39},
          doi = {10.1051/0004-6361/201629969},
archivePrefix = {arXiv},
       eprint = {1611.06400},
 primaryClass = {astro-ph.SR},
       adsurl = {https://ui.adsabs.harvard.edu/abs/2017A&A...599A..39A}
}

@ARTICLE{Busso2010ApJ...717L..47B,
       author = {{Busso}, M. and {Palmerini}, S. and {Maiorca}, E. and {Cristallo}, S. and {Straniero}, O. and {Abia}, C. and {Gallino}, R. and {La Cognata}, M.},
        title = "{On the Need for Deep-mixing in Asymptotic Giant Branch Stars of Low Mass}",
      journal = {\apjl},
         year = 2010,
        month = jul,
       volume = {717},
       number = {1},
        pages = {L47-L51},
          doi = {10.1088/2041-8205/717/1/L47},
archivePrefix = {arXiv},
       eprint = {1005.3549},
 primaryClass = {astro-ph.SR},
       adsurl = {https://ui.adsabs.harvard.edu/abs/2010ApJ...717L..47B}
}

@ARTICLE{Nollett2003ApJ...582.1036N,
       author = {{Nollett}, Kenneth M. and {Busso}, M. and {Wasserburg}, G.~J.},
        title = "{Cool Bottom Processes on the Thermally Pulsing Asymptotic Giant Branch and the Isotopic Composition of Circumstellar Dust Grains}",
      journal = {\apj},
         year = 2003,
        month = jan,
       volume = {582},
       number = {2},
        pages = {1036-1058},
          doi = {10.1086/344817},
archivePrefix = {arXiv},
       eprint = {astro-ph/0211271},
 primaryClass = {astro-ph},
       adsurl = {https://ui.adsabs.harvard.edu/abs/2003ApJ...582.1036N}
}

@ARTICLE{Denissenkov2003ApJ...593..509D,
       author = {{Denissenkov}, Pavel A. and {VandenBerg}, Don A.},
        title = "{Canonical Extra Mixing in Low-Mass Red Giants}",
      journal = {\apj},
         year = 2003,
        month = aug,
       volume = {593},
       number = {1},
        pages = {509-523},
          doi = {10.1086/376410},
       adsurl = {https://ui.adsabs.harvard.edu/abs/2003ApJ...593..509D}
}

@ARTICLE{Lind2024ARA&A..62..475L,
       author = {{Lind}, Karin and {Amarsi}, Anish M.},
        title = "{Three-Dimensional Nonlocal Thermodynamic Equilibrium Abundance Analyses of Late-Type Stars}",
      journal = {\araa},
         year = 2024,
        month = sep,
       volume = {62},
       number = {1},
        pages = {475-527},
          doi = {10.1146/annurev-astro-052722-103557},
archivePrefix = {arXiv},
       eprint = {2401.00697},
 primaryClass = {astro-ph.SR},
       adsurl = {https://ui.adsabs.harvard.edu/abs/2024ARA&A..62..475L}
}

@ARTICLE{iwamoto04,
       author = {{Iwamoto}, Nobuyuki and {Kajino}, Toshitaka and {Mathews}, Grant J. and {Fujimoto}, Masayuki Y. and {Aoki}, Wako},
        title = "{Flash-Driven Convective Mixing in Low-Mass, Metal-deficient Asymptotic Giant Branch Stars: A New Paradigm for Lithium Enrichment and a Possible s-Process}",
      journal = {\apj},
         year = 2004,
        month = feb,
       volume = {602},
       number = {1},
        pages = {377-388},
          doi = {10.1086/380989},
       adsurl = {https://ui.adsabs.harvard.edu/abs/2004ApJ...602..377I}
}

@ARTICLE{spyrou24,
       author = {{Spyrou}, A. and {M{\"u}cher}, D. and {Denissenkov}, P.~A. and {Herwig}, F. and {Good}, E.~C. and {Balk}, G. and {Berg}, H.~C. and {Bleuel}, D.~L. and {Clark}, J.~A. and {Dembski}, C. and {DeYoung}, P.~A. and {Greaves}, B. and {Guttormsen}, M. and {Harris}, C. and {Larsen}, A.~C. and {Liddick}, S.~N. and {Lyons}, S. and {Markova}, M. and {Mogannam}, M.~J. and {Nikas}, S. and {Owens-Fryar}, J. and {Palmisano-Kyle}, A. and {Perdikakis}, G. and {Pogliano}, F. and {Quintieri}, M. and {Richard}, A.~L. and {Santiago-Gonzalez}, D. and {Savard}, G. and {Smith}, M.~K. and {Sweet}, A. and {Tsantiri}, A. and {Wiedeking}, M.},
        title = "{First Study of the $^{139}$Ba (n ,{\ensuremath{\gamma}} )$^{140}$Ba Reaction to Constrain the Conditions for the Astrophysical i Process}",
      journal = {\prl},
         year = 2024,
        month = may,
       volume = {132},
       number = {20},
          eid = {202701},
        pages = {202701},
          doi = {10.1103/PhysRevLett.132.202701},
       adsurl = {https://ui.adsabs.harvard.edu/abs/2024PhRvL.132t2701S}
}

@ARTICLE{masseron10,
       author = {{Masseron}, T. and {Johnson}, J.~A. and {Plez}, B. and {van Eck}, S. and {Primas}, F. and {Goriely}, S. and {Jorissen}, A.},
        title = "{A holistic approach to carbon-enhanced metal-poor stars}",
      journal = {\aap},
         year = 2010,
        month = jan,
       volume = {509},
          eid = {A93},
        pages = {A93},
          doi = {10.1051/0004-6361/200911744},
archivePrefix = {arXiv},
       eprint = {0901.4737},
 primaryClass = {astro-ph.SR},
       adsurl = {https://ui.adsabs.harvard.edu/abs/2010A&A...509A..93M}
}

@ARTICLE{cristallo16,
       author = {{Cristallo}, S. and {Karinkuzhi}, D. and {Goswami}, A. and {Piersanti}, L. and {Gobrecht}, D.},
        title = "{Constraints of the Physics of Low-mass AGB Stars from CH and CEMP Stars}",
      journal = {\apj},
         year = 2016,
        month = dec,
       volume = {833},
       number = {2},
          eid = {181},
        pages = {181},
          doi = {10.3847/1538-4357/833/2/181},
archivePrefix = {arXiv},
       eprint = {1610.05475},
 primaryClass = {astro-ph.SR},
       adsurl = {https://ui.adsabs.harvard.edu/abs/2016ApJ...833..181C}
}

@ARTICLE{cristallo09a,
       author = {{Cristallo}, S. and {Straniero}, O. and {Gallino}, R. and {Piersanti}, L. and {Dom{\'\i}nguez}, I. and {Lederer}, M.~T.},
        title = "{Evolution, Nucleosynthesis, and Yields of Low-Mass Asymptotic Giant Branch Stars at Different Metallicities}",
      journal = {\apj},
         year = 2009,
        month = may,
       volume = {696},
       number = {1},
        pages = {797-820},
          doi = {10.1088/0004-637X/696/1/797},
archivePrefix = {arXiv},
       eprint = {0902.0243},
 primaryClass = {astro-ph.SR},
       adsurl = {https://ui.adsabs.harvard.edu/abs/2009ApJ...696..797C}
}

@ARTICLE{cristallo09b,
       author = {{Cristallo}, S. and {Piersanti}, L. and {Straniero}, O. and {Gallino}, R. and {Dom{\'\i}nguez}, I. and {K{\"a}ppeler}, F.},
        title = "{Asymptotic-Giant-Branch Models at Very Low Metallicity}",
      journal = {\pasa},
         year = 2009,
        month = aug,
       volume = {26},
       number = {3},
        pages = {139-144},
          doi = {10.1071/AS09003},
archivePrefix = {arXiv},
       eprint = {0904.4173},
 primaryClass = {astro-ph.SR},
       adsurl = {https://ui.adsabs.harvard.edu/abs/2009PASA...26..139C}
}

@ARTICLE{fujimoto2000,
       author = {{Fujimoto}, Masayuki Y. and {Ikeda}, Yasufumi and {Iben}, Jr., Icko},
        title = "{The Origin of Extremely Metal-poor Carbon Stars and the Search for Population III}",
      journal = {\apjl},
         year = 2000,
        month = jan,
       volume = {529},
       number = {1},
        pages = {L25-L28},
          doi = {10.1086/312453},
       adsurl = {https://ui.adsabs.harvard.edu/abs/2000ApJ...529L..25F}
}

@ARTICLE{panov10,
       author = {{Panov}, I.~V. and {Korneev}, I. Yu. and {Rauscher}, T. and {Mart{\'\i}nez-Pinedo}, G. and {Keli{\'c}-Heil}, A. and {Zinner}, N.~T. and {Thielemann}, F.-K.},
        title = "{Neutron-induced astrophysical reaction rates for translead nuclei}",
      journal = {\aap},
         year = 2010,
        month = apr,
       volume = {513},
          eid = {A61},
        pages = {A61},
          doi = {10.1051/0004-6361/200911967},
archivePrefix = {arXiv},
       eprint = {0911.2181},
 primaryClass = {astro-ph.SR},
       adsurl = {https://ui.adsabs.harvard.edu/abs/2010A&A...513A..61P}
}

@ARTICLE{marketin16,
       author = {{Marketin}, T. and {Huther}, L. and {Mart{\'\i}nez-Pinedo}, G.},
        title = "{Large-scale evaluation of {\ensuremath{\beta}} -decay rates of r -process nuclei with the inclusion of first-forbidden transitions}",
      journal = {\prc},
         year = 2016,
        month = feb,
       volume = {93},
       number = {2},
          eid = {025805},
        pages = {025805},
          doi = {10.1103/PhysRevC.93.025805},
archivePrefix = {arXiv},
       eprint = {1507.07442},
 primaryClass = {nucl-th},
       adsurl = {https://ui.adsabs.harvard.edu/abs/2016PhRvC..93b5805M}
}

@ARTICLE{kondev21,
       author = {{Kondev}, F.~G. and {Wang}, M. and {Huang}, W.~J. and {Naimi}, S. and {Audi}, G.},
        title = "{The NUBASE2020 evaluation of nuclear physics properties}",
      journal = {Chinese Physics C},
         year = 2021,
        month = mar,
       volume = {45},
       number = {3},
          eid = {030001},
        pages = {030001},
          doi = {10.1088/1674-1137/abddae},
       adsurl = {https://ui.adsabs.harvard.edu/abs/2021ChPhC..45c0001K}
}

@ARTICLE{rochman25,
       author = {{Rochman}, D. and {Koning}, A. and {Goriely}, S. and {Hilaire}, S.},
        title = "{TENDL-astro: A new nuclear data set for astrophysics interest}",
      journal = {\nphysa},
         year = 2025,
        month = jan,
       volume = {1053},
          eid = {122951},
        pages = {122951},
          doi = {10.1016/j.nuclphysa.2024.122951},
archivePrefix = {arXiv},
       eprint = {2510.11262},
 primaryClass = {nucl-th},
       adsurl = {https://ui.adsabs.harvard.edu/abs/2025NuPhA105322951R}
}

@ARTICLE{vescovi20,
       author = {{Vescovi}, Diego and {Cristallo}, Sergio and {Busso}, Maurizio and {Liu}, Nan},
        title = "{Magnetic-buoyancy-induced Mixing in AGB Stars: Presolar SiC Grains}",
      journal = {\apjl},
         year = 2020,
        month = jul,
       volume = {897},
       number = {2},
          eid = {L25},
        pages = {L25},
          doi = {10.3847/2041-8213/ab9fa1},
archivePrefix = {arXiv},
       eprint = {2006.13729},
 primaryClass = {astro-ph.SR},
       adsurl = {https://ui.adsabs.harvard.edu/abs/2020ApJ...897L..25V}
}

@ARTICLE{vescovi21b,
       author = {{Vescovi}, Diego},
        title = "{Mixing and Magnetic Fields in Asymptotic Giant Branch Stars in the Framework of FRUITY Models}",
      journal = {Universe},
         year = 2021,
        month = dec,
       volume = {8},
       number = {1},
          eid = {16},
        pages = {16},
          doi = {10.3390/universe8010016},
       adsurl = {https://ui.adsabs.harvard.edu/abs/2021Univ....8...16V}
}

@ARTICLE{vescovi21a,
       author = {{Vescovi}, D. and {Cristallo}, S. and {Palmerini}, S. and {Abia}, C. and {Busso}, M.},
        title = "{Magnetic-buoyancy-induced mixing in AGB stars: Fluorine nucleosynthesis at different metallicities}",
      journal = {\aap},
         year = 2021,
        month = aug,
       volume = {652},
          eid = {A100},
        pages = {A100},
          doi = {10.1051/0004-6361/202141173},
archivePrefix = {arXiv},
       eprint = {2106.08241},
 primaryClass = {astro-ph.SR},
       adsurl = {https://ui.adsabs.harvard.edu/abs/2021A&A...652A.100V}
}

@ARTICLE{straniero06,
       author = {{Straniero}, Oscar and {Gallino}, Roberto and {Cristallo}, Sergio},
        title = "{s process in low-mass asymptotic giant branch stars}",
      journal = {\nphysa},
         year = 2006,
        month = oct,
       volume = {777},
        pages = {311-339},
          doi = {10.1016/j.nuclphysa.2005.01.011},
archivePrefix = {arXiv},
       eprint = {astro-ph/0501405},
 primaryClass = {astro-ph},
       adsurl = {https://ui.adsabs.harvard.edu/abs/2006NuPhA.777..311S}
}

@ARTICLE{Arnaboldi2011Msngr.144...17A,
       author = {{Arnaboldi}, M. and {Retzlaff}, J. and {Slijkhuis}, R. and {Forch{\'\i}}, V. and {Nunes}, P. and {Sforna}, D. and {Zampieri}, S. and {Bierwirth}, T. and {Comer{\'o}n}, F. and {P{\'e}ron}, M. and {Romaniello}, M. and {Suchar}, D.},
        title = "{Phase 3 -- Handling Data Products from ESO Public Surveys, Large Programmes and Other Contributions}",
      journal = {The Messenger},
         year = 2011,
        month = jun,
       volume = {144},
        pages = {17-19},
       adsurl = {https://ui.adsabs.harvard.edu/abs/2011Msngr.144...17A}
}

@ARTICLE{mashonkina2012A&A...540A..98M,
       author = {{Mashonkina}, L. and {Ryabtsev}, A. and {Frebel}, A.},
        title = "{Non-LTE effects on the lead and thorium abundance determinations for cool stars}",
      journal = {\aap},
         year = 2012,
        month = apr,
       volume = {540},
          eid = {A98},
        pages = {A98},
          doi = {10.1051/0004-6361/201218790},
archivePrefix = {arXiv},
       eprint = {1202.2630},
 primaryClass = {astro-ph.GA},
       adsurl = {https://ui.adsabs.harvard.edu/abs/2012A&A...540A..98M}
}

@ARTICLE{hill2002A&A...387..560H,
       author = {{Hill}, V. and {Plez}, B. and {Cayrel}, R. and {Beers}, T.~C. and {Nordstr{\"o}m}, B. and {Andersen}, J. and {Spite}, M. and {Spite}, F. and {Barbuy}, B. and {Bonifacio}, P. and {Depagne}, E. and {Fran{\c{c}}ois}, P. and {Primas}, F.},
        title = "{First stars. I. The extreme r-element rich, iron-poor halo giant CS 31082-001. Implications for the r-process site(s) and radioactive cosmochronology}",
      journal = {\aap},
         year = 2002,
        month = may,
       volume = {387},
        pages = {560-579},
          doi = {10.1051/0004-6361:20020434},
archivePrefix = {arXiv},
       eprint = {astro-ph/0203462},
 primaryClass = {astro-ph},
       adsurl = {https://ui.adsabs.harvard.edu/abs/2002A&A...387..560H}
}

@ARTICLE{da_silva2025A&A...696A.122D,
       author = {{da Silva}, A.~R. and {Smiljanic}, R.},
        title = "{Discovery of a pair of very metal-poor stars enriched in neutron-capture elements: The proto-disk r-II star BPS CS 29529-0089 and the Gaia-Sausage-Enceladus r-I star TYC 9219-2422-1}",
      journal = {\aap},
         year = 2025,
        month = apr,
       volume = {696},
          eid = {A122},
        pages = {A122},
          doi = {10.1051/0004-6361/202453295},
archivePrefix = {arXiv},
       eprint = {2503.04926},
 primaryClass = {astro-ph.SR},
       adsurl = {https://ui.adsabs.harvard.edu/abs/2025A&A...696A.122D}
}

@ARTICLE{dutra-ferreira2016A&A...585A..75D,
       author = {{Dutra-Ferreira}, L. and {Pasquini}, L. and {Smiljanic}, R. and {Porto de Mello}, G.~F. and {Steffen}, M.},
        title = "{Consistent metallicity scale for cool dwarfs and giants. A benchmark test using the Hyades}",
      journal = {\aap},
         year = 2016,
        month = jan,
       volume = {585},
          eid = {A75},
        pages = {A75},
          doi = {10.1051/0004-6361/201526783},
archivePrefix = {arXiv},
       eprint = {1509.07725},
 primaryClass = {astro-ph.SR},
       adsurl = {https://ui.adsabs.harvard.edu/abs/2016A&A...585A..75D}
}

@ARTICLE{giribaldi2025,
       author = {{Giribaldi}, Riano E. and {Magrini}, Laura and {Rossi}, Martina and {Amarsi}, Anish M. and {Romano}, Donatella and {Massari}, Davide},
        title = "{The most metal-poor tail of the Galactic halo: Hypothesis for its origin from precise spectral analysis}",
      journal = {\aap},
         year = 2025,
        month = jun,
       volume = {698},
          eid = {A11},
        pages = {A11},
          doi = {10.1051/0004-6361/202453470},
archivePrefix = {arXiv},
       eprint = {2503.19472},
 primaryClass = {astro-ph.GA},
       adsurl = {https://ui.adsabs.harvard.edu/abs/2025A&A...698A..11G}
}

@ARTICLE{choplin2022A&A...667L..13C,
       author = {{Choplin}, A. and {Goriely}, S. and {Siess}, L.},
        title = "{Synthesis of thorium and uranium in asymptotic giant branch stars}",
      journal = {\aap},
         year = 2022,
        month = nov,
       volume = {667},
          eid = {L13},
        pages = {L13},
          doi = {10.1051/0004-6361/202244928},
archivePrefix = {arXiv},
       eprint = {2211.03824},
 primaryClass = {astro-ph.SR},
       adsurl = {https://ui.adsabs.harvard.edu/abs/2022A&A...667L..13C}
}

@ARTICLE{gallagher2020A&A...634A..55G,
       author = {{Gallagher}, A.~J. and {Bergemann}, M. and {Collet}, R. and {Plez}, B. and {Leenaarts}, J. and {Carlsson}, M. and {Yakovleva}, S.~A. and {Belyaev}, A.~K.},
        title = "{Observational constraints on the origin of the elements. II. 3D non-LTE formation of Ba II lines in the solar atmosphere}",
      journal = {\aap},
         year = 2020,
        month = feb,
       volume = {634},
          eid = {A55},
        pages = {A55},
          doi = {10.1051/0004-6361/201936104},
archivePrefix = {arXiv},
       eprint = {1910.03898},
 primaryClass = {astro-ph.SR},
       adsurl = {https://ui.adsabs.harvard.edu/abs/2020A&A...634A..55G}
}

@ARTICLE{magrini2023arXiv231208270M,
       author = {{Magrini}, Laura and {Bensby}, Thomas and {Brucalassi}, Anna and {Randich}, Sofia and {Jeffries}, Robin and {de Silva}, Gayandhi and {Skuladottir}, Asa and {Smiljanic}, Rodolfo and {Gonzalez}, Oscar and {Hill}, Vanessa and {Lagarde}, Nadege and {Tolstoy}, Eline and {Arroyo-Polonio}, Jose' Maria and {Baratella}, Martina and {Barnes}, John R. and {Battaglia}, Giuseppina and {Baumgardt}, Holger and {Bellazzini}, Michele and {Biazzo}, Katia and {Bragaglia}, Angela and {Carter}, Bradley and {Casali}, Giada and {Cescutti}, Gabriele and {Danielski}, Camilla and {Delgado Mena}, Elisa and {Drazdauskas}, Arnas and {Gieles}, Mark and {Giribaldi}, Riano and {Hawkins}, Keith and {Hoeijmakers}, H. Jens and {Jablonka}, Pascale and {Kamath}, Devika and {Louth}, Tom and {Fabiola Marino}, Anna and {Martell}, Sarah and {Merle}, Thibault and {Montet}, Benjamin and {Murphy}, Michael T. and {Nisini}, Brunella and {Nordlander}, Thomas and {D'Orazi}, Valentina and {Pino}, Lorenzo and {Romano}, Donatella and {Sacco}, Germano and {Sandford}, Nathan R. and {Sollima}, Antonio and {Spina}, Lorenzo and {Tautvaisiene}, Grazina and {Ting}, Yuan-Sen and {Tozzi}, Andrea and {Van der Swaelmen}, Mathieu and {Van Eck}, Sophie and {Watson}, Stephen and {Worley}, C. Clare and {Zocchi}, Alice},
        title = "{HRMOS White Paper: Science Motivation}",
      journal = {arXiv e-prints},
         year = 2023,
        month = dec,
          eid = {arXiv:2312.08270},
        pages = {arXiv:2312.08270},
          doi = {10.48550/arXiv.2312.08270},
archivePrefix = {arXiv},
       eprint = {2312.08270},
 primaryClass = {astro-ph.IM},
       adsurl = {https://ui.adsabs.harvard.edu/abs/2023arXiv231208270M}
}

@MISC{Linfor3D,
       author = {{Steffen}, Matthias and {Ludwig}, Hans-G\"unter and {Wedemeyer-B\"ohm}, Sven and {Gallagher}, andy},
        title = "{Linfor3D User Manual}",
         year = 2023,
        month = nov,
          eid = {7.3.0 / 0.9.0},
       adsurl = {https://www.aip.de/en/members/matthias-steffen/linfor3d-user-manual/}
}

@ARTICLE{sakari2018ApJ...868..110S,
       author = {{Sakari}, Charli M. and {Placco}, Vinicius M. and {Farrell}, Elizabeth M. and {Roederer}, Ian U. and {Wallerstein}, George and {Beers}, Timothy C. and {Ezzeddine}, Rana and {Frebel}, Anna and {Hansen}, Terese and {Holmbeck}, Erika M. and {Sneden}, Christopher and {Cowan}, John J. and {Venn}, Kim A. and {Davis}, Christopher Evan and {Matijevi{\v{c}}}, Gal and {Wyse}, Rosemary F.~G. and {Bland-Hawthorn}, Joss and {Chiappini}, Cristina and {Freeman}, Kenneth C. and {Gibson}, Brad K. and {Grebel}, Eva K. and {Helmi}, Amina and {Kordopatis}, Georges and {Kunder}, Andrea and {Navarro}, Julio and {Reid}, Warren and {Seabroke}, George and {Steinmetz}, Matthias and {Watson}, Fred},
        title = "{The R-Process Alliance: First Release from the Northern Search for r-process-enhanced Metal-poor Stars in the Galactic Halo}",
      journal = {\apj},
         year = 2018,
        month = dec,
       volume = {868},
       number = {2},
          eid = {110},
        pages = {110},
          doi = {10.3847/1538-4357/aae9df},
archivePrefix = {arXiv},
       eprint = {1809.09156},
 primaryClass = {astro-ph.SR},
       adsurl = {https://ui.adsabs.harvard.edu/abs/2018ApJ...868..110S}
}

@ARTICLE{beers2005ARA&A..43..531B,
       author = {{Beers}, Timothy C. and {Christlieb}, Norbert},
        title = "{The Discovery and Analysis of Very Metal-Poor Stars in the Galaxy}",
      journal = {\araa},
         year = 2005,
        month = sep,
       volume = {43},
       number = {1},
        pages = {531-580},
          doi = {10.1146/annurev.astro.42.053102.134057},
       adsurl = {https://ui.adsabs.harvard.edu/abs/2005ARA&A..43..531B}
}

@ARTICLE{gull2018ApJ...862..174G,
       author = {{Gull}, Maude and {Frebel}, Anna and {Cain}, Madelyn G. and {Placco}, Vinicius M. and {Ji}, Alexander P. and {Abate}, Carlo and {Ezzeddine}, Rana and {Karakas}, Amanda I. and {Hansen}, Terese T. and {Sakari}, Charli and {Holmbeck}, Erika M. and {Santucci}, Rafael M. and {Casey}, Andrew R. and {Beers}, Timothy C.},
        title = "{The R-Process Alliance: Discovery of the First Metal-poor Star with a Combined r- and s-process Element Signature}",
      journal = {\apj},
         year = 2018,
        month = aug,
       volume = {862},
       number = {2},
          eid = {174},
        pages = {174},
          doi = {10.3847/1538-4357/aacbc3},
archivePrefix = {arXiv},
       eprint = {1806.00645},
 primaryClass = {astro-ph.SR},
       adsurl = {https://ui.adsabs.harvard.edu/abs/2018ApJ...862..174G}
}

@ARTICLE{sbordone2020A&A...641A.135S,
       author = {{Sbordone}, L. and {Hansen}, C.~J. and {Monaco}, L. and {Cristallo}, S. and {Bonifacio}, P. and {Caffau}, E. and {Villanova}, S. and {Amigo}, P.},
        title = "{A wide angle view of the Sagittarius dwarf spheroidal galaxy. II. A CEMP-r/s star in the Sagittarius dwarf spheroidal galaxy}",
      journal = {\aap},
         year = 2020,
        month = sep,
       volume = {641},
          eid = {A135},
        pages = {A135},
          doi = {10.1051/0004-6361/202037908},
archivePrefix = {arXiv},
       eprint = {2005.03027},
 primaryClass = {astro-ph.SR},
       adsurl = {https://ui.adsabs.harvard.edu/abs/2020A&A...641A.135S}
}

@ARTICLE{choplin2021A&A...648A.119C,
       author = {{Choplin}, A. and {Siess}, L. and {Goriely}, S.},
        title = "{The intermediate neutron capture process. I. Development of the i-process in low-metallicity low-mass AGB stars}",
      journal = {\aap},
         year = 2021,
        month = apr,
       volume = {648},
          eid = {A119},
        pages = {A119},
          doi = {10.1051/0004-6361/202040170},
archivePrefix = {arXiv},
       eprint = {2102.08840},
 primaryClass = {astro-ph.SR},
       adsurl = {https://ui.adsabs.harvard.edu/abs/2021A&A...648A.119C}
}

@ARTICLE{bergemann2012,
       author = {{Bergemann}, Maria and {Lind}, K. and {Collet}, R. and {Magic}, Z. and {Asplund}, M.},
        title = "{Non-LTE line formation of Fe in late-type stars - I. Standard stars with 1D and <3D> model atmospheres}",
      journal = {\mnras},
         year = 2012,
        month = nov,
       volume = {427},
       number = {1},
        pages = {27-49},
          doi = {10.1111/j.1365-2966.2012.21687.x},
archivePrefix = {arXiv},
       eprint = {1207.2455},
 primaryClass = {astro-ph.SR},
       adsurl = {https://ui.adsabs.harvard.edu/abs/2012MNRAS.427...27B}
}

@ARTICLE{semenova2020,
       author = {{Semenova}, Ekaterina and {Bergemann}, Maria and {Deal}, Morgan and {Serenelli}, Aldo and {Hansen}, Camilla Juul and {Gallagher}, Andrew J. and {Bayo}, Amelia and {Bensby}, Thomas and {Bragaglia}, Angela and {Carraro}, Giovanni and {Morbidelli}, Lorenzo and {Pancino}, Elena and {Smiljanic}, Rodolfo},
        title = "{The Gaia-ESO survey: 3D NLTE abundances in the open cluster NGC 2420 suggest atomic diffusion and turbulent mixing are at the origin of chemical abundance variations}",
      journal = {\aap},
         year = 2020,
        month = nov,
       volume = {643},
          eid = {A164},
        pages = {A164},
          doi = {10.1051/0004-6361/202038833},
archivePrefix = {arXiv},
       eprint = {2007.09153},
 primaryClass = {astro-ph.SR},
       adsurl = {https://ui.adsabs.harvard.edu/abs/2020A&A...643A.164S}
}

@ARTICLE{melendez2014ApJ...791...14M,
       author = {{Mel{\'e}ndez}, Jorge and {Ram{\'\i}rez}, Iv{\'a}n and {Karakas}, Amanda I. and {Yong}, David and {Monroe}, TalaWanda R. and {Bedell}, Megan and {Bergemann}, Maria and {Asplund}, Martin and {Tucci Maia}, Marcelo and {Bean}, Jacob and {do Nascimento}, Jos{\'e}-Dias, Jr. and {Bazot}, Michael and {Alves-Brito}, Alan and {Freitas}, Fabr{\'\i}cio C. and {Castro}, Matthieu},
        title = "{18 Sco: A Solar Twin Rich in Refractory and Neutron-capture Elements. Implications for Chemical Tagging}",
      journal = {\apj},
         year = 2014,
        month = aug,
       volume = {791},
       number = {1},
          eid = {14},
        pages = {14},
          doi = {10.1088/0004-637X/791/1/14},
archivePrefix = {arXiv},
       eprint = {1406.5244},
 primaryClass = {astro-ph.SR},
       adsurl = {https://ui.adsabs.harvard.edu/abs/2014ApJ...791...14M}
}

@ARTICLE{vald32015PhyS...90e4005R,
       author = {{Ryabchikova}, T. and {Piskunov}, N. and {Kurucz}, R.~L. and {Stempels}, H.~C. and {Heiter}, U. and {Pakhomov}, Yu and {Barklem}, P.~S.},
        title = "{A major upgrade of the VALD database}",
      journal = {\physscr},
         year = 2015,
        month = may,
       volume = {90},
       number = {5},
          eid = {054005},
        pages = {054005},
          doi = {10.1088/0031-8949/90/5/054005},
       adsurl = {https://ui.adsabs.harvard.edu/abs/2015PhyS...90e4005R}
}

@ARTICLE{masseron2014A&A...571A..47M,
       author = {{Masseron}, T. and {Plez}, B. and {Van Eck}, S. and {Colin}, R. and {Daoutidis}, I. and {Godefroid}, M. and {Coheur}, P. -F. and {Bernath}, P. and {Jorissen}, A. and {Christlieb}, N.},
        title = "{CH in stellar atmospheres: an extensive linelist}",
      journal = {\aap},
         year = 2014,
        month = nov,
       volume = {571},
          eid = {A47},
        pages = {A47},
          doi = {10.1051/0004-6361/201423956},
archivePrefix = {arXiv},
       eprint = {1410.4005},
 primaryClass = {astro-ph.SR},
       adsurl = {https://ui.adsabs.harvard.edu/abs/2014A&A...571A..47M}
}

@ARTICLE{brooke2014ApJS..210...23B,
       author = {{Brooke}, James S.~A. and {Ram}, Ram S. and {Western}, Colin M. and {Li}, Gang and {Schwenke}, David W. and {Bernath}, Peter F.},
        title = "{Einstein A Coefficients and Oscillator Strengths for the A $^{2}${\ensuremath{\Pi}}-X $^{2}${\ensuremath{\Sigma}}$^{+}$ (Red) and B$^{2}${\ensuremath{\Sigma}}$^{+}$-X$^{2}${\ensuremath{\Sigma}}$^{+}$ (Violet) Systems and Rovibrational Transitions in the X $^{2}${\ensuremath{\Sigma}}$^{+}$ State of CN}",
      journal = {\apjs},
         year = 2014,
        month = feb,
       volume = {210},
       number = {2},
          eid = {23},
        pages = {23},
          doi = {10.1088/0067-0049/210/2/23},
       adsurl = {https://ui.adsabs.harvard.edu/abs/2014ApJS..210...23B}
}

@ARTICLE{sneden2014ApJS..214...26S,
       author = {{Sneden}, Christopher and {Lucatello}, Sara and {Ram}, Ram S. and {Brooke}, James S.~A. and {Bernath}, Peter},
        title = "{Line Lists for the A $^{2}${\ensuremath{\Pi}}-X $^{2}${\ensuremath{\Sigma}}$^{+}$ (Red) and B $^{2}${\ensuremath{\Sigma}}$^{+}$-X $^{2}${\ensuremath{\Sigma}}$^{+}$ (Violet) Systems of CN, $^{13}$C$^{14}$N, and $^{12}$C$^{15}$N, and Application to Astronomical Spectra}",
      journal = {\apjs},
         year = 2014,
        month = oct,
       volume = {214},
       number = {2},
          eid = {26},
        pages = {26},
          doi = {10.1088/0067-0049/214/2/26},
archivePrefix = {arXiv},
       eprint = {1408.3828},
 primaryClass = {astro-ph.SR},
       adsurl = {https://ui.adsabs.harvard.edu/abs/2014ApJS..214...26S}
}

@ARTICLE{brooke2013JQSRT.124...11B,
       author = {{Brooke}, James S.~A. and {Bernath}, Peter F. and {Schmidt}, Timothy W. and {Bacskay}, George B.},
        title = "{Line strengths and updated molecular constants for the C$_{2}$ Swan system}",
      journal = {\jqsrt},
         year = 2013,
        month = jul,
       volume = {124},
        pages = {11-20},
          doi = {10.1016/j.jqsrt.2013.02.025},
archivePrefix = {arXiv},
       eprint = {1212.2102},
 primaryClass = {astro-ph.SR},
       adsurl = {https://ui.adsabs.harvard.edu/abs/2013JQSRT.124...11B}
}

@ARTICLE{Ram2014ApJS..211....5R,
       author = {{Ram}, Ram S. and {Brooke}, James S.~A. and {Bernath}, Peter F. and {Sneden}, Christopher and {Lucatello}, Sara},
        title = "{Improved Line Data for the Swan System $^{12}$C$^{13}$C Isotopologue}",
      journal = {\apjs},
         year = 2014,
        month = mar,
       volume = {211},
       number = {1},
          eid = {5},
        pages = {5},
          doi = {10.1088/0067-0049/211/1/5},
       adsurl = {https://ui.adsabs.harvard.edu/abs/2014ApJS..211....5R}
}

@ARTICLE{skory2003ApJS..148..599S,
       author = {{Skory}, S. and {Weck}, P.~F. and {Stancil}, P.~C. and {Kirby}, K.},
        title = "{New Theoretical Line List for the B' $^{ 2}${\ensuremath{\Sigma}}$^{+}$<--X $^{ 2}${\ensuremath{\Sigma}}$^{+}$ System of $^{24}$MgH}",
      journal = {\apjs},
         year = 2003,
        month = oct,
       volume = {148},
       number = {2},
        pages = {599-606},
          doi = {10.1086/376834},
       adsurl = {https://ui.adsabs.harvard.edu/abs/2003ApJS..148..599S}
}

@ARTICLE{giribaldi2023A&A...679A.110G,
       author = {{Giribaldi}, R.~E. and {Van Eck}, S. and {Merle}, T. and {Jorissen}, A. and {Krynski}, P. and {Planquart}, L. and {Valentini}, M. and {Chiappini}, C. and {Van Winckel}, H.},
        title = "{TITANS metal-poor reference stars. II. Red giants and CEMP stars}",
      journal = {\aap},
         year = 2023,
        month = nov,
       volume = {679},
          eid = {A110},
        pages = {A110},
          doi = {10.1051/0004-6361/202347208},
archivePrefix = {arXiv},
       eprint = {2308.10118},
 primaryClass = {astro-ph.SR},
       adsurl = {https://ui.adsabs.harvard.edu/abs/2023A&A...679A.110G}
}

@INPROCEEDINGS{dekker2000,
       author = {{Dekker}, Hans and {D'Odorico}, Sandro and {Kaufer}, Andreas and {Delabre}, Bernard and {Kotzlowski}, Heinz},
        title = "{Design, construction, and performance of UVES, the echelle spectrograph for the UT2 Kueyen Telescope at the ESO Paranal Observatory}",
    booktitle = {Optical and IR Telescope Instrumentation and Detectors},
         year = 2000,
       editor = {{Iye}, Masanori and {Moorwood}, Alan F.},
       series = {Society of Photo-Optical Instrumentation Engineers (SPIE) Conference Series},
       volume = {4008},
        month = aug,
        pages = {534-545},
          doi = {10.1117/12.395512},
       adsurl = {https://ui.adsabs.harvard.edu/abs/2000SPIE.4008..534D}
}

@ARTICLE{giribaldi2023A&A...673A..18G,
       author = {{Giribaldi}, R.~E. and {Smiljanic}, R.},
        title = "{Chronology of the chemical enrichment of the old Galactic stellar populations}",
      journal = {\aap},
         year = 2023,
        month = may,
       volume = {673},
          eid = {A18},
        pages = {A18},
          doi = {10.1051/0004-6361/202245404},
archivePrefix = {arXiv},
       eprint = {2302.09640},
 primaryClass = {astro-ph.GA},
       adsurl = {https://ui.adsabs.harvard.edu/abs/2023A&A...673A..18G}
}

@ARTICLE{helmi2018,
       author = {{Helmi}, Amina and {Babusiaux}, Carine and {Koppelman}, Helmer H. and
         {Massari}, Davide and {Veljanoski}, Jovan and {Brown}, Anthony G.~A.},
        title = "{The merger that led to the formation of the Milky Way's inner stellar halo and thick disk}",
      journal = {Nature},
         year = 2018,
        month = oct,
       volume = {563},
       number = {7729},
        pages = {85-88},
          doi = {10.1038/s41586-018-0625-x},
archivePrefix = {arXiv},
       eprint = {1806.06038},
 primaryClass = {astro-ph.GA},
       adsurl = {https://ui.adsabs.harvard.edu/abs/2018Natur.563...85H}
}

@ARTICLE{bergemann2008A&A...492..823B,
       author = {{Bergemann}, M. and {Gehren}, T.},
        title = "{NLTE abundances of Mn in a sample of metal-poor stars}",
      journal = {\aap},
         year = 2008,
        month = dec,
       volume = {492},
       number = {3},
        pages = {823-831},
          doi = {10.1051/0004-6361:200810098},
archivePrefix = {arXiv},
       eprint = {0811.0681},
 primaryClass = {astro-ph},
       adsurl = {https://ui.adsabs.harvard.edu/abs/2008A&A...492..823B}
}

@ARTICLE{Sitnova2022,
       author = {{Sitnova}, T.~M. and {Yakovleva}, S.~A. and {Belyaev}, A.~K. and {Mashonkina}, L.~I.},
        title = "{Non-LTE abundances of zinc in different spectral type stars and the Galactic [Zn/Fe] trend based on quantum-mechanical data on inelastic processes in zinc-hydrogen collisions}",
      journal = {\mnras},
         year = 2022,
        month = sep,
       volume = {515},
       number = {1},
        pages = {1510-1523},
          doi = {10.1093/mnras/stac1813},
archivePrefix = {arXiv},
       eprint = {2205.05819},
 primaryClass = {astro-ph.SR},
       adsurl = {https://ui.adsabs.harvard.edu/abs/2022MNRAS.515.1510S}
}

@ARTICLE{Mashonkina2022,
       author = {{Mashonkina}, L. and {Pakhomov}, Yu V. and {Sitnova}, T. and {Jablonka}, P. and {Yakovleva}, S.~A. and {Belyaev}, A.~K.},
        title = "{The formation of the Milky Way halo and its dwarf satellites: A NLTE-1D abundance analysis. V. The Sextans galaxy}",
      journal = {\mnras},
         year = 2022,
        month = jan,
       volume = {509},
       number = {3},
        pages = {3626-3642},
          doi = {10.1093/mnras/stab3189},
archivePrefix = {arXiv},
       eprint = {2110.09402},
 primaryClass = {astro-ph.SR},
       adsurl = {https://ui.adsabs.harvard.edu/abs/2022MNRAS.509.3626M}
}

@ARTICLE{lodders2021SSRv..217...44L,
       author = {{Lodders}, Katharina},
        title = "{Relative Atomic Solar System Abundances, Mass Fractions, and Atomic Masses of the Elements and Their Isotopes, Composition of the Solar Photosphere, and Compositions of the Major Chondritic Meteorite Groups}",
      journal = {\ssr},
         year = 2021,
        month = apr,
       volume = {217},
       number = {3},
          eid = {44},
        pages = {44},
          doi = {10.1007/s11214-021-00825-8},
       adsurl = {https://ui.adsabs.harvard.edu/abs/2021SSRv..217...44L}
}

@ARTICLE{nordlander2017A&A...607A..75N,
       author = {{Nordlander}, T. and {Lind}, K.},
        title = "{Non-LTE aluminium abundances in late-type stars}",
      journal = {\aap},
         year = 2017,
        month = nov,
       volume = {607},
          eid = {A75},
        pages = {A75},
          doi = {10.1051/0004-6361/201730427},
archivePrefix = {arXiv},
       eprint = {1708.01949},
 primaryClass = {astro-ph.SR},
       adsurl = {https://ui.adsabs.harvard.edu/abs/2017A&A...607A..75N}
}

@ARTICLE{amarsi2017MNRAS.464..264A,
       author = {{Amarsi}, A.~M. and {Asplund}, M.},
        title = "{The solar silicon abundance based on 3D non-LTE calculations}",
      journal = {\mnras},
         year = 2017,
        month = jan,
       volume = {464},
       number = {1},
        pages = {264-273},
          doi = {10.1093/mnras/stw2445},
archivePrefix = {arXiv},
       eprint = {1609.07283},
 primaryClass = {astro-ph.SR},
       adsurl = {https://ui.adsabs.harvard.edu/abs/2017MNRAS.464..264A}
}

@ARTICLE{Koutsouridou2025A&A...699A..32K,
       author = {{Koutsouridou}, I. and {Sk{\'u}lad{\'o}ttir}, {\'A}. and {Salvadori}, S.},
        title = "{Large databases of metal-poor stars corrected for three-dimensional and/or non-local thermodynamic equilibrium effects}",
      journal = {\aap},
         year = 2025,
        month = jul,
       volume = {699},
          eid = {A32},
        pages = {A32},
          doi = {10.1051/0004-6361/202554228},
archivePrefix = {arXiv},
       eprint = {2505.13607},
 primaryClass = {astro-ph.GA},
       adsurl = {https://ui.adsabs.harvard.edu/abs/2025A&A...699A..32K}
}

@ARTICLE{Eitner2023A&A...677A.151E,
       author = {{Eitner}, P. and {Bergemann}, M. and {Ruiter}, A.~J. and {Avril}, O. and {Seitenzahl}, I.~R. and {Gent}, M.~R. and {C{\^o}t{\'e}}, B.},
        title = "{Observational constraints on the origin of the elements. V. NLTE abundance ratios of [Ni/Fe] in Galactic stars and enrichment by sub-Chandrasekhar mass supernovae}",
      journal = {\aap},
         year = 2023,
        month = sep,
       volume = {677},
          eid = {A151},
        pages = {A151},
          doi = {10.1051/0004-6361/202244286},
archivePrefix = {arXiv},
       eprint = {2206.10258},
 primaryClass = {astro-ph.GA},
       adsurl = {https://ui.adsabs.harvard.edu/abs/2023A&A...677A.151E}
}

@ARTICLE{bergemann2010MNRAS.401.1334B,
       author = {{Bergemann}, Maria and {Pickering}, Juliet C. and {Gehren}, Thomas},
        title = "{NLTE analysis of CoI/CoII lines in spectra of cool stars with new laboratory hyperfine splitting constants}",
      journal = {\mnras},
         year = 2010,
        month = jan,
       volume = {401},
       number = {2},
        pages = {1334-1346},
          doi = {10.1111/j.1365-2966.2009.15736.x},
archivePrefix = {arXiv},
       eprint = {0909.2178},
 primaryClass = {astro-ph.SR},
       adsurl = {https://ui.adsabs.harvard.edu/abs/2010MNRAS.401.1334B}
}

@ARTICLE{bergemann2010A&A...522A...9B,
       author = {{Bergemann}, M. and {Cescutti}, G.},
        title = "{Chromium: NLTE abundances in metal-poor stars and nucleosynthesis in the Galaxy}",
      journal = {\aap},
         year = 2010,
        month = nov,
       volume = {522},
          eid = {A9},
        pages = {A9},
          doi = {10.1051/0004-6361/201014250},
archivePrefix = {arXiv},
       eprint = {1006.0243},
 primaryClass = {astro-ph.SR},
       adsurl = {https://ui.adsabs.harvard.edu/abs/2010A&A...522A...9B}
}

@ARTICLE{bergemann2011,
       author = {{Bergemann}, Maria},
        title = "{Ionization balance of Ti in the photospheres of the Sun and four late-type stars}",
      journal = {MNRAS},
         year = 2011,
        month = may,
       volume = {413},
       number = {3},
        pages = {2184-2198},
          doi = {10.1111/j.1365-2966.2011.18295.x},
archivePrefix = {arXiv},
       eprint = {1101.0828},
 primaryClass = {astro-ph.SR},
       adsurl = {https://ui.adsabs.harvard.edu/abs/2011MNRAS.413.2184B}
}

@ARTICLE{heiter2021A&A...645A.106H,
       author = {{Heiter}, U. and {Lind}, K. and {Bergemann}, M. and {Asplund}, M. and {Mikolaitis}, {\v{S}}. and {Barklem}, P.~S. and {Masseron}, T. and {de Laverny}, P. and {Magrini}, L. and {Edvardsson}, B. and {J{\"o}nsson}, H. and {Pickering}, J.~C. and {Ryde}, N. and {Bayo Ar{\'a}n}, A. and {Bensby}, T. and {Casey}, A.~R. and {Feltzing}, S. and {Jofr{\'e}}, P. and {Korn}, A.~J. and {Pancino}, E. and {Damiani}, F. and {Lanzafame}, A. and {Lardo}, C. and {Monaco}, L. and {Morbidelli}, L. and {Smiljanic}, R. and {Worley}, C. and {Zaggia}, S. and {Randich}, S. and {Gilmore}, G.~F.},
        title = "{Atomic data for the Gaia-ESO Survey}",
      journal = {\aap},
         year = 2021,
        month = jan,
       volume = {645},
          eid = {A106},
        pages = {A106},
          doi = {10.1051/0004-6361/201936291},
archivePrefix = {arXiv},
       eprint = {2011.02049},
 primaryClass = {astro-ph.IM},
       adsurl = {https://ui.adsabs.harvard.edu/abs/2021A&A...645A.106H}
}

@ARTICLE{Massari2019,
       author = {{Massari}, D. and {Koppelman}, H.~H. and {Helmi}, A.},
        title = "{Origin of the system of globular clusters in the Milky Way}",
      journal = {A\&A},
         year = 2019,
        month = oct,
       volume = {630},
          eid = {L4},
        pages = {L4},
          doi = {10.1051/0004-6361/201936135},
archivePrefix = {arXiv},
       eprint = {1906.08271},
 primaryClass = {astro-ph.GA},
       adsurl = {https://ui.adsabs.harvard.edu/abs/2019A&A...630L...4M}
}

@ARTICLE{gallart2019NatAs...3..932G,
       author = {{Gallart}, Carme and {Bernard}, Edouard J. and {Brook}, Chris B. and {Ruiz-Lara}, Tom{\'a}s and {Cassisi}, Santi and {Hill}, Vanessa and {Monelli}, Matteo},
        title = "{Uncovering the birth of the Milky Way through accurate stellar ages with Gaia}",
      journal = {Nature Astronomy},
         year = 2019,
        month = jul,
       volume = {3},
        pages = {932-939},
          doi = {10.1038/s41550-019-0829-5},
archivePrefix = {arXiv},
       eprint = {1901.02900},
 primaryClass = {astro-ph.GA},
       adsurl = {https://ui.adsabs.harvard.edu/abs/2019NatAs...3..932G}
}

@ARTICLE{Myeong2019,
       author = {{Myeong}, G.~C. and {Vasiliev}, E. and {Iorio}, G. and {Evans}, N.~W. and {Belokurov}, V.},
        title = "{Evidence for two early accretion events that built the Milky Way stellar halo}",
      journal = {MNRAS},
         year = 2019,
        month = sep,
       volume = {488},
       number = {1},
        pages = {1235-1247},
          doi = {10.1093/mnras/stz1770},
archivePrefix = {arXiv},
       eprint = {1904.03185},
 primaryClass = {astro-ph.GA},
       adsurl = {https://ui.adsabs.harvard.edu/abs/2019MNRAS.488.1235M}
}

@ARTICLE{blanco-cuaresma2014,
       author = {{Blanco-Cuaresma}, S. and {Soubiran}, C. and {Heiter}, U. and
         {Jofr{\'e}}, P.},
        title = "{Determining stellar atmospheric parameters and chemical abundances of FGK stars with iSpec}",
      journal = {A\&A},
         year = 2014,
        month = sep,
       volume = {569},
          eid = {A111},
        pages = {A111},
          doi = {10.1051/0004-6361/201423945},
archivePrefix = {arXiv},
       eprint = {1407.2608},
 primaryClass = {astro-ph.IM},
       adsurl = {https://ui.adsabs.harvard.edu/abs/2014A&A...569A.111B}
}

@ARTICLE{gustafson2008,
   author = {{Gustafsson}, B. and {Edvardsson}, B. and {Eriksson}, K. and 
	{J{\o}rgensen}, U.~G. and {Nordlund}, {\AA}. and {Plez}, B.},
    title = "{A grid of MARCS model atmospheres for late-type stars. I. Methods and general properties}",
  journal = {A\&A},
archivePrefix = "arXiv",
   eprint = {0805.0554},
     year = 2008,
    month = aug,
   volume = 486,
    pages = {951-970},
      doi = {10.1051/0004-6361:200809724},
   adsurl = {http://adsabs.harvard.edu/abs/2008A%26A...486..951G}
}

@ARTICLE{amarsi2018,
       author = {{Amarsi}, A.~M. and {Nordlander}, T. and {Barklem}, P.~S. and {Asplund}, M. and {Collet}, R. and {Lind}, K.},
        title = "{Effective temperature determinations of late-type stars based on 3D non-LTE Balmer line formation}",
      journal = {A\&A},
         year = 2018,
        month = jul,
       volume = {615},
          eid = {A139},
        pages = {A139},
          doi = {10.1051/0004-6361/201732546},
archivePrefix = {arXiv},
       eprint = {1804.02305},
 primaryClass = {astro-ph.SR},
       adsurl = {https://ui.adsabs.harvard.edu/abs/2018A&A...615A.139A}
}

@ARTICLE{koppelman2019A&A...631L...9K,
       author = {{Koppelman}, Helmer H. and {Helmi}, Amina and {Massari}, Davide and {Price-Whelan}, Adrian M. and {Starkenburg}, Tjitske K.},
        title = "{Multiple retrograde substructures in the Galactic halo: A shattered view of Galactic history}",
      journal = {\aap},
         year = 2019,
        month = nov,
       volume = {631},
          eid = {L9},
        pages = {L9},
          doi = {10.1051/0004-6361/201936738},
archivePrefix = {arXiv},
       eprint = {1909.08924},
 primaryClass = {astro-ph.GA},
       adsurl = {https://ui.adsabs.harvard.edu/abs/2019A&A...631L...9K}
}

@ARTICLE{mashonkina2007,
       author = {{Mashonkina}, L. and {Korn}, A.~J. and {Przybilla}, N.},
        title = "{A non-LTE study of neutral and singly-ionized calcium in late-type stars}",
      journal = {A\&A},
         year = 2007,
        month = jan,
       volume = {461},
       number = {1},
        pages = {261-275},
          doi = {10.1051/0004-6361:20065999},
archivePrefix = {arXiv},
       eprint = {astro-ph/0609527},
 primaryClass = {astro-ph},
       adsurl = {https://ui.adsabs.harvard.edu/abs/2007A&A...461..261M}
}

@ARTICLE{bergemann2017_,
       author = {{Bergemann}, Maria and {Collet}, Remo and {Amarsi}, Anish M. and {Kovalev}, Mikhail and {Ruchti}, Greg and {Magic}, Zazralt},
        title = "{Non-local Thermodynamic Equilibrium Stellar Spectroscopy with 1D and <3D> Models. I. Methods and Application to Magnesium Abundances in Standard Stars}",
      journal = {ApJ},
         year = 2017,
        month = sep,
       volume = {847},
       number = {1},
          eid = {15},
        pages = {15},
          doi = {10.3847/1538-4357/aa88cb},
archivePrefix = {arXiv},
       eprint = {1612.07355},
 primaryClass = {astro-ph.SR},
       adsurl = {https://ui.adsabs.harvard.edu/abs/2017ApJ...847...15B}
}

@Misc{NLTE_MPIA,
author = {{Kovalev} and
{S.~Brinkmann} and {M.~Bergemann} and {MPIA IT-department}},
HOWPUBLISHED = {{NLTE MPIA web server, [Online]. Available:
{{http://nlte.mpia.de}}
Max Planck Institute for Astronomy,
Heidelberg.}},
year = {2018},
}

@ARTICLE{bisterzo2010,
       author = {{Bisterzo}, S. and {Gallino}, R. and {Straniero}, O. and {Cristallo}, S. and {K{\"a}ppeler}, F.},
        title = "{s-Process in low-metallicity stars - I. Theoretical predictions}",
      journal = {\mnras},
         year = 2010,
        month = may,
       volume = {404},
       number = {3},
        pages = {1529-1544},
          doi = {10.1111/j.1365-2966.2010.16369.x},
archivePrefix = {arXiv},
       eprint = {1001.5376},
 primaryClass = {astro-ph.SR},
       adsurl = {https://ui.adsabs.harvard.edu/abs/2010MNRAS.404.1529B}
}

@ARTICLE{bisterzo2011,
       author = {{Bisterzo}, S. and {Gallino}, R. and {Straniero}, O. and {Cristallo}, S. and {K{\"a}ppeler}, F.},
        title = "{The s-process in low-metallicity stars - II. Interpretation of high-resolution spectroscopic observations with asymptotic giant branch models}",
      journal = {\mnras},
         year = 2011,
        month = nov,
       volume = {418},
       number = {1},
        pages = {284-319},
          doi = {10.1111/j.1365-2966.2011.19484.x},
archivePrefix = {arXiv},
       eprint = {1108.0500},
 primaryClass = {astro-ph.SR},
       adsurl = {https://ui.adsabs.harvard.edu/abs/2011MNRAS.418..284B}
}

@ARTICLE{bisterzo2012,
       author = {{Bisterzo}, S. and {Gallino}, R. and {Straniero}, O. and {Cristallo}, S. and {K{\"a}ppeler}, F.},
        title = "{The s-process in low-metallicity stars - III. Individual analysis of CEMP-s and CEMP-s/r with asymptotic giant branch models}",
      journal = {\mnras},
         year = 2012,
        month = may,
       volume = {422},
       number = {1},
        pages = {849-884},
          doi = {10.1111/j.1365-2966.2012.20670.x},
archivePrefix = {arXiv},
       eprint = {1201.6198},
 primaryClass = {astro-ph.SR},
       adsurl = {https://ui.adsabs.harvard.edu/abs/2012MNRAS.422..849B}
}

@ARTICLE{astropy:2018,
       author = {{Astropy Collaboration} and {Price-Whelan}, A.~M. and {Sip{\H{o}}cz}, B.~M. and {G{\"u}nther}, H.~M. and {Lim}, P.~L. and {Crawford}, S.~M. and {Conseil}, S. and {Shupe}, D.~L. and {Craig}, M.~W. and {Dencheva}, N. and {Ginsburg}, A. and {VanderPlas}, J.~T. and {Bradley}, L.~D. and {P{\'e}rez-Su{\'a}rez}, D. and {de Val-Borro}, M. and {Aldcroft}, T.~L. and {Cruz}, K.~L. and {Robitaille}, T.~P. and {Tollerud}, E.~J. and {Ardelean}, C. and {Babej}, T. and {Bach}, Y.~P. and {Bachetti}, M. and {Bakanov}, A.~V. and {Bamford}, S.~P. and {Barentsen}, G. and {Barmby}, P. and {Baumbach}, A. and {Berry}, K.~L. and {Biscani}, F. and {Boquien}, M. and {Bostroem}, K.~A. and {Bouma}, L.~G. and {Brammer}, G.~B. and {Bray}, E.~M. and {Breytenbach}, H. and {Buddelmeijer}, H. and {Burke}, D.~J. and {Calderone}, G. and {Cano Rodr{\'\i}guez}, J.~L. and {Cara}, M. and {Cardoso}, J.~V.~M. and {Cheedella}, S. and {Copin}, Y. and {Corrales}, L. and {Crichton}, D. and {D'Avella}, D. and {Deil}, C. and {Depagne}, {\'E}. and {Dietrich}, J.~P. and {Donath}, A. and {Droettboom}, M. and {Earl}, N. and {Erben}, T. and {Fabbro}, S. and {Ferreira}, L.~A. and {Finethy}, T. and {Fox}, R.~T. and {Garrison}, L.~H. and {Gibbons}, S.~L.~J. and {Goldstein}, D.~A. and {Gommers}, R. and {Greco}, J.~P. and {Greenfield}, P. and {Groener}, A.~M. and {Grollier}, F. and {Hagen}, A. and {Hirst}, P. and {Homeier}, D. and {Horton}, A.~J. and {Hosseinzadeh}, G. and {Hu}, L. and {Hunkeler}, J.~S. and {Ivezi{\'c}}, {\v{Z}}. and {Jain}, A. and {Jenness}, T. and {Kanarek}, G. and {Kendrew}, S. and {Kern}, N.~S. and {Kerzendorf}, W.~E. and {Khvalko}, A. and {King}, J. and {Kirkby}, D. and {Kulkarni}, A.~M. and {Kumar}, A. and {Lee}, A. and {Lenz}, D. and {Littlefair}, S.~P. and {Ma}, Z. and {Macleod}, D.~M. and {Mastropietro}, M. and {McCully}, C. and {Montagnac}, S. and {Morris}, B.~M. and {Mueller}, M. and {Mumford}, S.~J. and {Muna}, D. and {Murphy}, N.~A. and {Nelson}, S. and {Nguyen}, G.~H. and {Ninan}, J.~P. and {N{\"o}the}, M. and {Ogaz}, S. and {Oh}, S. and {Parejko}, J.~K. and {Parley}, N. and {Pascual}, S. and {Patil}, R. and {Patil}, A.~A. and {Plunkett}, A.~L. and {Prochaska}, J.~X. and {Rastogi}, T. and {Reddy Janga}, V. and {Sabater}, J. and {Sakurikar}, P. and {Seifert}, M. and {Sherbert}, L.~E. and {Sherwood-Taylor}, H. and {Shih}, A.~Y. and {Sick}, J. and {Silbiger}, M.~T. and {Singanamalla}, S. and {Singer}, L.~P. and {Sladen}, P.~H. and {Sooley}, K.~A. and {Sornarajah}, S. and {Streicher}, O. and {Teuben}, P. and {Thomas}, S.~W. and {Tremblay}, G.~R. and {Turner}, J.~E.~H. and {Terr{\'o}n}, V. and {van Kerkwijk}, M.~H. and {de la Vega}, A. and {Watkins}, L.~L. and {Weaver}, B.~A. and {Whitmore}, J.~B. and {Woillez}, J. and {Zabalza}, V. and {Astropy Contributors}},
        title = "{The Astropy Project: Building an Open-science Project and Status of the v2.0 Core Package}",
      journal = {AJ},
         year = 2018,
        month = sep,
       volume = {156},
       number = {3},
          eid = {123},
        pages = {123},
          doi = {10.3847/1538-3881/aabc4f},
archivePrefix = {arXiv},
       eprint = {1801.02634},
 primaryClass = {astro-ph.IM},
       adsurl = {https://ui.adsabs.harvard.edu/abs/2018AJ....156..123A}
}

@ARTICLE{diMatteo2019A&A...632A...4D,
       author = {{Di Matteo}, P. and {Haywood}, M. and {Lehnert}, M.~D. and {Katz}, D. and {Khoperskov}, S. and {Snaith}, O.~N. and {G{\'o}mez}, A. and {Robichon}, N.},
        title = "{The Milky Way has no in-situ halo other than the heated thick disc. Composition of the stellar halo and age-dating the last significant merger with Gaia DR2 and APOGEE}",
      journal = {\aap},
         year = 2019,
        month = dec,
       volume = {632},
          eid = {A4},
        pages = {A4},
          doi = {10.1051/0004-6361/201834929},
archivePrefix = {arXiv},
       eprint = {1812.08232},
 primaryClass = {astro-ph.GA},
       adsurl = {https://ui.adsabs.harvard.edu/abs/2019A&A...632A...4D}
}

@misc{buder2021galah,
        author = {{Buder}, Sven and {Sharma}, Sanjib and {Kos}, Janez and {Amarsi}, Anish M. and {Nordlander}, Thomas and {Lind}, Karin and {Martell}, Sarah L. and {Asplund}, Martin and {Bland-Hawthorn}, Joss and {Casey}, Andrew R. and {de Silva}, Gayandhi M. and {D'Orazi}, Valentina and {Freeman}, Ken C. and {Hayden}, Michael R. and {Lewis}, Geraint F. and {Lin}, Jane and {Schlesinger}, Katharine J. and {Simpson}, Jeffrey D. and {Stello}, Dennis and {Zucker}, Daniel B. and {Zwitter}, Toma{\v{z}} and {Beeson}, Kevin L. and {Buck}, Tobias and {Casagrande}, Luca and {Clark}, Jake T. and {{\v{C}}otar}, Klemen and {da Costa}, Gary S. and {de Grijs}, Richard and {Feuillet}, Diane and {Horner}, Jonathan and {Kafle}, Prajwal R. and {Khanna}, Shourya and {Kobayashi}, Chiaki and {Liu}, Fan and {Montet}, Benjamin T. and {Nandakumar}, Govind and {Nataf}, David M. and {Ness}, Melissa K. and {Spina}, Lorenzo and {Tepper-Garc{\'\i}a}, Thor and {Ting}, Yuan-Sen and {Traven}, Gregor and {Vogrin{\v{c}}i{\v{c}}}, Rok and {Wittenmyer}, Robert A. and {Wyse}, Rosemary F.~G. and {{\v{Z}}erjal}, Maru{\v{s}}a and {GALAH Collaboration}},
        title = "{The GALAH+ survey: Third data release}",
      journal = {\mnras},
         year = 2021,
        month = sep,
       volume = {506},
       number = {1},
        pages = {150-201},
          doi = {10.1093/mnras/stab1242},
archivePrefix = {arXiv},
       eprint = {2011.02505},
 primaryClass = {astro-ph.GA},
       adsurl = {https://ui.adsabs.harvard.edu/abs/2021MNRAS.506..150B}
}

@ARTICLE{Belokurov2018,
       author = {{Belokurov}, V. and {Erkal}, D. and {Evans}, N.~W. and {Koposov}, S.~E. and
         {Deason}, A.~J.},
        title = "{Co-formation of the disc and the stellar halo}",
      journal = {MNRAS},
         year = 2018,
        month = jul,
       volume = {478},
       number = {1},
        pages = {611-619},
          doi = {10.1093/mnras/sty982},
archivePrefix = {arXiv},
       eprint = {1802.03414},
 primaryClass = {astro-ph.GA},
       adsurl = {https://ui.adsabs.harvard.edu/abs/2018MNRAS.478..611B}
}

@ARTICLE{giribaldi2021A&A...650A.194G,
       author = {{Giribaldi}, R.~E. and {da Silva}, A.~R. and {Smiljanic}, R. and {Cornejo Espinoza}, D.},
        title = "{TITANS metal-poor reference stars. I. Accurate effective temperatures and surface gravities for dwarfs and subgiants from 3D non-LTE H{\ensuremath{\alpha}} profiles and Gaia parallaxes}",
      journal = {\aap},
         year = 2021,
        month = jun,
       volume = {650},
          eid = {A194},
        pages = {A194},
          doi = {10.1051/0004-6361/202140751},
archivePrefix = {arXiv},
       eprint = {2104.13931},
 primaryClass = {astro-ph.SR},
       adsurl = {https://ui.adsabs.harvard.edu/abs/2021A&A...650A.194G}
}

@ARTICLE{lucatello2005ApJ...625..833L,
       author = {{Lucatello}, Sara and {Gratton}, Raffaele G. and {Beers}, Timothy C. and {Carretta}, Eugenio},
        title = "{Observational Evidence for a Different Initial Mass Function in the Early Galaxy}",
      journal = {\apj},
         year = 2005,
        month = jun,
       volume = {625},
       number = {2},
        pages = {833-837},
          doi = {10.1086/428105},
archivePrefix = {arXiv},
       eprint = {astro-ph/0412423},
 primaryClass = {astro-ph},
       adsurl = {https://ui.adsabs.harvard.edu/abs/2005ApJ...625..833L}
}

@ARTICLE{hansen2016A&A...588A...3H,
       author = {{Hansen}, T.~T. and {Andersen}, J. and {Nordstr{\"o}m}, B. and {Beers}, T.~C. and {Placco}, V.~M. and {Yoon}, J. and {Buchhave}, L.~A.},
        title = "{The role of binaries in the enrichment of the early Galactic halo. III. Carbon-enhanced metal-poor stars - CEMP-s stars}",
      journal = {\aap},
         year = 2016,
        month = apr,
       volume = {588},
          eid = {A3},
        pages = {A3},
          doi = {10.1051/0004-6361/201527409},
archivePrefix = {arXiv},
       eprint = {1601.03385},
 primaryClass = {astro-ph.SR},
       adsurl = {https://ui.adsabs.harvard.edu/abs/2016A&A...588A...3H}
}

@ARTICLE{hampel2016ApJ...831..171H,
       author = {{Hampel}, Melanie and {Stancliffe}, Richard J. and {Lugaro}, Maria and {Meyer}, Bradley S.},
        title = "{The Intermediate Neutron-capture Process and Carbon-enhanced Metal-poor Stars}",
      journal = {\apj},
         year = 2016,
        month = nov,
       volume = {831},
       number = {2},
          eid = {171},
        pages = {171},
          doi = {10.3847/0004-637X/831/2/171},
archivePrefix = {arXiv},
       eprint = {1608.08634},
 primaryClass = {astro-ph.SR},
       adsurl = {https://ui.adsabs.harvard.edu/abs/2016ApJ...831..171H}
}

@ARTICLE{simmerer2004ApJ...617.1091S,
       author = {{Simmerer}, Jennifer and {Sneden}, Christopher and {Cowan}, John J. and {Collier}, Jason and {Woolf}, Vincent M. and {Lawler}, James E.},
        title = "{The Rise of the s-Process in the Galaxy}",
      journal = {\apj},
         year = 2004,
        month = dec,
       volume = {617},
       number = {2},
        pages = {1091-1114},
          doi = {10.1086/424504},
archivePrefix = {arXiv},
       eprint = {astro-ph/0410396},
 primaryClass = {astro-ph},
       adsurl = {https://ui.adsabs.harvard.edu/abs/2004ApJ...617.1091S}
}

@ARTICLE{roederer2016ApJ...821...37R,
       author = {{Roederer}, Ian U. and {Karakas}, Amanda I. and {Pignatari}, Marco and {Herwig}, Falk},
        title = "{The Diverse Origins of Neutron-capture Elements in the Metal-poor Star HD 94028: Possible Detection of Products of I-Process Nucleosynthesis}",
      journal = {\apj},
         year = 2016,
        month = apr,
       volume = {821},
       number = {1},
          eid = {37},
        pages = {37},
          doi = {10.3847/0004-637X/821/1/37},
archivePrefix = {arXiv},
       eprint = {1603.00036},
 primaryClass = {astro-ph.SR},
       adsurl = {https://ui.adsabs.harvard.edu/abs/2016ApJ...821...37R}
}

@ARTICLE{aoki2007ApJ...655..492A,
       author = {{Aoki}, Wako and {Beers}, Timothy C. and {Christlieb}, Norbert and {Norris}, John E. and {Ryan}, Sean G. and {Tsangarides}, Stelios},
        title = "{Carbon-enhanced Metal-poor Stars. I. Chemical Compositions of 26 Stars}",
      journal = {\apj},
         year = 2007,
        month = jan,
       volume = {655},
       number = {1},
        pages = {492-521},
          doi = {10.1086/509817},
archivePrefix = {arXiv},
       eprint = {astro-ph/0609702},
 primaryClass = {astro-ph},
       adsurl = {https://ui.adsabs.harvard.edu/abs/2007ApJ...655..492A}
}

@ARTICLE{prsa2016AJ....152...41P,
       author = {{Pr{\v{s}}a}, Andrej and {Harmanec}, Petr and {Torres}, Guillermo and {Mamajek}, Eric and {Asplund}, Martin and {Capitaine}, Nicole and {Christensen-Dalsgaard}, J{\o}rgen and {Depagne}, {\'E}ric and {Haberreiter}, Margit and {Hekker}, Saskia and {Hilton}, James and {Kopp}, Greg and {Kostov}, Veselin and {Kurtz}, Donald W. and {Laskar}, Jacques and {Mason}, Brian D. and {Milone}, Eugene F. and {Montgomery}, Michele and {Richards}, Mercedes and {Schmutz}, Werner and {Schou}, Jesper and {Stewart}, Susan G.},
        title = "{Nominal Values for Selected Solar and Planetary Quantities: IAU 2015 Resolution B3}",
      journal = {\aj},
         year = 2016,
        month = aug,
       volume = {152},
       number = {2},
          eid = {41},
        pages = {41},
          doi = {10.3847/0004-6256/152/2/41},
archivePrefix = {arXiv},
       eprint = {1605.09788},
 primaryClass = {astro-ph.SR},
       adsurl = {https://ui.adsabs.harvard.edu/abs/2016AJ....152...41P}
}

@BOOK{2mass2003tmc..book.....C,
       author = {{Cutri}, R.~M. and {Skrutskie}, M.~F. and {van Dyk}, S. and {Beichman}, C.~A. and {Carpenter}, J.~M. and {Chester}, T. and {Cambresy}, L. and {Evans}, T. and {Fowler}, J. and {Gizis}, J. and {Howard}, E. and {Huchra}, J. and {Jarrett}, T. and {Kopan}, E.~L. and {Kirkpatrick}, J.~D. and {Light}, R.~M. and {Marsh}, K.~A. and {McCallon}, H. and {Schneider}, S. and {Stiening}, R. and {Sykes}, M. and {Weinberg}, M. and {Wheaton}, W.~A. and {Wheelock}, S. and {Zacarias}, N.},
        title = "{2MASS All Sky Catalog of point sources.}",
         year = 2003,
       adsurl = {https://ui.adsabs.harvard.edu/abs/2003tmc..book.....C}
}

@ARTICLE{fitzpatrick1999PASP..111...63F,
       author = {{Fitzpatrick}, Edward L.},
        title = "{Correcting for the Effects of Interstellar Extinction}",
      journal = {\pasp},
         year = 1999,
        month = jan,
       volume = {111},
       number = {755},
        pages = {63-75},
          doi = {10.1086/316293},
archivePrefix = {arXiv},
       eprint = {astro-ph/9809387},
 primaryClass = {astro-ph},
       adsurl = {https://ui.adsabs.harvard.edu/abs/1999PASP..111...63F}
}

@ARTICLE{casagrande2018MNRAS.479L.102C,
       author = {{Casagrande}, L. and {VandenBerg}, Don A.},
        title = "{On the use of Gaia magnitudes and new tables of bolometric corrections}",
      journal = {\mnras},
         year = 2018,
        month = sep,
       volume = {479},
       number = {1},
        pages = {L102-L107},
          doi = {10.1093/mnrasl/sly104},
archivePrefix = {arXiv},
       eprint = {1806.01953},
 primaryClass = {astro-ph.SR},
       adsurl = {https://ui.adsabs.harvard.edu/abs/2018MNRAS.479L.102C}
}

@ARTICLE{Schlafly2011ApJ...737..103S,
       author = {{Schlafly}, Edward F. and {Finkbeiner}, Douglas P.},
        title = "{Measuring Reddening with Sloan Digital Sky Survey Stellar Spectra and Recalibrating SFD}",
      journal = {\apj},
         year = 2011,
        month = aug,
       volume = {737},
       number = {2},
          eid = {103},
        pages = {103},
          doi = {10.1088/0004-637X/737/2/103},
archivePrefix = {arXiv},
       eprint = {1012.4804},
 primaryClass = {astro-ph.GA},
       adsurl = {https://ui.adsabs.harvard.edu/abs/2011ApJ...737..103S}
}

@ARTICLE{ruchti2013MNRAS.429..126R,
       author = {{Ruchti}, Gregory R. and {Bergemann}, Maria and {Serenelli}, Aldo and {Casagrande}, Luca and {Lind}, Karin},
        title = "{Unveiling systematic biases in the 1D LTE excitation-ionization balance of Fe for FGK stars: a novel approach to determination of stellar parameters}",
      journal = {\mnras},
         year = 2013,
        month = feb,
       volume = {429},
       number = {1},
        pages = {126-134},
          doi = {10.1093/mnras/sts319},
archivePrefix = {arXiv},
       eprint = {1210.7998},
 primaryClass = {astro-ph.SR},
       adsurl = {https://ui.adsabs.harvard.edu/abs/2013MNRAS.429..126R}
}

@ARTICLE{wiedeking2025NatRP...7..696W,
       author = {{Wiedeking}, Mathis and {Goriely}, Stephane and {Guttormsen}, Magne and {Herwig}, Falk and {Larsen}, Ann-Cecilie and {Liddick}, Sean N. and {M{\"u}cher}, Dennis and {Richard}, Andrea L. and {Siem}, Sunniva and {Spyrou}, Artemis},
        title = "{Unlocking i-process nucleosynthesis by bridging stellar and nuclear physics}",
      journal = {Nature Reviews Physics},
         year = 2025,
        month = dec,
       volume = {7},
       number = {12},
        pages = {696-712},
          doi = {10.1038/s42254-025-00885-7},
       adsurl = {https://ui.adsabs.harvard.edu/abs/2025NatRP...7..696W}
}

@ARTICLE{lai2007ApJ...667.1185L,
       author = {{Lai}, David K. and {Johnson}, Jennifer A. and {Bolte}, Michael and {Lucatello}, Sara},
        title = "{Carbon and Strontium Abundances of Metal-poor Stars}",
      journal = {\apj},
         year = 2007,
        month = oct,
       volume = {667},
       number = {2},
        pages = {1185-1195},
          doi = {10.1086/520949},
archivePrefix = {arXiv},
       eprint = {0706.3043},
 primaryClass = {astro-ph},
       adsurl = {https://ui.adsabs.harvard.edu/abs/2007ApJ...667.1185L}
}

@ARTICLE{isotopic2005JPCRD..34...57B,
       author = {{B{\"o}hlke}, J.~K. and {de Laeter}, J.~R. and {De Bi{\`e}vre}, P. and {Hidaka}, H. and {Peiser}, H.~S. and {Rosman}, K.~J.~R. and {Taylor}, P.~D.~P.},
        title = "{Isotopic Compositions of the Elements, 2001}",
      journal = {Journal of Physical and Chemical Reference Data},
         year = 2005,
        month = mar,
       volume = {34},
       number = {1},
        pages = {57-67},
          doi = {10.1063/1.1836764},
       adsurl = {https://ui.adsabs.harvard.edu/abs/2005JPCRD..34...57B}
}

@ARTICLE{aoki2006ApJ...650L.127A,
       author = {{Aoki}, Wako and {Bisterzo}, Sara and {Gallino}, Roberto and {Beers}, Timothy C. and {Norris}, John E. and {Ryan}, Sean G. and {Tsangarides}, Stelios},
        title = "{Carbon-enhanced Metal-poor Stars: Osmium and Iridium Abundances in the Neutron-Capture-enhanced Subgiants CS 31062-050 and LP 625-44}",
      journal = {\apjl},
         year = 2006,
        month = oct,
       volume = {650},
       number = {2},
        pages = {L127-L130},
          doi = {10.1086/508878},
archivePrefix = {arXiv},
       eprint = {astro-ph/0609138},
 primaryClass = {astro-ph},
       adsurl = {https://ui.adsabs.harvard.edu/abs/2006ApJ...650L.127A}
}

@ARTICLE{johnson2004ApJ...605..462J,
       author = {{Johnson}, Jennifer A. and {Bolte}, Michael},
        title = "{The s-Process in Metal-Poor Stars: Abundances for 22 Neutron-Capture Elements in CS 31062-050}",
      journal = {\apj},
         year = 2004,
        month = apr,
       volume = {605},
       number = {1},
        pages = {462-471},
          doi = {10.1086/382147},
archivePrefix = {arXiv},
       eprint = {astro-ph/0402003},
 primaryClass = {astro-ph},
       adsurl = {https://ui.adsabs.harvard.edu/abs/2004ApJ...605..462J}
}

@ARTICLE{aoki2002PASJ...54..933A,
       author = {{Aoki}, Wako and {Norris}, John E. and {Ryan}, Sean G. and {Beers}, Timothy C. and {Ando}, Hiroyasu},
        title = "{Subaru/HDS Study of the Extremely Metal-Poor Star CS 29498-043: Abundance Analysis Details and Comparison with Other Carbon-Rich Objects}",
      journal = {\pasj},
         year = 2002,
        month = dec,
       volume = {54},
        pages = {933-949},
          doi = {10.1093/pasj/54.6.933},
       adsurl = {https://ui.adsabs.harvard.edu/abs/2002PASJ...54..933A}
}

@ARTICLE{cowan1977ApJ...212..149C,
       author = {{Cowan}, J.~J. and {Rose}, W.~K.},
        title = "{Production of $^{14}$C and neutrons in red giants.}",
      journal = {\apj},
         year = 1977,
        month = feb,
       volume = {212},
        pages = {149-158},
          doi = {10.1086/155030},
       adsurl = {https://ui.adsabs.harvard.edu/abs/1977ApJ...212..149C}
}

@ARTICLE{clarkson2018MNRAS.474L..37C,
       author = {{Clarkson}, O. and {Herwig}, F. and {Pignatari}, M.},
        title = "{Pop III i-process nucleosynthesis and the elemental abundances of SMSS J0313-6708 and the most iron-poor stars}",
      journal = {\mnras},
         year = 2018,
        month = feb,
       volume = {474},
       number = {1},
        pages = {L37-L41},
          doi = {10.1093/mnrasl/slx190},
archivePrefix = {arXiv},
       eprint = {1710.01763},
 primaryClass = {astro-ph.SR},
       adsurl = {https://ui.adsabs.harvard.edu/abs/2018MNRAS.474L..37C}
}

@ARTICLE{Banerjee2018ApJ...865..120B,
       author = {{Banerjee}, Projjwal and {Qian}, Yong-Zhong and {Heger}, Alexander},
        title = "{New Neutron-capture Site in Massive Pop III and Pop II Stars as a Source for Heavy Elements in the Early Galaxy}",
      journal = {\apj},
         year = 2018,
        month = oct,
       volume = {865},
       number = {2},
          eid = {120},
        pages = {120},
          doi = {10.3847/1538-4357/aadb8c},
archivePrefix = {arXiv},
       eprint = {1711.05964},
 primaryClass = {astro-ph.SR},
       adsurl = {https://ui.adsabs.harvard.edu/abs/2018ApJ...865..120B}
}

@ARTICLE{casagrande2021,
       author = {{Casagrande}, Luca and {Lin}, Jane and {Rains}, Adam D. and {Liu}, Fan and {Buder}, Sven and {Horner}, Jonathan and {Asplund}, Martin and {Lewis}, Geraint F. and {Martell}, Sarah L. and {Nordlander}, Thomas and {Stello}, Dennis and {Ting}, Yuan-Sen and {Wittenmyer}, Robert A. and {Bland-Hawthorn}, Joss and {Casey}, Andrew R. and {De Silva}, Gayandhi M. and {D'Orazi}, Valentina and {Freeman}, Ken C. and {Hayden}, Michael R. and {Kos}, Janez and {Lind}, Karin and {Schlesinger}, Katharine J. and {Sharma}, Sanjib and {Simpson}, Jeffrey D. and {Zucker}, Daniel B. and {Zwitter}, Toma{\v{z}}},
        title = "{The GALAH survey: effective temperature calibration from the InfraRed Flux Method in the Gaia system}",
      journal = {\mnras},
         year = 2021,
        month = oct,
       volume = {507},
       number = {2},
        pages = {2684-2696},
          doi = {10.1093/mnras/stab2304},
archivePrefix = {arXiv},
       eprint = {2011.02517},
 primaryClass = {astro-ph.SR},
       adsurl = {https://ui.adsabs.harvard.edu/abs/2021MNRAS.507.2684C}
}

@ARTICLE{NI,
   author = {{Nilsson}, H. and {Ivarsson}, S.},
    title = "{Experimental oscillator strengths and hyperfine constants in Nb ii}",
  journal = {\aap},
     year = 2008,
    month = dec,
   volume = 492,
    pages = {609-616},
      doi = {10.1051/0004-6361:200811019},
   adsurl = {http://adsabs.harvard.edu/abs/2008A%26A...492..609N},
     note = {(NI)},
  VALD_ref_number = {365}
}

@ARTICLE{de_munshi2015PhRvA..91d0501D,
       author = {{De Munshi}, D. and {Dutta}, T. and {Rebhi}, R. and {Mukherjee}, M.},
        title = "{Precision measurement of branching fractions of $^{138}$Ba$^{+}$ : Testing many-body theories below the 1\% level}",
      journal = {\pra},
         year = 2015,
        month = apr,
       volume = {91},
       number = {4},
          eid = {040501},
        pages = {040501},
          doi = {10.1103/PhysRevA.91.040501},
archivePrefix = {arXiv},
       eprint = {1411.5041},
 primaryClass = {physics.atom-ph},
       adsurl = {https://ui.adsabs.harvard.edu/abs/2015PhRvA..91d0501D}
}

@ARTICLE{dutta2016NatSR...629772D,
       author = {{Dutta}, Tarun and {de Munshi}, Debashis and {Yum}, Dahyun and {Rebhi}, Riadh and {Mukherjee}, Manas},
        title = "{An exacting transition probability measurement - a direct test of atomic many-body theories}",
      journal = {Scientific Reports},
         year = 2016,
        month = jul,
       volume = {6},
          eid = {29772},
        pages = {29772},
          doi = {10.1038/srep29772},
archivePrefix = {arXiv},
       eprint = {1604.01488},
 primaryClass = {physics.atom-ph},
       adsurl = {https://ui.adsabs.harvard.edu/abs/2016NatSR...629772D}
}

@ARTICLE{anders1982GeCoA..46.2363A,
       author = {{Anders}, E. and {Ebihara}, M.},
        title = "{Solar-system abundances of the elements}",
      journal = {\gca},
         year = 1982,
        month = nov,
       volume = {46},
       number = {11},
        pages = {2363-2380},
          doi = {10.1016/0016-7037(82)90208-3},
       adsurl = {https://ui.adsabs.harvard.edu/abs/1982GeCoA..46.2363A}
}

@ARTICLE{Hannaford1985A&A...143..447H,
       author = {{Hannaford}, P. and {Lowe}, R.~M. and {Biemont}, E. and {Grevesse}, N.},
        title = "{Radiative lifetimes for Nb II and the problem of the solar abundance of niobium}",
      journal = {\aap},
         year = 1985,
        month = feb,
       volume = {143},
       number = {2},
        pages = {447-450},
       adsurl = {https://ui.adsabs.harvard.edu/abs/1985A&A...143..447H}
}

@ARTICLE{heiter2015A&A...582A..49H,
       author = {{Heiter}, U. and {Jofr{\'e}}, P. and {Gustafsson}, B. and {Korn}, A.~J. and {Soubiran}, C. and {Th{\'e}venin}, F.},
        title = "{Gaia FGK benchmark stars: Effective temperatures and surface gravities}",
      journal = {\aap},
         year = 2015,
        month = oct,
       volume = {582},
          eid = {A49},
        pages = {A49},
          doi = {10.1051/0004-6361/201526319},
archivePrefix = {arXiv},
       eprint = {1506.06095},
 primaryClass = {astro-ph.SR},
       adsurl = {https://ui.adsabs.harvard.edu/abs/2015A&A...582A..49H}
}

@ARTICLE{belokurov2020MNRAS.494.3880B,
       author = {{Belokurov}, Vasily and {Sanders}, Jason L. and {Fattahi}, Azadeh and {Smith}, Martin C. and {Deason}, Alis J. and {Evans}, N. Wyn and {Grand}, Robert J.~J.},
        title = "{The biggest splash}",
      journal = {\mnras},
         year = 2020,
        month = may,
       volume = {494},
       number = {3},
        pages = {3880-3898},
          doi = {10.1093/mnras/staa876},
archivePrefix = {arXiv},
       eprint = {1909.04679},
 primaryClass = {astro-ph.GA},
       adsurl = {https://ui.adsabs.harvard.edu/abs/2020MNRAS.494.3880B}
}

@ARTICLE{bergemann2013ApJ...764..115B,
       author = {{Bergemann}, Maria and {Kudritzki}, Rolf-Peter and {W{\"u}rl}, Matthias and {Plez}, Bertrand and {Davies}, Ben and {Gazak}, Zach},
        title = "{Red Supergiant Stars as Cosmic Abundance Probes. II. NLTE Effects in J-band Silicon Lines}",
      journal = {\apj},
         year = 2013,
        month = feb,
       volume = {764},
       number = {2},
          eid = {115},
        pages = {115},
          doi = {10.1088/0004-637X/764/2/115},
archivePrefix = {arXiv},
       eprint = {1212.2649},
 primaryClass = {astro-ph.SR},
       adsurl = {https://ui.adsabs.harvard.edu/abs/2013ApJ...764..115B}
}

@ARTICLE{giribaldi2025A&A...702A..65G,
       author = {{Giribaldi}, R.~E. and {Magrini}, L. and {Schiappacasse-Ulloa}, J. and {Randich}, S. and {Merle}, T.},
        title = "{Barium isotopic ratios in metal-poor stars: Calibrating the method with globular clusters: I. Dwarf and giant stars in NGC 6752}",
      journal = {\aap},
         year = 2025,
        month = oct,
       volume = {702},
          eid = {A65},
        pages = {A65},
          doi = {10.1051/0004-6361/202556407},
archivePrefix = {arXiv},
       eprint = {2508.14208},
 primaryClass = {astro-ph.SR},
       adsurl = {https://ui.adsabs.harvard.edu/abs/2025A&A...702A..65G}
}

@ARTICLE{bailer-jones2021AJ....161..147B,
       author = {{Bailer-Jones}, C.~A.~L. and {Rybizki}, J. and {Fouesneau}, M. and {Demleitner}, M. and {Andrae}, R.},
        title = "{Estimating Distances from Parallaxes. V. Geometric and Photogeometric Distances to 1.47 Billion Stars in Gaia Early Data Release 3}",
      journal = {\aj},
         year = 2021,
        month = mar,
       volume = {161},
       number = {3},
          eid = {147},
        pages = {147},
          doi = {10.3847/1538-3881/abd806},
archivePrefix = {arXiv},
       eprint = {2012.05220},
 primaryClass = {astro-ph.SR},
       adsurl = {https://ui.adsabs.harvard.edu/abs/2021AJ....161..147B}
}

@ARTICLE{Bonaca2017,
       author = {{Bonaca}, Ana and {Conroy}, Charlie and {Wetzel}, Andrew and {Hopkins}, Philip F. and {Kere{\v{s}}}, Du{\v{s}}an},
        title = "{Gaia Reveals a Metal-rich, in situ Component of the Local Stellar Halo}",
      journal = {\apj},
         year = 2017,
        month = aug,
       volume = {845},
       number = {2},
          eid = {101},
        pages = {101},
          doi = {10.3847/1538-4357/aa7d0c},
archivePrefix = {arXiv},
       eprint = {1704.05463},
 primaryClass = {astro-ph.GA},
       adsurl = {https://ui.adsabs.harvard.edu/abs/2017ApJ...845..101B}
}

@ARTICLE{Koppelman2020,
       author = {{Koppelman}, Helmer H. and {Bos}, Roy O.~Y. and {Helmi}, Amina},
        title = "{The messy merger of a large satellite and a Milky Way-like galaxy}",
      journal = {\aap},
         year = 2020,
        month = oct,
       volume = {642},
          eid = {L18},
        pages = {L18},
          doi = {10.1051/0004-6361/202038652},
archivePrefix = {arXiv},
       eprint = {2006.07620},
 primaryClass = {astro-ph.GA},
       adsurl = {https://ui.adsabs.harvard.edu/abs/2020A&A...642L..18K}
}

@ARTICLE{Jean-Baptiste2017,
       author = {{Jean-Baptiste}, I. and {Di Matteo}, P. and {Haywood}, M. and {G{\'o}mez}, A. and {Montuori}, M. and {Combes}, F. and {Semelin}, B.},
        title = "{On the kinematic detection of accreted streams in the Gaia era: a cautionary tale}",
      journal = {\aap},
         year = 2017,
        month = aug,
       volume = {604},
          eid = {A106},
        pages = {A106},
          doi = {10.1051/0004-6361/201629691},
archivePrefix = {arXiv},
       eprint = {1611.07193},
 primaryClass = {astro-ph.GA},
       adsurl = {https://ui.adsabs.harvard.edu/abs/2017A&A...604A.106J}
}

@ARTICLE{gerber2023,
       author = {{Gerber}, Jeffrey M. and {Magg}, Ekaterina and {Plez}, Bertrand and {Bergemann}, Maria and {Heiter}, Ulrike and {Olander}, Terese and {Hoppe}, Richard},
        title = "{Non-LTE radiative transfer with Turbospectrum}",
      journal = {\aap},
         year = 2023,
        month = jan,
       volume = {669},
          eid = {A43},
        pages = {A43},
          doi = {10.1051/0004-6361/202243673},
archivePrefix = {arXiv},
       eprint = {2206.00967},
 primaryClass = {astro-ph.SR},
       adsurl = {https://ui.adsabs.harvard.edu/abs/2023A&A...669A..43G}
}

@ARTICLE{giribaldi2019A&A...624A..10G,
       author = {{Giribaldi}, R.~E. and {Ubaldo-Melo}, M.~L. and {Porto de Mello}, G.~F. and {Pasquini}, L. and {Ludwig}, H. -G. and {Ulmer-Moll}, S. and {Lorenzo-Oliveira}, D.},
        title = "{Accurate effective temperature from H{\ensuremath{\alpha}} profiles}",
      journal = {\aap},
         year = 2019,
        month = apr,
       volume = {624},
          eid = {A10},
        pages = {A10},
          doi = {10.1051/0004-6361/201833763},
archivePrefix = {arXiv},
       eprint = {1811.12274},
 primaryClass = {astro-ph.SR},
       adsurl = {https://ui.adsabs.harvard.edu/abs/2019A&A...624A..10G}
}

@ARTICLE{2021A&A...649A...4L,
       author = {{Lindegren}, L. and {Bastian}, U. and {Biermann}, M. and {Bombrun}, A. and {de Torres}, A. and {Gerlach}, E. and {Geyer}, R. and {Hern{\'a}ndez}, J. and {Hilger}, T. and {Hobbs}, D. and {Klioner}, S.~A. and {Lammers}, U. and {McMillan}, P.~J. and {Ramos-Lerate}, M. and {Steidelm{\"u}ller}, H. and {Stephenson}, C.~A. and {van Leeuwen}, F.},
        title = "{Gaia Early Data Release 3. Parallax bias versus magnitude, colour, and position}",
      journal = {\aap},
         year = 2021,
        month = may,
       volume = {649},
          eid = {A4},
        pages = {A4},
          doi = {10.1051/0004-6361/202039653},
archivePrefix = {arXiv},
       eprint = {2012.01742},
 primaryClass = {astro-ph.IM},
       adsurl = {https://ui.adsabs.harvard.edu/abs/2021A&A...649A...4L}
}

@ARTICLE{storm2024,
       author = {{Storm}, N. and {Barklem}, P.~S. and {Yakovleva}, S.~A. and {Belyaev}, A.~K. and {Palmeri}, P. and {Quinet}, P. and {Lodders}, K. and {Bergemann}, M. and {Hoppe}, R.},
        title = "{3D NLTE modelling of Y and Eu. Centre-to-limb variation and solar abundances}",
      journal = {\aap},
         year = 2024,
        month = mar,
       volume = {683},
          eid = {A200},
        pages = {A200},
          doi = {10.1051/0004-6361/202348971},
archivePrefix = {arXiv},
       eprint = {2401.13450},
 primaryClass = {astro-ph.SR},
       adsurl = {https://ui.adsabs.harvard.edu/abs/2024A&A...683A.200S}
}

@ARTICLE{Prantzos20,
       author = {{Prantzos}, N. and {Abia}, C. and {Cristallo}, S. and {Limongi}, M. and {Chieffi}, A.},
        title = "{Chemical evolution with rotating massive star yields II. A new assessment of the solar s- and r-process components}",
      journal = {\mnras},
         year = 2020,
        month = jan,
       volume = {491},
       number = {2},
        pages = {1832-1850},
          doi = {10.1093/mnras/stz3154},
archivePrefix = {arXiv},
       eprint = {1911.02545},
 primaryClass = {astro-ph.GA},
       adsurl = {https://ui.adsabs.harvard.edu/abs/2020MNRAS.491.1832P}
}

@ARTICLE{choplin24,
       author = {{Choplin}, A. and {Siess}, L. and {Goriely}, S. and {Martinet}, S.},
        title = "{The intermediate neutron capture process. V. The i-process in AGB stars with overshoot}",
      journal = {\aap},
         year = 2024,
        month = apr,
       volume = {684},
          eid = {A206},
        pages = {A206},
          doi = {10.1051/0004-6361/202348957},
archivePrefix = {arXiv},
       eprint = {2402.10284},
 primaryClass = {astro-ph.SR},
       adsurl = {https://ui.adsabs.harvard.edu/abs/2024A&A...684A.206C}
}

@ARTICLE{choplin2022A&A...667A.155C,
       author = {{Choplin}, A. and {Siess}, L. and {Goriely}, S.},
        title = "{The intermediate neutron capture process. III. The i-process in AGB stars of different masses and metallicities without overshoot}",
      journal = {\aap},
         year = 2022,
        month = nov,
       volume = {667},
          eid = {A155},
        pages = {A155},
          doi = {10.1051/0004-6361/202244360},
archivePrefix = {arXiv},
       eprint = {2209.10303},
 primaryClass = {astro-ph.SR},
       adsurl = {https://ui.adsabs.harvard.edu/abs/2022A&A...667A.155C}
}

@article{Martinet24,
	Author = {S. Martinet and A. Choplin and S. Goriely and L. Siess},
	Journal = {Astronomy and Astrophysics},
	Pages = {A8},
	Title = {The intermediate neutron capture process IV. Impact of nuclear model and parameter uncertainties},
	Volume = {684},
	Year = {2024}}

@ARTICLE{matsuno2019ApJ...874L..35M,
       author = {{Matsuno}, Tadafumi and {Aoki}, Wako and {Suda}, Takuma},
        title = "{Origin of the Excess of High-energy Retrograde Stars in the Galactic Halo}",
      journal = {\apjl},
         year = 2019,
        month = apr,
       volume = {874},
       number = {2},
          eid = {L35},
        pages = {L35},
          doi = {10.3847/2041-8213/ab0ec0},
archivePrefix = {arXiv},
       eprint = {1903.09456},
 primaryClass = {astro-ph.GA},
       adsurl = {https://ui.adsabs.harvard.edu/abs/2019ApJ...874L..35M}
}

@ARTICLE{asplund2021A&A...653A.141A,
       author = {{Asplund}, M. and {Amarsi}, A.~M. and {Grevesse}, N.},
        title = "{The chemical make-up of the Sun: A 2020 vision}",
      journal = {\aap},
         year = 2021,
        month = sep,
       volume = {653},
          eid = {A141},
        pages = {A141},
          doi = {10.1051/0004-6361/202140445},
archivePrefix = {arXiv},
       eprint = {2105.01661},
 primaryClass = {astro-ph.SR},
       adsurl = {https://ui.adsabs.harvard.edu/abs/2021A&A...653A.141A}
}

@ARTICLE{asplund2009ARA&A..47..481A,
       author = {{Asplund}, Martin and {Grevesse}, Nicolas and {Sauval}, A. Jacques and {Scott}, Pat},
        title = "{The Chemical Composition of the Sun}",
      journal = {\araa},
         year = 2009,
        month = sep,
       volume = {47},
       number = {1},
        pages = {481-522},
          doi = {10.1146/annurev.astro.46.060407.145222},
archivePrefix = {arXiv},
       eprint = {0909.0948},
 primaryClass = {astro-ph.SR},
       adsurl = {https://ui.adsabs.harvard.edu/abs/2009ARA&A..47..481A}
}

@ARTICLE{matsuno2024A&A...688A..72M,
       author = {{Matsuno}, T. and {Amarsi}, A.~M. and {Carlos}, M. and {Nissen}, P.~E.},
        title = "{3D non-local thermodynamic equilibrium magnesium abundances reveal a distinct halo population}",
      journal = {\aap},
         year = 2024,
        month = aug,
       volume = {688},
          eid = {A72},
        pages = {A72},
          doi = {10.1051/0004-6361/202450057},
archivePrefix = {arXiv},
       eprint = {2405.13486},
 primaryClass = {astro-ph.SR},
       adsurl = {https://ui.adsabs.harvard.edu/abs/2024A&A...688A..72M}
}

@ARTICLE{bonifacio2021A&A...651A..79B,
       author = {{Bonifacio}, P. and {Monaco}, L. and {Salvadori}, S. and {Caffau}, E. and {Spite}, M. and {Sbordone}, L. and {Spite}, F. and {Ludwig}, H. -G. and {Di Matteo}, P. and {Haywood}, M. and {Fran{\c{c}}ois}, P. and {Koch-Hansen}, A.~J. and {Christlieb}, N. and {Zaggia}, S.},
        title = "{TOPoS. VI. The metal-weak tail of the metallicity distribution functions of the Milky Way and the Gaia-Sausage-Enceladus structure}",
      journal = {\aap},
         year = 2021,
        month = jul,
       volume = {651},
          eid = {A79},
        pages = {A79},
          doi = {10.1051/0004-6361/202140816},
archivePrefix = {arXiv},
       eprint = {2105.08360},
 primaryClass = {astro-ph.GA},
       adsurl = {https://ui.adsabs.harvard.edu/abs/2021A&A...651A..79B}
}

\begin{appendix} 
\newpage

\section{Chemo-dynamic analysis}
\label{sec:dynamic}

We computed the orbital parameters of TYC~6044-714-1 using the code \texttt{GalPot}\footnote{\url{https://github.com/PaulMcMillan-Astro/GalPot}}, with coordinates, proper motions, parallax, and radial velocity from \textit{Gaia} DR3 \citep{Gaia(c)} as inputs. Our algorithm follows that of \cite{giribaldi2023A&A...673A..18G} and \cite{giribaldi2025}; we refer the reader to those papers for details.

We used the samples in the two papers above as references. 
These populations comprise stars from the Gaia-Enceladus merger \citep{helmi2018,Belokurov2018,gallart2019NatAs...3..932G}, the Splash or heated thick disc \citep[e.g.,][]{belokurov2020MNRAS.494.3880B,Bonaca2017,diMatteo2019A&A...632A...4D}, Erebus \citep{giribaldi2023A&A...673A..18G}, and the metal-poor tail of the Milky Way \citep{giribaldi2025}.
Figure~\ref{fig:lindblad} shows the location of TYC~6044-714-1 and the concentrations of these populations in the Lindblad diagram, which plots space binding energy versus angular momentum (\(L_Z\)). Among them, Gaia-Enceladus is the only population composed of stars formed outside the Milky Way. Erebus was initially hypothesised to be of external origin because most of its stars rotate contrary to the sense of the Galactic discs \citep[called the Thamnos II population in][]{koppelman2019A&A...631L...9K}. Subsequently, the possibility that this and other retrograde populations formed in situ was reconsidered \citep{Koppelman2020} based on numerical simulations \citep[e.g.,][]{Jean-Baptiste2017}. Using stellar ages, \cite{giribaldi2023A&A...673A..18G} demonstrated that Erebus stars are too old and metal-rich to originate from a satellite galaxy, thus confirming it as a Milky Way in situ population. This opens the possibility that other retrograde populations (e.g. Sequoia) may also have formed within the Milky Way.

We define the  metal-poor tail as all stars with [Fe/H] $< -1.8$~dex. As the dispersion in Fig~\ref{fig:lindblad} shows, these are not orbitally linked. 
However, in the [Mg/Fe] vs. [Fe/H] plane, they form a knee-shaped sequence consistent with predictions from an evolutionary model of the Milky Way halo \citep[see Fig.~10 in][]{giribaldi2025}, supporting the hypothesis of in situ formation.

Figure~\ref{fig:mgfe} displays the stars of the populations described above in the [Mg/Fe] versus [Fe/H] plane. We note that these Mg abundances include 1D~non-LTE corrections, which, according to Fig. A.7 in \cite{giribaldi2025}, mainly\footnote{Stars of Erebus, Splash, and Gaia-Enceladus are at the main sequence turn-off, where 3D~non-LTE corrections to the  1D~non-LTE Mg abundances significantly affect only stars with [Mg/Fe] $\lesssim 0.15$ at [Fe/H] $> -1.5$~dex \citep{matsuno2024A&A...688A..72M}.} affect the metal-poor tail. Based on comparisons with 3D~non-LTE, they show that 1D~non-LTE [Mg/Fe] of stars with [Fe/H] $\lesssim -2$~dex requires a correction of about $+0.2$~dex. 
Therefore, in Fig.~\ref{fig:mgfe}, the trends of the metal-poor tail (solid line) and Erebus (dashed line) converge when Mg abundances approach their true values.
Figure~\ref{fig:mgfe} shows that TYC~6044-714-1 lies within the trend of the metal-poor tail in the [Mg/Fe] versus [Fe/H] plane, suggesting it is likely a star of the in situ halo.
Figure~\ref{fig:lindblad} shows that it moves as fast as the majority of stars in the discs, but in the opposite direction. Additionally, it is nearly as tightly bound to the Galaxy as the Sun, appearing near the border of the diagram.

Based on its retrograde orbit and moderate binding energy, the possibility that it belongs to the Sequoia population \citep{Myeong2019} can be considered (see its proximity to the high-probability box in the figure). Chemically, Sequoia stars have been shown to form a sequence approximately 0.15~dex below that of Enceladus in the [Mg/Fe] versus  [Fe/H] plane  \citep{matsuno2019ApJ...874L..35M}. However, both the Sequoia and Enceladus sequences are clearly visible only for stars with [Fe/H] $> -2$~dex.
This is likely due to evolutionary reasons. It has been estimated that the Milky Way was 4–5 times more massive than Gaia-Enceladus at the time of their merger \citep{helmi2018}, while Gaia-Enceladus may have been about 10 times more massive than the hypothetical Sequoia progenitor \citep{Myeong2019}. A lower-mass galaxy is less efficient at forming stars, so the presence of Gaia-Enceladus and Sequoia stars is expected to be much less frequent than that of Milky Way in situ stars. The bias-corrected metallicity distribution function of Gaia-Enceladus \citep{bonifacio2021A&A...651A..79B} confirms only a marginal fraction of stars with [Fe/H] $< -2$~dex, implying that the fraction of Sequoia stars is not expected to be higher. Based on this reasoning, we propose that TYC~6044-714-1 likely formed as a typical star in the Milky Way in situ halo {more than 11 Gyr ago, as most dwarf stars in this population \cite[][Fig.~8]{giribaldi2025}.}

\begin{figure}
    \centering
    \includegraphics[width=1\linewidth]{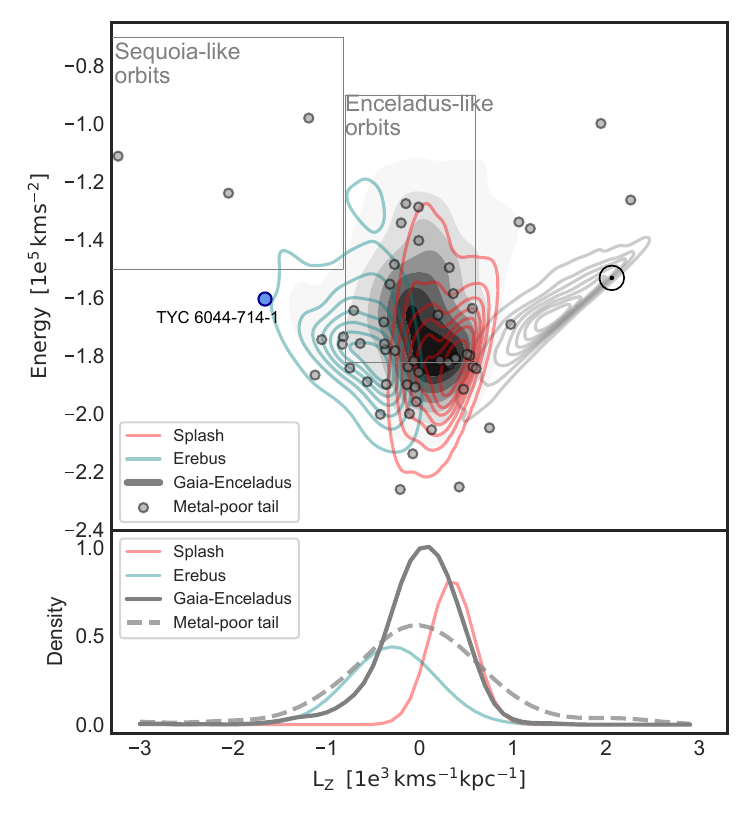}
    \caption{\tiny Linblad diagram. \textit{Top panel:} Contours represent the cumulative distributions of the Erebus, Splash, and Gaia-Enceladus populations in \cite{giribaldi2023A&A...673A..18G}; 
    the latter is represented by the grey gradient.
    The cumulative distribution of the Milky Way discs is also shown at the right side, as well as the Sun's location represented by the symbol $\odot$.
    Grey circles represent the stars of the metal-poor tail in \cite{giribaldi2025}.
    The boxes represent the high probability areas to find Gaia-Enceladus and Sequoia stars according to \cite{Massari2019}.
    \textit{Bottom panel:} Probability density distributions of the populations in the top panel, same colour coding is used.
    }
    \label{fig:lindblad}
\end{figure}

\begin{figure}
    \centering
    \includegraphics[width=1\linewidth]{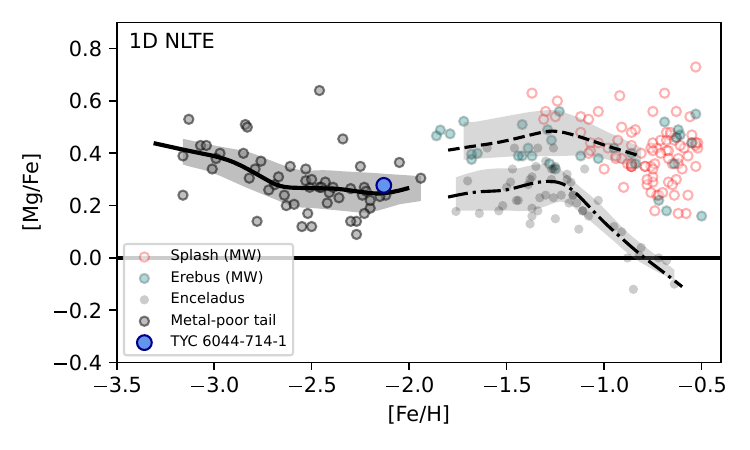}
    \caption{\tiny [Mg/Fe] vs. [Fe/H] diagram. Stars of the Erebus, Splash, Gaia-Encledus, and the metal-poor tail populations in \cite{giribaldi2023A&A...673A..18G} and \cite{giribaldi2025} are represented according to the legends. 
    Solid, dashed, and dash-dotted lines represent LOWESS regressions of each population except the Splash one.
    }
    \label{fig:mgfe}
\end{figure}

{

\section{Validation of surface gravity}
\label{sec:logg_sec}

\begin{figure}[!htbp]
    \centering
    \includegraphics[width=0.99\linewidth]{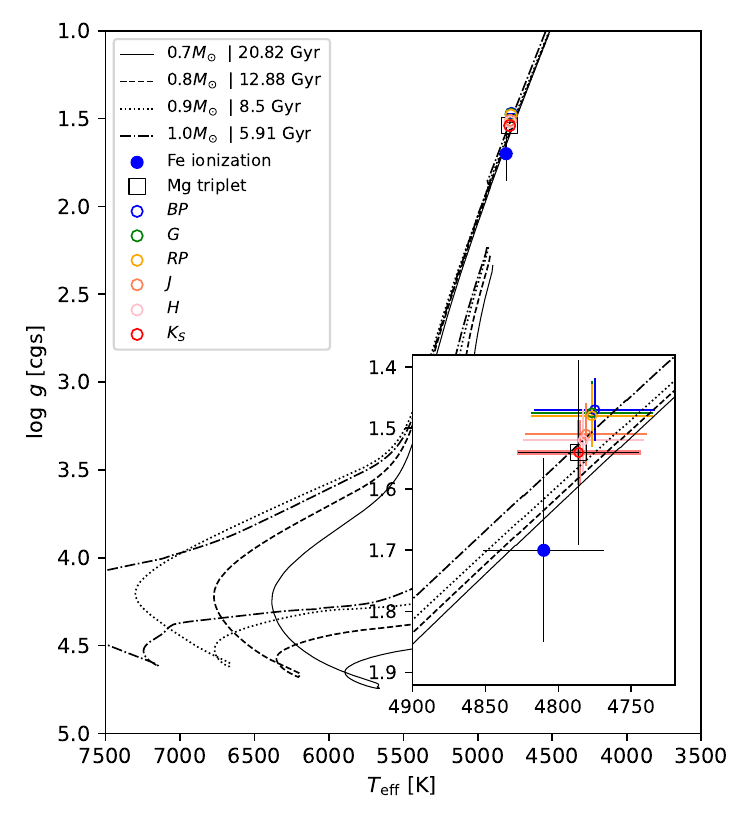}
    \caption{\tiny {Kiel diagram of \teff\ versus \logg. Surface gravities derived from three different methods are compared (see main text). FuNS evolutionary tracks for [Fe/H]$=-2.27$~dex and various initial masses are shown for reference. For each track, the age corresponding to the point closest to the observed \logg\ is indicated (these values are independent of the method used to derive \logg). The inset shows a zoomed-in view of the region highlighted in the main panel.
    }}
    \label{fig:kiel}
\end{figure}

Figure~\ref{fig:kiel} compares the surface gravities derived from the 1D non-LTE excitation balance of Fe lines (blue point) with those obtained from two other alternative methods. The application of those methods is described below, and corresponding results are listed in Table~\ref{tab:loggs}. \teff\ values in the figure were shifted to account for \logg\ variations relative to our fiducial \logg\ = 1.70~dex according to the grids in Fig.~5 of \cite{giribaldi2023A&A...679A.110G}, i.e. \teff$-15$~K per \logg$-0.10$~dex. 

Surface gravity from the \textit{Mg triplet} was determined as described in \cite{giribaldi2023A&A...679A.110G} from the Mg triplet lines 5172.68 and 5183.60~\AA, fixing \teff, [Fe/H], $v_{mic}$ in Table~\ref{tab:parameters}, and [Mg/Fe] in Table~\ref{tab:abundances}; see line fits in Fig.~\ref{fig:Mg_triplet}.
Errors relative to noise, \teff, and [Mg/Fe] errors are the most relevant, they account $\pm$0.10, $\pm$0.10, and $\pm$0.04~dex, respectively. We add them in quadrature to compute the total error given in the table.

Surface gravity from \textit{Parallax} was determined via the classical relation 
\begin{equation}
    \label{eq:fundamental}
    \mathrm{log}\,g = \mathrm{log} \,g_{\odot} + \mathrm{log} \frac{M}{M_\odot} + 4 \, \mathrm{log} \frac{T_{\mathrm{eff}}}{T_{\mathrm{eff} \odot}} +0.4 (M_{\mathrm{bol}} - M_{\mathrm{bol} \odot})
\end{equation}

\noindent where $M$ is the stellar mass and $M_{\rm bol}$ is the bolometric magnitude. We adopted $0.80 \pm 0.05\,M_\odot$ and the standard\footnote{IAU 2015 RESOLUTION B3} parameters \teff$_\odot = 5772$~K, \logg $= 4.44$~dex, and $M_{bol \odot} = 4.74$  \citep{prsa2016AJ....152...41P}.
$M_{\rm bol}$ was computed 
from the Gaia and 2MASS \citep{2mass2003tmc..book.....C}
magnitudes $BP$, $G$, $RP$, $J$, $H$, and $K_S$
with the routine {\it bcutil.py}\footnote{https://github.com/casaluca/bolometric-corrections} of \cite{casagrande2018MNRAS.479L.102C}.
Magnitudes were extinction corrected adopting the reddening value $E(B-V) = 0.0443$ from \cite{Schlafly2011ApJ...737..103S}\footnote{Online tool \url{https://irsa.ipac.caltech.edu/applications/DUST/}}, which was transformed to 
into the Gaia and 2MASS systems assuming
the extinction law of \cite{fitzpatrick1999PASP..111...63F} as normalised per \cite{Schlafly2011ApJ...737..103S}\footnote{Using the code {\it Colte} \citep{casagrande2021} at \url{https://github.com/casaluca/colte}.}.
Absolute magnitudes were determined adopting a distance of $3022 \pm 130$ parsec from \cite{bailer-jones2021AJ....161..147B}, which is based on Gaia 3D parallax and includes the zero-point correction +0.021 mas \citep{2021A&A...649A...4L}.
The main sources of uncertainty are the distance, mass, \teff, and magnitude errors, which account by 0.040, 0.026, 0.015, and 0.010, respectively; errors of the evolutionary model are neglected.
Magnitudes, extinctions, and bolometric corrections are listed in Table~\ref{tab:loggs}.

Figure~\ref{fig:kiel} shows that \logg\ from Fe lines, Mg triplet, and parallax using 2MASS bands are compatible within 1$\sigma$ errors.  
Equation~\ref{eq:fundamental} implicitly assumes that the bolometric magnitude, as a proxy for luminosity, reflects the intrinsic properties of the star.
In CEMP stars with strong blanketing, the flux can be redistributed toward redder wavelengths, decreasing magnitudes in the red bands and leading to systematic \logg\ underestimates.
However, the consistency of the outcomes of the methods implies that this is effect is negligible for TYC~6044-714-1.
A comparison with FuNS evolutionary tracks in Fig.~\ref{fig:kiel} indicates that the evolutionary path of a $\sim$0.8$M_\odot$ star provides an age diagnosis (13~Gyr) compatible to that in Sect.~\ref{sec:dynamic}, which suggests that TYC~6044-714-1 likely formed as a typical low-mass star more than 11~Gyr ago.

\begin{table}
\caption{Surface gravity values from diverse methods}
\label{tab:loggs}
\centering
\tiny 
\begin{threeparttable}
\begin{tabular}{l|cccc}
\hline\hline
Method & \logg & magnitude & extinction & BC\\
\hline
Fe ionisation & $1.70 \pm 0.15$ & --- & --- & --- \\
Mg triplet & $1.54 \pm 0.15$ & ---& --- & --- \\
Parallax ($BP$) & $1.47 \pm 0.05$ & $12.326 \pm 0.0006$ &  0.120 & $-0.710$ \\
Parallax ($G$)  & $1.48 \pm 0.05$ & $11.056 \pm 0.0002$ & 0.094 & $-0.207$\\
Parallax ($RP$) & $1.48 \pm 0.05$ & $11.343 \pm 0.0003$ & 0.070 & 0.461 \\
Parallax ($J$)  & $1.51 \pm 0.05$ & $9.59 \pm 0.027$ & 0.032 & 1.342\\
Parallax ($H$) & $1.52 \pm 0.05$ & $9.009 \pm 0.022$  & 0.020 & 1.872\\
Parallax ($K_S$) & $1.54 \pm 0.05$ & $8.926 \pm 0.021$ & 0.014 & 1.993\\
\hline
\end{tabular}
\begin{tablenotes}
\item{} Notes. Last column indicates the bolometric correction. 
\end{tablenotes}
\end{threeparttable}
\end{table}

\begin{figure*}[!htbp]
    \centering
    \includegraphics[width=0.75\linewidth]{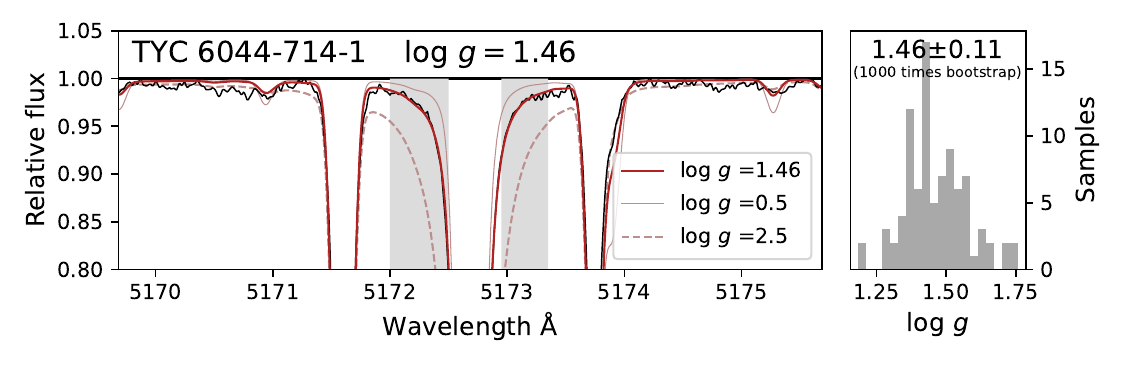}
    \includegraphics[width=0.75\linewidth]{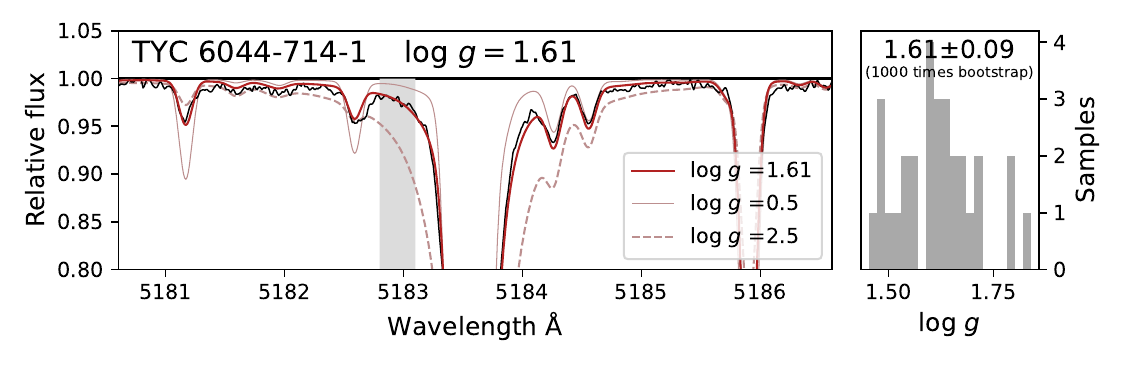}
    \caption{\tiny {Mg triplet line fitting. The observed spectrum is shown in black, while the best-fitting synthetic 1D LTE profiles are overplotted as thicker red lines. The shaded regions indicate the wavelength intervals used in the fitting procedure. The right-hand panels display the distributions of \logg\ values associated with the pixels within these regions. The most probable \logg\ and its uncertainties are derived via bootstrapping.}}
    \label{fig:Mg_triplet}
\end{figure*}

}

\section{Abundances of niobium and thorium}
\label{sec:Nb}

\begin{figure}[!htbp]
    \centering
    \includegraphics[width=0.7\linewidth]{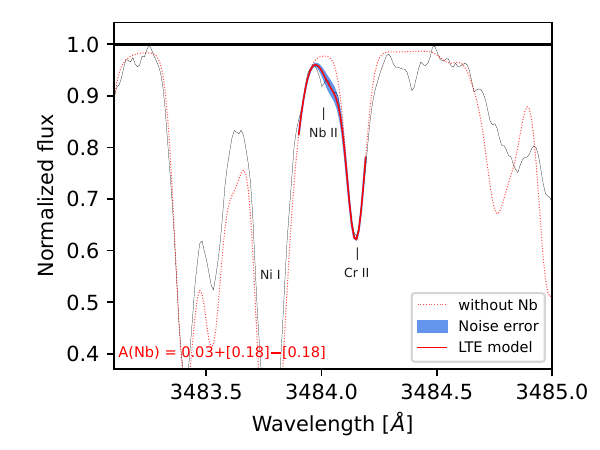}
    \includegraphics[width=0.7\linewidth]{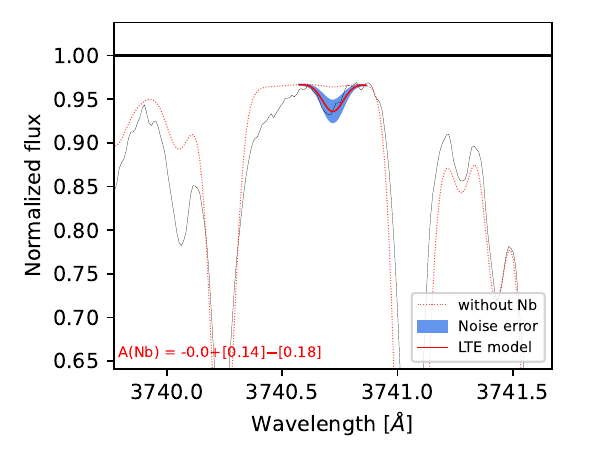}
    \caption{\tiny Nb spectral line synthesis. Symbols and colours  are the same as to those in Fig.~\ref{fig:ba}, but without 1D non-LTE line models.
    }
    \label{fig:Nb}
\end{figure}

\begin{figure*}[!htbp]
    \centering
    \includegraphics[width=0.75\linewidth]{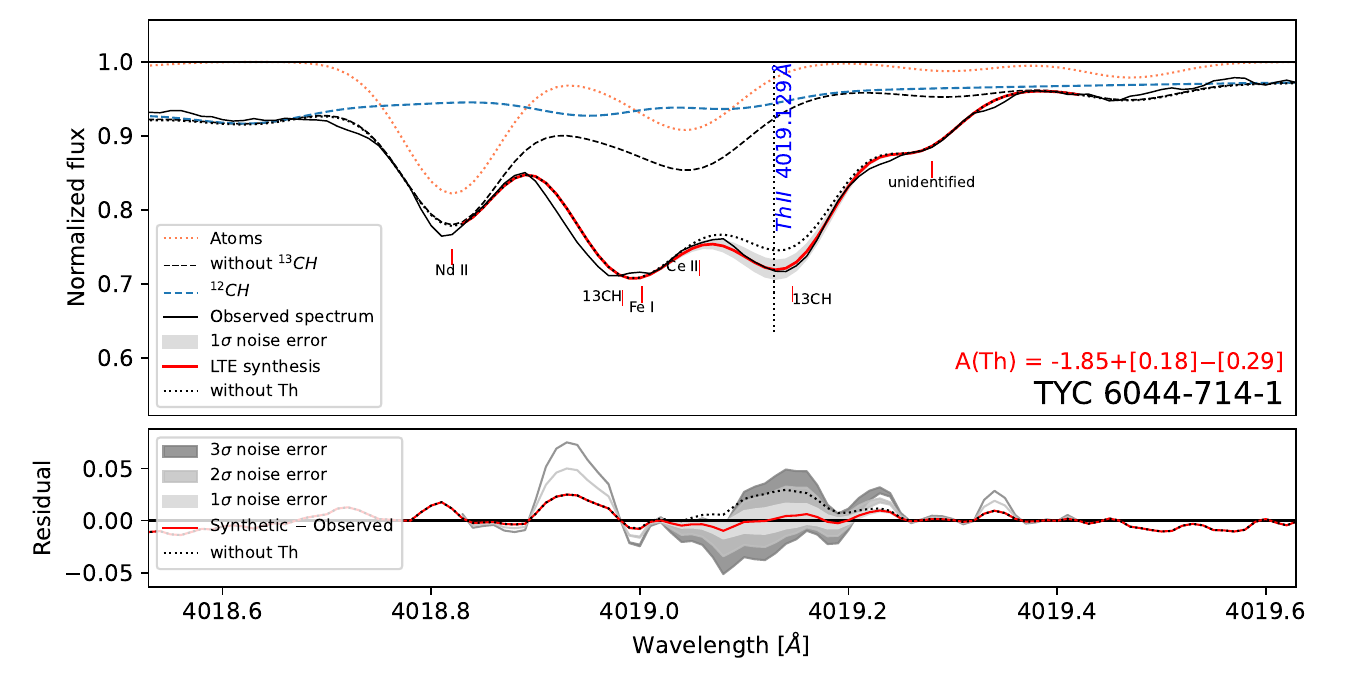}
    
    \caption{\tiny Modelling of the feature around the thorium line at $\lambda$4019.129~\AA. 
    The core of the Th line is marked by the dotted vertical line. Lines of species with significant contributions to the feature are marked in the plot.
    The abundance retrieved from the spectral synthesis including all species (red solid line) is noted at the right bottom part of the plot.
    A synthetic spectrum without Th is represented by the dotted black line.
    The bottom panel displays residuals and 1, 2, and 3$\sigma$ flux errors according to the legends.
    }
    \label{fig:Th}
\end{figure*}

We measured the Nb abundance from the \ion{Nb}{ii} lines at 3484.046 and 3740.72~\AA\ (Fig.~\ref{fig:Nb}). The log~$gf$ values of the lines ($-1.060$ and $-0.310$) were revised by \cite{NI} and agree with former ones in \cite{Hannaford1985A&A...143..447H}, which were used to successfully reproduce the solar meteoritic abundance A(Nb) = 1.41~dex \citep{anders1982GeCoA..46.2363A}. 
Despite some blending, both lines yield consistent abundances (see Figure~\ref{fig:Nb}). 
Biases arising from 1D~LTE modelling and noise effects cannot be ruled out as contributing to an overestimation. {The noise-related uncertainty per line is approximately 0.2~dex, which is adopted in Table~\ref{tab:abundances} to account for potential biases. Elements with $38 \leq Z \leq 56$ may be distinctive between s+r and i+r+s scenarios (Sect.~\ref{sec:patterns}).
Among the six elements in this range with detected lines, only Nb supports the latter, suggesting that its abundance may be affected by noise-induced bias.}

We fitted the feature containing the \ion{Th}{ii} line at 4019.129~\AA. In TYC~6044-714-1, the pseudo-continuum around this line is dominated by $^{12}\mathrm{CH}$ (Figure~\ref{fig:Th}),  therefore we normalised its spectrum to the synthetic flux using the blue and red limits at 4018.528 and 4019.629~\AA. An unidentified line at 4019.28~\AA\ was modelled with a Gaussian.
Although the Th contribution is subtle, the high S/N allows a 2$\sigma$ measurement, as seen in the residuals. The derived abundance depends on the proper modelling of the $^{13}\mathrm{CH}$ feature near 4019.0~\AA, the \ion{Ce}{ii} line at 4019.057~\AA, and the $^{13}\mathrm{CH}$ feature aligned with Th.

\section{Extra Figures}

\begin{figure*}[!htbp]
\centering
\includegraphics[width=0.7\linewidth]{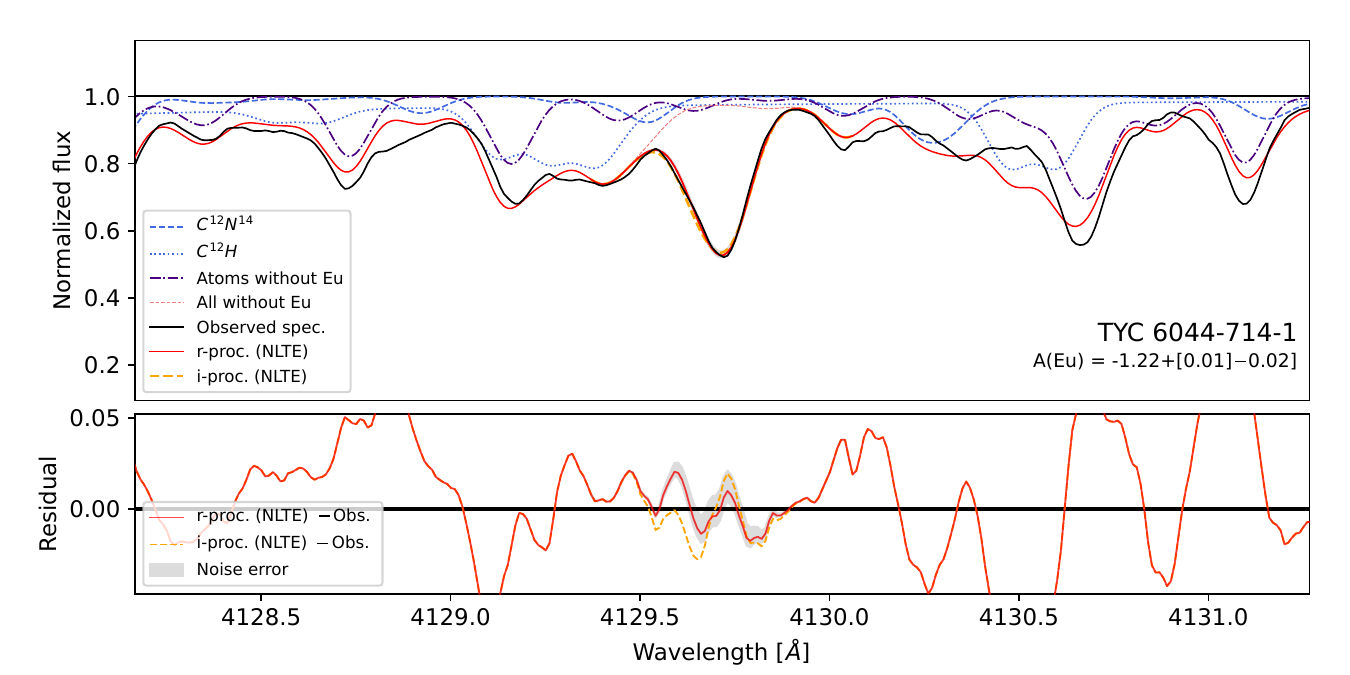}\\
\includegraphics[width=0.7\linewidth]{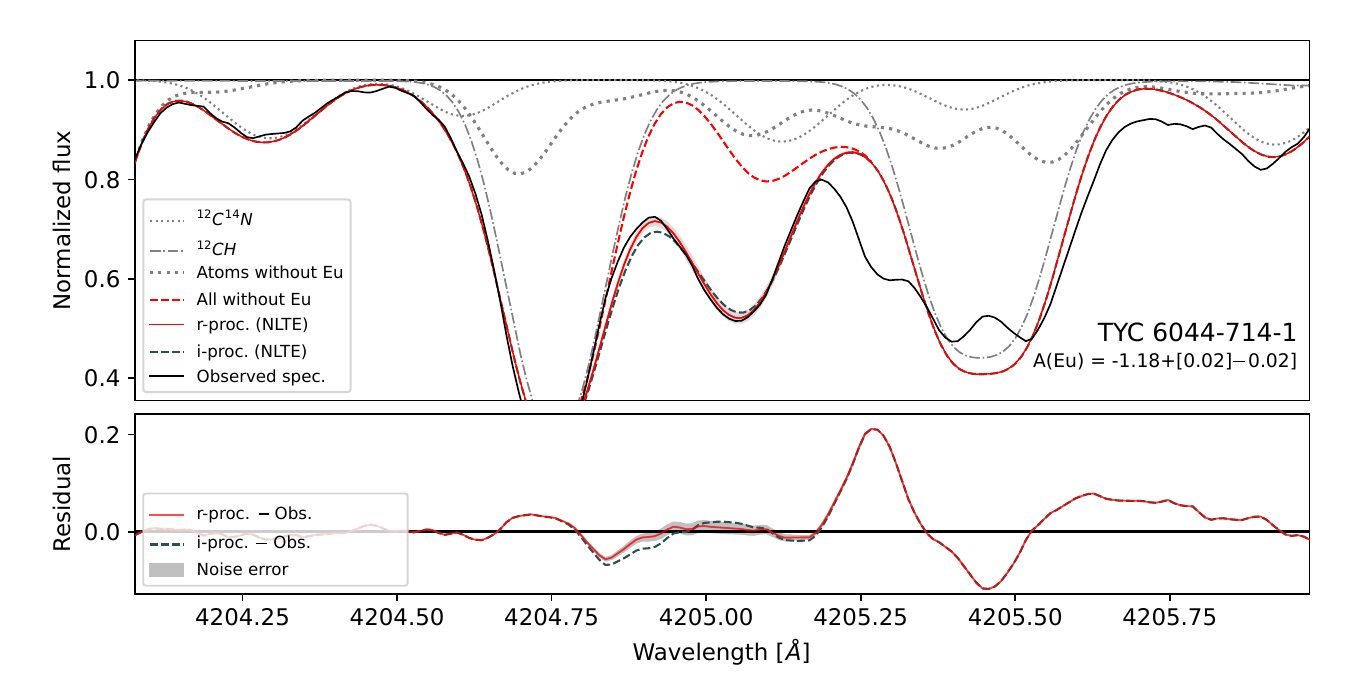}
\caption{{\it Top panel:} Fit of the Eu line at 4129~\AA.
Line profiles from the r- and i-processes from the isotope fractions in Table~\ref{tab:ratios} are represented in different colours as indicated in the legends. Flux contributions from CN and CH molecule features are also plotted. 
{\it Bottom panel:} Similar to top panel but for the Eu line at 4205~\AA.
\label{fig:Eu}}
\end{figure*}   

\begin{figure*}[!htbp]
    \centering
    \includegraphics[width=1\linewidth]{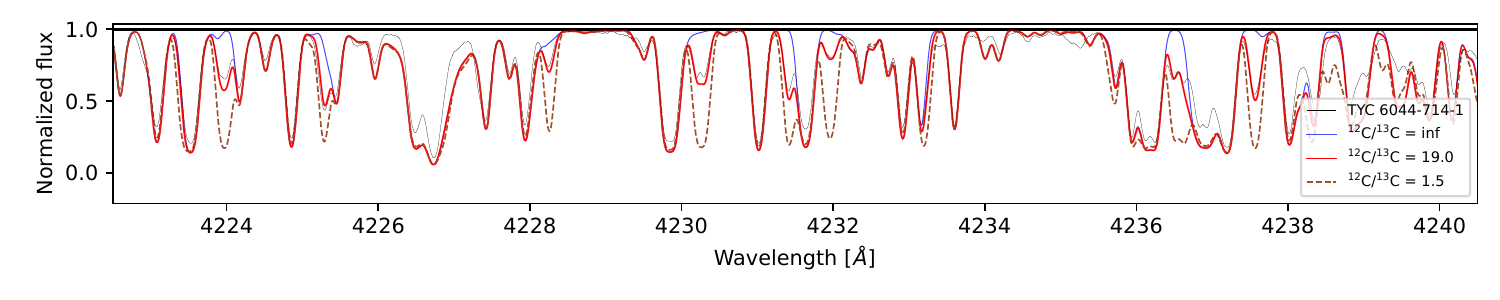}
    \caption{\tiny Section of the band used to determine $^{12}$C/$^{13}$C. The observational spectrum is represented by the black line. Synthetic with extreme $^{12}$C/$^{13}$C fractions are represented by the red dashed and blue lines according to the legends. The synthetic spectrum most similar to the observational one is represented by the red solid line.}
    \label{fig:c13}
\end{figure*}

\begin{figure*}[!htbp]
    \centering
    \includegraphics[width=1\linewidth]{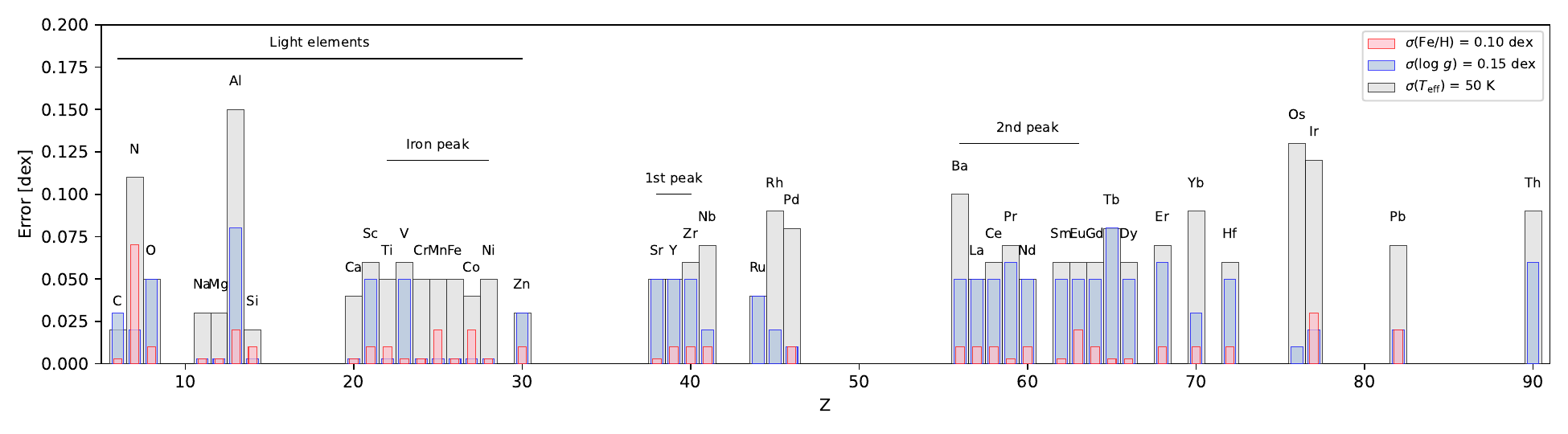}
    \caption{\tiny Elemental abundance variations induced by typical errors of \teff, \logg, and [Fe/H].}
    \label{fig:errors}
\end{figure*}

\begin{figure*}[!htbp]
    \centering
    \includegraphics[width=0.7\linewidth]{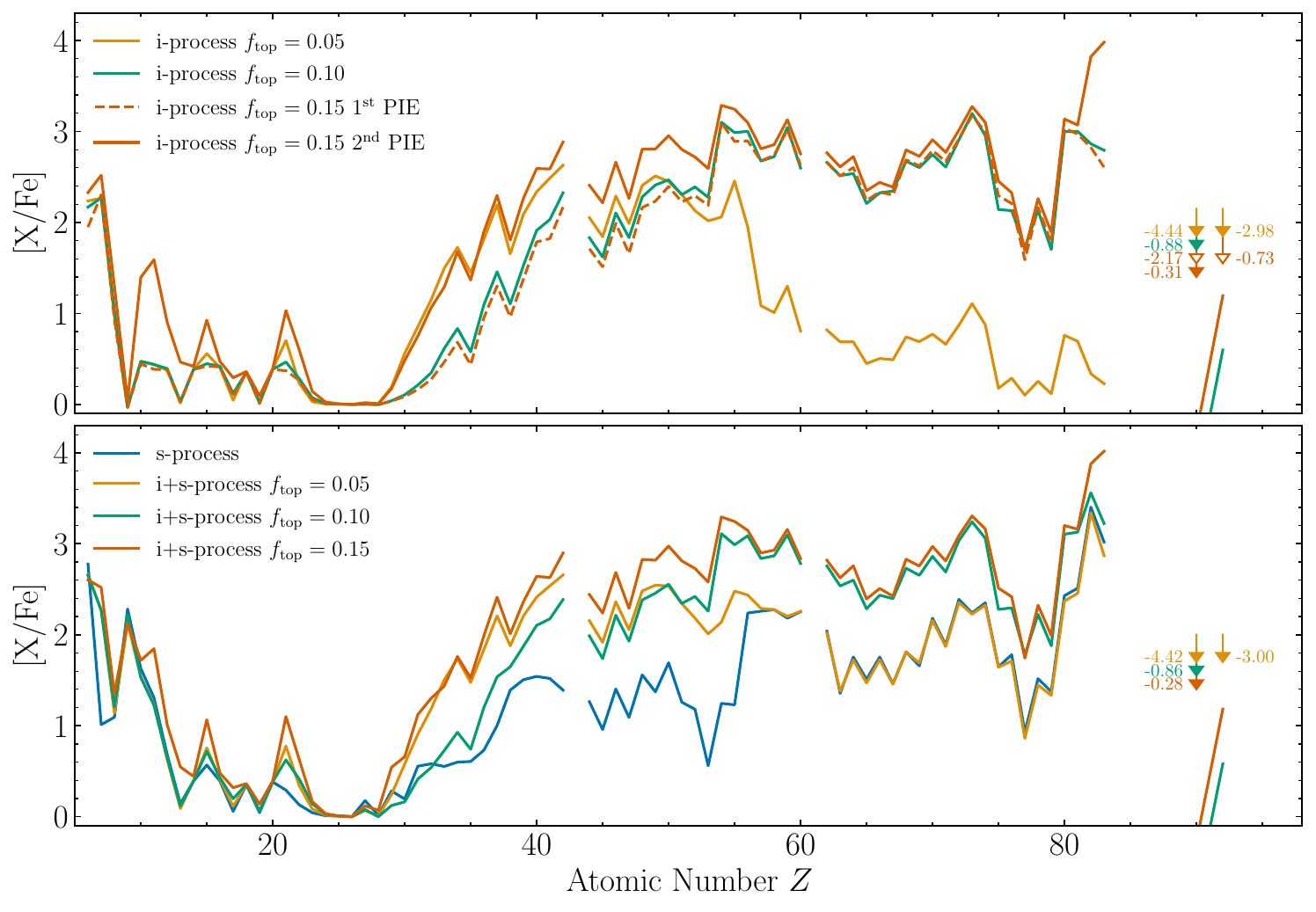}
    \caption{\tiny Final surface [X/Fe] ratios predicted by our AGB models for a $1.5\,M_\odot$ star at $\mathrm{[Fe/H]} = -2.27$ (see text for details).  
    \textit{Upper panel}: Surface abundance patterns obtained immediately after PIEs for models with increasing overshooting efficiency at the top of the convective thermal pulse. 
    The yellow, green, and orange curves correspond to $f_{\mathrm{top}} = 0.05$, $0.10$, and $0.15$, respectively. 
    For $f_{\mathrm{top}} = 0.15$, two PIEs occur; the surface composition after the first PIE is shown as a dashed orange line, while the composition after the second PIE is shown as a solid orange line.  
    \textit{Lower panel}: final surface abundances after the full AGB evolution. The pure s-process model (blue line) results from 15 standard third dredge-up episodes associated with the formation of a $^{13}$C pocket. 
    Mixed i+s models are shown for increasing $f_{\mathrm{top}}$ values and include an initial enrichment caused by one or two PIEs and the associated deep third dredge-up, followed by standard s-process nucleosynthesis during the subsequent 12 thermal pulses.  
    When computing elemental abundances, all unstable isotopes with half-lives shorter than 10~Myr are assumed to have fully decayed into their stable daughter nuclei.
}
    \label{fig:ele_fe}
\end{figure*}

\begin{figure*}[!htbp]
\centering
\includegraphics[width=0.70\linewidth]{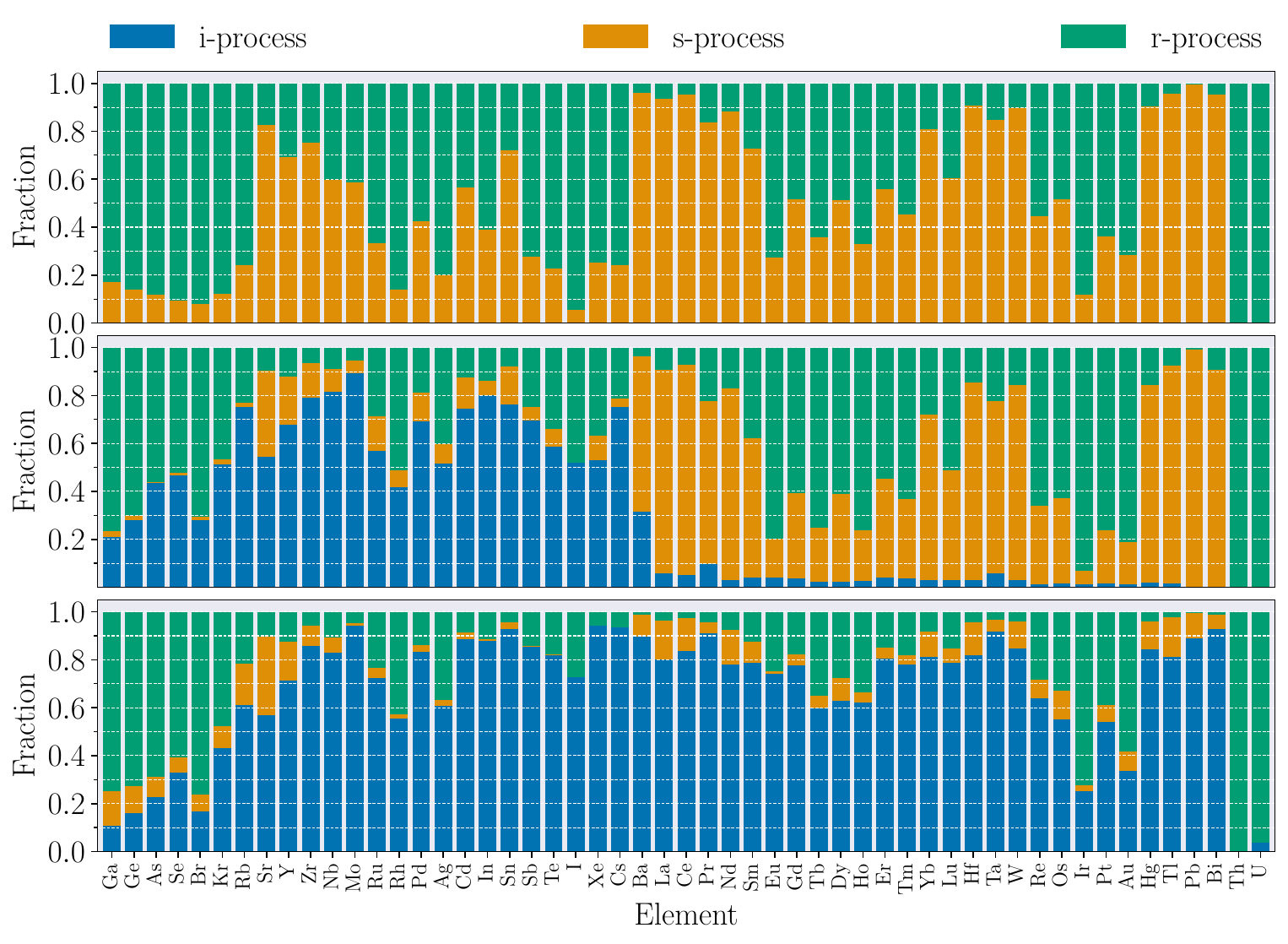}
    \caption{Nucleosynthesis fractions for the s+r model (upper panel) and for the mixed i+s+r models with $f_{\mathrm{top}} = 0.05$ (middle panel) and $0.15$ (lower panel), corresponding to the weakest and strongest i-process cases considered (see Fig.~\ref{fig:ele_fit_r}).
    }
    \label{fig:percentage}
\end{figure*}

\begin{figure*}[!htbp]
\centering
\includegraphics[width=0.75\linewidth]{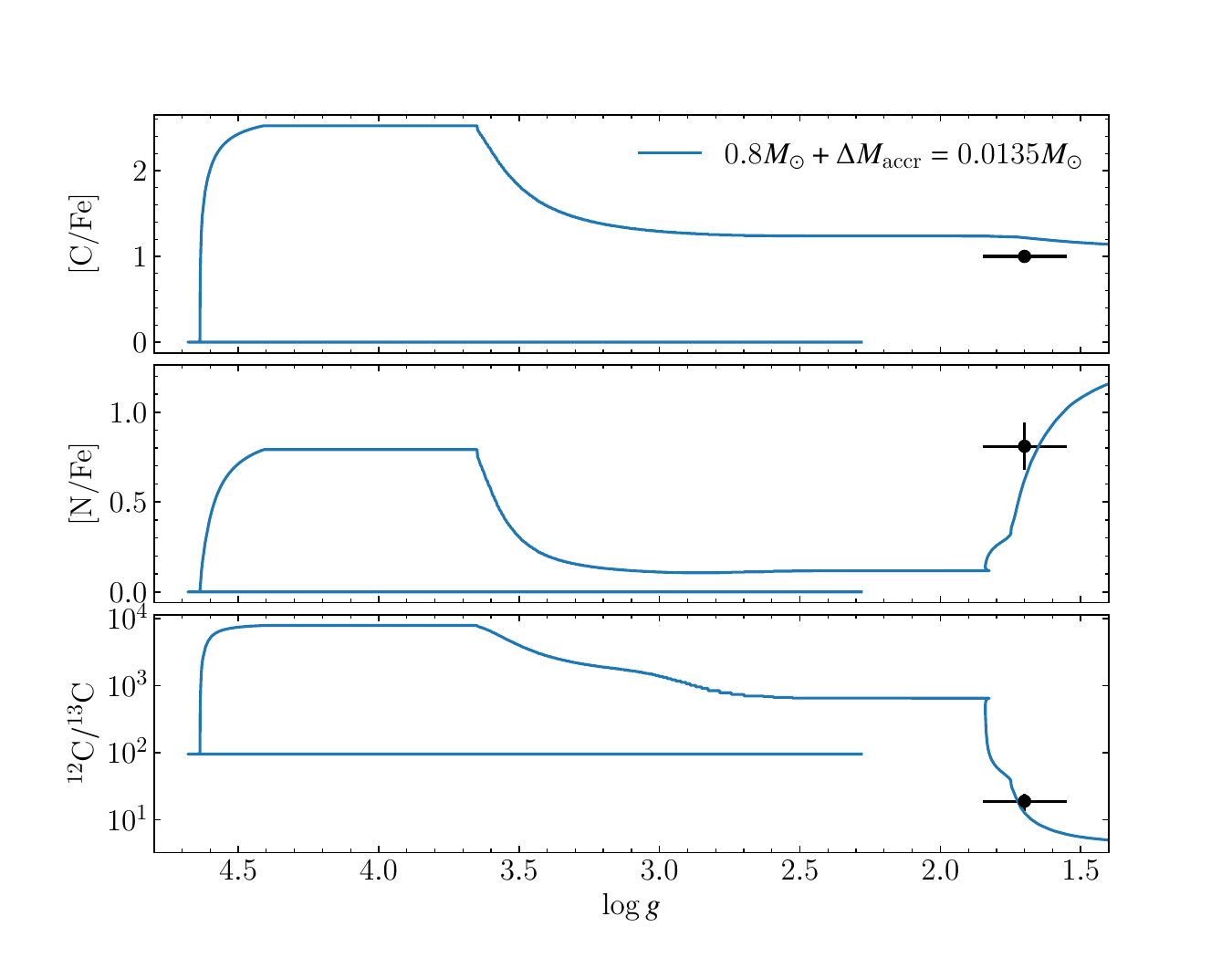}
\caption{Evolution of surface [C/Fe], [N/Fe], and the $^{12}$C/$^{13}$C ratio as a function of surface gravity ($\log g$) for the secondary star model. The star ($0.8\,M_\odot$) is assumed to have accreted $0.0135\,M_\odot$ from the primary AGB companion while on the Main Sequence, a mass required to match the dilution factor ($d_s = 0.027$) derived from the abundance fit of heavy elements (Fig.~\ref{fig:ele_fit_r}). The resulting $0.8135\,M_\odot$ star then evolves through the RGB experiencing mass-loss. Non-convective extra-mixing ($v_{\rm mix} = 1~\mathrm{cm~s^{-1}}$, $T_{\rm mix} = 23$~MK) is implemented after the luminosity bump. The observed values for TYC~6044-714-1 at its current evolutionary stage are indicated by symbols.}
    \label{fig:cn_logg}
\end{figure*}

\end{appendix}

\end{document}